\documentclass[twocolumn,pre,aps,caption,floatfix,superscriptaddress]{revtex4}
\usepackage{graphicx,float}
\usepackage{latexsym}
\usepackage{epstopdf}
\usepackage{amssymb,amsmath}
\usepackage{color}
\usepackage{multirow}
\usepackage[linktoc=all]{hyperref}
\usepackage[caption=false]{subfig}
\makeatletter
\let\Hy@linktoc\Hy@linktoc@none
\makeatother
\usepackage{verbatim}
\usepackage{bm}
\begin{document}

\title{Entanglement dynamics in short and long-range harmonic oscillators}
\author{M. Ghasemi Nezhadhaghighi}
\affiliation{Department of Physics, College of Sciences, Shiraz University, Shiraz 71454, Iran}
\author{ M. A. Rajabpour}\email{rajabpour@ursa.ifsc.usp.br}
\address{Instituto de F\'{\i}sica de S\~{a}o Carlos, Universidade de S\~{a}o Paulo, Caixa Postal 369, 13560-970 S\~{a}o Carlos, SP, Brazil}
\address{Instituto de F\'{\i}sica, Universidade Federal Fluminense,
Av. Gal. Milton Tavares de Souza s$/$n, Gragoat\'{a}, 24210-346, Niter\'{o}i, RJ, Brazil.}

\begin{abstract}  
We study the time evolution of the entanglement entropy in the short and 
long-range coupled harmonic oscillators that have well-defined continuum 
limit field theories. We first introduce a method to calculate the 
entanglement evolution in generic coupled harmonic oscillators after 
quantum quench. Then we study the entanglement evolution after quantum 
quench in harmonic systems that the couplings decay effectively as 
$1/r^{d+\alpha}$ with the distance $r$. After quenching the mass from 
non-zero value to zero we calculate numerically the time evolution of von
Neumann and R\'enyi entropies.  We show that for $1<\alpha<2$ we have a 
linear growth of entanglement and then saturation independent of the 
initial state. For $0<\alpha<1$ depending on the initial state we can 
have logarithmic growth or just fluctuation of entanglement. We also 
calculate the mutual information dynamics of two separated individual 
harmonic oscillators. Our findings suggest that in our system there 
is no particular connection between having a linear growth of entanglement
after quantum quench and having a maximum group velocity or generalized 
Lieb-Robinson bound.

\end{abstract}
\pacs{ 
      {75.10.Pq}
      {03.67.Mn}
      {05.70.Ln} 
     } 
\maketitle
\tableofcontents
\section{Introduction}
 
There has been considerable interest in 
the entanglement properties of quantum many body systems from 
different point of views \textit{i.e.} quantum field theory \cite{holzhey1994geometric}, 
quantum phase transition \cite{vidal2003entanglement} and
quantum information theory \cite{vedral2002role}. 
Among the various measures of quantifying quantum entanglement entropy,
the von Neumann and R\'enyi entropies, have been studied in many different
locally coupled systems 
\cite{calabrese2004entanglement,amico2008,calabrese2009entanglement,Eisert2010,mallayya,ghosh}.
At the quantum critical point, conformal field theory (CFT) predicts that the
entanglement entropy in one spatial
dimension scales logarithmically with respect to the size of the
subsystem $l$. If the total system is infinitely long, 
$S_l = \frac{c}{3}\log l$ where $c$ is the central charge of the 
CFT \cite{holzhey1994geometric}. Away from the critical point, the entropy 
for blocks larger than the spatial correlation length $\xi \sim m^{-1}$ saturates to a value
$S_l = -\frac{c}{3}\log m$ where $m$ is the energy gap of
the system \cite{calabrese2009entanglement}. 
For general free quantum field theories in the case of higher dimensions, 
it has been shown that the von Neumann entropy satisfies the area law.
Based on the area law the entanglement entropy is
proportional to the interface area between the subsystem and its complement
\cite{srednicki1993entropy,braunstein}.

Although many studies have dealt with the short-range systems and numerous
results have been discovered in the last few years, a less investigated situation is 
how (one measure of) quantum entanglement scales with the subsystem size when 
the couplings in the model are long-range. The entanglement entropy of 
one dimensional models with long-range couplings have been studied so far in
the following examples: the Lipkin-Meshkov-Glick 
(LMG) model which in that all spins interact among themselves \cite{Lattore2005,Orus2008}, 
the long-range Ising type model \cite{Dur2005}, the anti-ferromagnetic long-range Ising
chain \cite{Koffel2012} and free fermions with long-range unshielded Coulomb
interaction \cite{Eisert2006}. 
One of the most important features of the long-range systems is
the presence of a power-law like dispersion relation, 
\textit{i.e.} $\omega^2(k) \sim |k|^\alpha$ with $0 < \alpha < 2$ \cite{rajabpour2014}. 
The same is true for those harmonic oscillators that have $1/|r|^{d+\alpha}$ like couplings.
It is widely believed that it should be true 
also for spin systems with long-range interactions \cite{dutta2001,Blanchard2013}. 

In the sense of universality, quantum phase transition in spin systems with long-range interaction
 is generally described, via thermal phase transitions in an equivalent 
 classical ($d+1$)-dimensional spin model with long-range interaction in 
$d$-spatial dimensions and short-range
interaction in the ($d+1$)-th dimension. 
In the specific case where a long-range couplings falling off as 
$1/|r|^{d+\alpha}$, if $\alpha \geq 2$ the corresponding classical ($d+1$)-dimensional spin model describes
the short-range quantum models. It is important to note that 
based on the results were given in Ref. \cite{dutta2001}, 
the upper critical dimension for such systems is given by $d_u = 3\alpha /2$. 
For example, 
the quantum transitions of one-dimensional
quantum transverse Ising model with long-range interaction with values of 
$\alpha\leq 2/3$ are
described by long-range mean-field theory \cite{dutta2001}.
Therefore, the scaling limit of such systems with long-range couplings ``particularly
in some regions'' can be described by 
free fractional field theories.
Recently, in 
 Refs. \cite{Nezhadhaghighi2012} and \cite{Nezhadhaghighi2013}, we studied
the entanglement entropy of a block of long-range coupled harmonic oscillators. 
A crucial point is that we showed the entanglement of the
gapless $1d$ system is logarithmically dependent on the 
subsystem size and we measured the prefactor of the logarithm
in different situations. In higher dimensions we show the 
von Neumann entropy of a partition scales with the surface area.

In principle our formalism is based on the first quantization 
of the free fractional field theories. This method is first discussed by
 Bombelli, et.al \cite{bombelli1986} which they compute the entanglement entropy of
free field theory by using the discrete version of the field
theory which is simply coupled harmonic oscillators
 making it easy to numerically evaluate. 
It is worth mentioning that 
\cite{shiba2013volume} and \cite{pang2014holographic}, studied 
a certain class of non-local field theories whose ground
state entanglement entropy follows a volume law.

The past few years have witnessed a renewed interest in the
the experimental methods that have been
proposed in the study of isolated quantum systems \textit{i.e.}
optically trapped ultra-cold atomic gases,
which they rely on probing the non-equilibrium
properties of the system \cite{ultracoldatomexperiment}. 
From a theoretical point of view, the non-equilibrium dynamics after a 
sudden change (quench) of a parameter
in the hamiltonian of the quantum system is one of the most remarkable 
aspects of these studies. 
For example, consider a quantum system with hamiltonian $H(m)$ depending on
a parameter $m$. The system is prepared
in a pure state (eigenstate) of a given hamiltonian $H(m_0)$. 
Then, at time $t = 0$ the parameter is suddenly quenched 
from $m_0$ to $m$. 
After the quench, there is an
extensive excess in energy 
which appear as quasiparticles that propagate in time 
\cite{calabrese2005evolution}. 

The qualitative, and quantitative, features found for
the non-equilibrium dynamics of quantum systems
 may be understood by the time evolution of the
entanglement entropy. In this investigation one asks how the entanglement
evolves in time after the quantum quench. 
The most remarkable
results that emerged from theoretical investigations is that;  
 if the hamiltonian $H$ governing the time evolution is at a critical point,
 the entanglement entropy $S_A(t)$ grows linearly with time $t$, 
up to $t \sim l/2$ where $l$ is the subsystem size. 
Thereafter it immediately saturates \cite{calabrese2005evolution}. 
The same linear behavior for the growth of entanglement entropy,
 has been also shown for a free quantum field theory in $d+1$-dimensions 
\cite{abajo2010holographic,liu2014entanglement}.

As it is well known, a lot of numerical and
theoretical works have been done seeking to understand
the mechanism of the time evolution of the quantum entanglement in
quantum systems \cite{plenio2004dynamics,perales2005manipulating,
ghahari2007dynamics,kim2013ballistic,jurcevic2014observation,
unanyan2010entanglement,de2006entanglement, canovi2014dynamics,
alba2014entanglement}. Recent theoretical works have investigated 
the existence of quasiparticles or long-
wavelength propagating modes such as acoustic sound following a quantum
quench \cite{calabrese2005evolution,
eisert2006general}. In real physical systems, with local interaction,
 one might expect that, there
is an upper limit on the velocity of propagating information (such as energy quasiparticles).
 This limit is a consequence of the Lieb-Robinson (LR)
bounds \cite{lieb1972finite}. Based
on the LR theorem, the propagation of perturbations, 
 cannot spread faster than Lieb-Robinson velocity. 
 The existence of such bound corresponds with a horizon like region defined by
this velocity, which inside of it the correlations are non-zero and
 outside they are exponentially suppressed \cite{cramer2008locality}.
 This effect in non-relativistic quantum many-body systems is usually called
quasi-locality which plays the same role as the velocity of light
in the Lorentz invariant field theories.
 Lieb-Robinson bounds can be used to show that there exists an upper bound
for the amount of entanglement that can be produced in a finite time after quench
 which increases linearly in time \cite{bravyi2006lieb,eisert2006general}.

An important question to ask about entanglement dynamics is what
kind of scaling with time can one expect for the entanglement entropy of quantum 
systems with long-range couplings? 
The first numerical studies of entanglement dynamics 
generated by long-range interactions appeared in 
 Refs. \cite{schachenmayer2013entanglement}. 
It is worth mentioning that there are some considerable experimental results 
available about the 
propagation of the correlations through the quantum many-body system 
with long-range interactions \cite{richerme2014non,jurcevic2014observation}. 
 Specifically,
 in Ref. \cite{schachenmayer2013entanglement} they explored 
 entanglement entropy and mutual information dynamics after a quantum quench in
the transverse field Ising model with long-range
interactions, which  can be described by the hamiltonian
$H = \sum_{i<j}J_{i,j}\hat{\sigma}^x_i\hat{\sigma}^x_j +B\sum _i \hat{\sigma}^z_i $
 and the couplings decay with distance like 
$J_{i,j}\propto |i-j|^{-\sigma}$ ($\sigma >0$). Interestingly, 
for those long-range interactions with
$\sigma>1$, they found that the behavior of the time evolution of
entanglement entropy is qualitatively similar
to nearest-neighbor interactions. 
Remarkably, they also found that, for interaction exponent $\sigma <1$, 
the growth of entanglement is only logarithmic.

It is worth mentioning that there are also some interesting results 
available to extend the Lieb and Robinson theory for 
quantum systems with power-law ($\sim 1/r^{\sigma}$) decaying
 couplings with $\sigma >d$ \cite{nachtergaele2006lieb,hastings2006spectral}. 
Unlike the systems with only short-range interactions,
the locality picture for quantum systems with long range couplings 
when the interaction exhibits a power-law decay with an exponent $\sigma <d$, 
is no-longer true. There is no concrete result for the
special behavior of the growth and propagation of
entanglement entropy in quantum systems with long-range couplings.
Currently, the question of extension of Lieb-Robinson theorem to different quantum systems 
with long-range interactions is a subject of active theoretical 
and numerical debate \cite{hauke2013spread,eisert2013breakdown,gong2014persistence,
metivier2014spreading}.

The purpose of this manuscript is to describe an efficient
computational method to study the time evolution of
the entanglement entropy for coupled long-range harmonic oscillators in any dimension.
 The structure of the rest of the paper is as follows. In section 
 \ref{definition sec} we will first introduce our system. 
 In section \ref{qq section} we introduce a method
 to numerically evaluate the time evolution of 
von-Neumann and R\'enyi entropies in harmonic oscillators and
 we outline the setup for the global quench which is then
applied to different situations in the following sections. In
section \ref{SRHO EEDynamics sec} we numerically study the entanglement entropy dynamics
for harmonic oscillators with short-range couplings. 
In section \ref{LRHO EEDynamics sec} we evaluated the same analysis when the harmonic system
is long-range. In section \ref{initial state sec} we analyzed the initial state effect on the
entanglement dynamics. In section \ref{mutual section}, we studied different
aspects of the mutual information propagation in long-range
harmonic oscillators after global quench. 
Finally, in section \ref{conc sec} we provide a conclusion. 
To be self-explanatory, in the appendix
we will give more details about our numerical methods.

\section{Basic Definitions}\label{definition sec}

Consider a one-dimensional system of $N$
bosonic oscillators. They are coupled by a quadratic hamiltonian of the form
\begin{equation}\label{harmonicOsc}
\mathcal{H}=\frac{1}{2}\sum_{n=1}^{N}\pi_n^2+\frac{1}{2} 
\sum_{n,n^\prime=1}^{N}\phi_{n} K_{nn^\prime}\phi_{n^\prime}~,
\end{equation}
where the kernel matrix ${K}$ is real and positive semi-definite due to the
hermiticity and $\lbrace \phi_i  \rbrace$ is a scalar field. The normalized
 ground state wave function for such system is given as gaussian state
 \begin{equation} \label{GroundSwave} 
\Psi_0(\lbrace \phi \rbrace) \propto 
(\mathrm{det} \Gamma)^{\frac{1}{4}}\exp \lbrace-\frac{1}{2} 
\sum_{n,n^\prime=1}^{N}\phi_{n} 
\Gamma_{nn^\prime}\phi_{n^\prime} \rbrace,
\end{equation}
where $\lbrace \phi \rbrace$ denotes the collection of all $\phi$'s, 
one for each oscillator and $\Gamma = {K}^{1/2}$. 
Having the solution of the
ground state wave function one can calculate the reduced
density matrix of a block of $l$ oscillators. 
Consider a system which is divided into two subsystems $A$ and $B$ with sizes
$l$ and $N-l$ respectively. 
Suppose the whole system is in a pure quantum 
state, with density matrix $\rho$. One can obtain $A$'s reduced density matrix
 by tracing out the remaining degrees of
freedom
$\rho_A = \mathrm{tr}_B \rho$.

There are several measures of entanglement between parties of a
 closed system, examples being the von Neumann and R\'enyi entropies. 
The entanglement entropy associated to the local density matrix $\rho_A$
 is just the von Neumann entropy
\begin{equation} \label{vonNeumaan} 
S_A = -\mathrm{tr}(\rho_A \log(\rho_A))~.
\end{equation}
The von Neumann entropy is the most well-known member of a
more general family of entanglement entropies, the so-called
Re\'nyi entropies, defined as
\begin{equation} \label{Renyi} 
S_n(A) = \frac{1}{1-n}\log(\mathrm{tr} \rho_A^n), \hspace{0.5cm}
 n\geq 0, \hspace{0.5cm} n \neq 1~.
\end{equation}
The R\'enyi entropy reduces to the von Neumann entropy when $n\rightarrow 1$ 
($S_A= \lim _{n\rightarrow 1}S_n(A)$).

In the next section we will present a detailed study of the 
time evolution of von Neumann and R\'enyi entropies that results after 
a quantum quench in a quantum system.

\section{Quantum quench}\label{qq section}
The purpose of this section is to describe an efficient
 method to study the time-dependent behavior of the 
entanglement entropy after a global changing of a parameter in the system
in any spatial dimension, also known as quantum quench.

Here we consider first that the system is prepared in the ground state of a 
given gapped hamiltonian $H(m)$, where $m$ is tuneable
parameter. At time $t = 0$
the parameter $m_0$ is changed suddenly to a different value
$m$. For $t>0$ the system is allowed to evolve with a 
different hamiltonian $H(m)$. In this study we will consider 
  $m = 0$. To study the entanglement dynamics for the 
systems with global quench the only thing that
we need is to start with an arbitrary gaussian state and then see its 
evolution with respect to the new hamiltonian. Since the hamiltonian of 
the system is quadratic, one can expect that, the time-evolved 
 state remains gaussian in the form
  \begin{equation} \label{GroundSwave} 
\Psi(\lbrace \phi \rbrace,t) \propto (\mathrm{det} 
\tilde{A})^{\frac{1}{4}}\exp \lbrace- \frac{1}{2}
\sum_{n,n^\prime=1}^{N}\phi_{n} 
A_{nn^\prime} (t)\phi_{n^\prime} \rbrace,
\end{equation}
where a tilde denotes the real part. It is easy to show that
 $A(t)$ obeys the Riccati equation
\begin{eqnarray}
i\frac{\partial A}{\partial t} = A^2 -{K}, \hspace{1cm} A(0) = \Gamma.
\end{eqnarray}
Its solution is given by
\begin{eqnarray}\label{A matrix with time}
A(t) = K^{1/2}\frac{\cos(t K^{1/2})\Gamma +i K^{1/2}\sin(t K^{1/2})}
{\cos(t K^{1/2})K^{1/2} +i\sin(t K^{1/2})\Gamma},
\end{eqnarray}
where $K$ is the hamiltonian after quench and the matrix 
$\Gamma = K_0^{1/2}$ is prepared at $t=0$. 

Having the above equations one can simply use the following techniques 
to calculate the dynamics of entanglement entropy. Suppose we cut a 
harmonic chain at the boundaries between two subsystems $A$
and $B$. Now consider
\begin{eqnarray}\label{X_A P_A}
A^{-1}(t)=\left(
\begin{array}{cc}
  X_{A} & X_{AB} \\
  X^{ T}_{AB} & X_{B}
 \end{array}\right), \hspace{0.5cm} 
 A(t)=\left(
\begin{array}{cc}
  P_{A} & P_{AB} \\
  P^{ T}_{AB} & P_{B} .
 \end{array}\right)~,
\end{eqnarray}
where $X_A$ ($P_A$) is $l \times l$ and $X_B$ ($P_B$) is $(N-l) \times (N-l)$ matrix. 
Then one can write the 
reduced density matrix as \cite{unanyan2010entanglement}
\begin{eqnarray}
\rho_B(\lbrace\phi^1\rbrace_B; \lbrace\phi^2\rbrace_B) & \propto 
	\exp \Biggl\{ -{1\over 2}(\lbrace\phi^1\rbrace_B~ 
	\lbrace\phi^2\rbrace_B) \times \nonumber \\ 
&\left( \begin{array}{cc}
  \mathcal{C} & 2\mathcal{D} \\
  2\mathcal{D^*} & \mathcal{C^*}
   \end{array}\right) 
   \left(\begin{array}{cc}
  \lbrace\phi^1\rbrace_B \\
  \lbrace\phi^2\rbrace_B
 \end{array}\right) \Biggr\}~,
\label{gaussdm}
\end{eqnarray}
where we have
\begin{eqnarray}\label{C and D}
\mathcal{C}&=&P_B-\frac{P^T_{AB}(\tilde{P}_A)^{-1}P_{AB}}{2}~, \nonumber \\
\mathcal{D}&=&-\frac{P^T_{AB}(\tilde{P}_A)^{-1}P^*_{AB}}{4}~,
\end{eqnarray}
with $\tilde{M}=(M+M^*)/2$. 

We will now follow the method was first introduced in \cite{bombelli1986}
and then elaborated in \cite{Callan1994}, to get the
von Neumann and R\'enyi
entropies. Using Eq. (\ref{gaussdm}) 
and the method followed in \cite{Callan1994}
we investigated the trace of $n$-th power
 of the reduced density matrix as a functional integral:
\begin{eqnarray}
{\rm tr}\rho_B^n &=
\int \left[\prod_{m\in(B)} d\phi_m\right] \exp
 \Biggl\{ -(\lbrace\phi^1\rbrace_B \cdots \lbrace\phi^n\rbrace_B) 
 \times \nonumber \\ 
 &{\cal M}_n
 \left( \begin{array}{c}
  \lbrace\phi^1\rbrace_B \\
  \vdots\\
  \lbrace\phi^n\rbrace_B
 \end{array} \right) \Biggr\}~,
 \label{gaussdm 1}
\end{eqnarray}
where ${\cal M}_n$ is the matrix
\begin{equation}
{\cal M}_n = 
\left( \begin{array}{cccccc}
  \tilde{\mathcal{C}} & \mathcal{D} & 0 & \ldots & 0 & \mathcal{D}^*\\
  \mathcal{D}^* & \tilde{\mathcal{C}} & \mathcal{D} & \ldots & 0 & 0 \\
  0 & \mathcal{D}^* & \tilde{\mathcal{C}} & \ldots & 0 & 0  \\
  \vdots & \vdots & \vdots && \vdots & \vdots \\
  0 & 0 & 0 & \ldots & \tilde{\mathcal{C}} &\mathcal{D} \\ \\
\mathcal{D} & 0 & 0 & \ldots & \mathcal{D}^* & \tilde{\mathcal{C}}
 \end{array}\right).
\end{equation}
Two matrices $\mathcal{C}$ and $\mathcal{D}$ are defined in the Eq. (\ref{C and D}). 
 It is worth mentioning that the Eq. (\ref{gaussdm 1})
 is a gaussian functional integral, therefore it
 is possible to find the result
 $\mathrm{tr}\rho ^n = 1/\sqrt{\mathrm{det} \mathcal{M}_n}$ but
 there is no straight-forward way to compute this determinant
 as a function of $n$, however, one can find the result by rescaling the 
 density matrix. Consider the new matrix
\begin{equation}
{\cal M}'_n = 
\left(\begin{array}{cccccc}
  2 & -\mathcal{E} & 0 & \ldots & 0 & -\mathcal{E}^*\\
  -\mathcal{E}^* & 2 & -\mathcal{E} & \ldots & 0 & 0 \\
  0 & -\mathcal{E}^* & 2 & \ldots & 0 & 0  \\
  \vdots & \vdots & \vdots && \vdots & \vdots \\
  0 & 0 & 0 & \ldots & 2 &-\mathcal{E} \\ \\
-\mathcal{E} & 0 & 0 & \ldots & -\mathcal{E}^* & 2
 \end{array}\right),
\end{equation}
where $\mathcal{E}=-2 (\tilde{\mathcal{C}})^{-1}\mathcal{D}$. 
By taking into account
$\mathrm{det} \mathcal{M}^\prime_n = 2^n \mathrm{det} \mathcal{M}_n/ (\mathrm{det} \tilde{\mathcal{C}})^n$
 the trace of the power $n$ of the reduced density matrix is then
\begin{eqnarray}\label{trace}
{\rm tr}\rho_B^n \propto\frac{1}{\sqrt{\det \mathcal{M}'_n}}.
\end{eqnarray}
Let us note that it is possible to consider the matrix $\mathcal{E}$ 
as a parameter to explicitly diagonalize
 the matrix $\mathcal{M}'_n$ and give the result as  
\begin{eqnarray}\label{trace}
\det \mathcal{M}'_n=\prod_{j=1}^n\Big{(} 2-2\mathcal{E}
\omega_j-2\mathcal{E}^*\omega_j^{n-1}\Big{)}~,
\end{eqnarray}
with $\omega_j=e^{2\pi ij/n}$. One can also write the above equation
 in the form
\begin{eqnarray}\label{trace}
\det \mathcal{M}'_n=\prod_{j=1}^n\Big{(} 2-2|\mathcal{E}| \cos(\frac{2\pi j}{n}+\theta)\Big{)}~,
\end{eqnarray}
where $\mathcal{E}=|\mathcal{E}|e^{i\theta}$. Define $|\mathcal{E}|=\frac{2\xi}{\xi^2+1}$ and use the equality
\begin{eqnarray}\label{produc equality}
\prod_{j=0}^{n-1}\Big{(}1+\xi^2&-2\xi\cos(\theta+\frac{2\pi j}{n})\Big{)}=\nonumber \\
& 1-2\xi^n\cos (n\theta)+\xi^{2n}~.
\end{eqnarray}
As has been pointed out above, 
 the operator $\mathcal{E}$ was considered as a number to explicitly 
 diagonalize the matrix $\mathcal{M}'_n$ and 
 calculate the Eq. (\ref{trace}). But $\mathcal{E}$
 is actually an operator, not a number, but we may
diagonalize it and apply the above argument for all the eigenvalues. 
Finally we will have
\begin{eqnarray}\label{trace final}
\det \mathcal{M}'_n=2^n\prod_{\mathcal{E}_j} 
\frac{1-2\xi^n(\mathcal{E}_j)\cos (n\theta_j)+\xi^{2n}(\mathcal{E}_j)}
{(1+\xi^2(\mathcal{E}_j))^n}~,
\end{eqnarray}
with $\mathcal{E}_je^{i\theta_j}$ as the eigenvalue of the matrix 
$\mathcal{E}$. We can now calculate the entanglement entropy
by the following formula
\begin{eqnarray}\label{entanglment entropy}
S_A=(-\frac{d}{d n}+1)\log {\rm tr}\rho_A^n|_{_{n=1}}.
\end{eqnarray}
Using Eq. (\ref{trace}), one can write the entropy as a
sum over contributions from each eigenvalue of $\mathcal{E}$
\begin{eqnarray}\label{entanglment entropy formula}
S_A=-\sum_{j=1}^l \Biggl{[} \frac{\xi_j\log \xi_j \cos\theta_j-\xi_j\theta_j
\sin\theta_j-\xi^2_j\log \xi_j}
{1-2\xi_j\cos\theta_j+\xi^2_j}\nonumber\\
+\frac{1}{2}\log(1-2\xi_j\cos\theta_j+\xi^2_j)\Biggr{]}~,
\end{eqnarray}
where $l$ is the size of the subsystem $A$. We should note that
 Eq. (\ref{entanglment entropy formula}) at $t=0$ follows 
\begin{eqnarray}\label{renyi entropy formula}
S_A=-\sum_{j=1}^l \left[ \ln (1-\xi _j) +\frac{\xi_j}{1-\xi_j}\ln \xi_j \right] ~,
\end{eqnarray}
in agreement with Ref. \cite{Callan1994}.

It is also straightforward to write the R\'enyi entropy
$S_n$ in terms of $\xi_j$ and $\theta_j$ as:
\begin{eqnarray}\label{renyi entropy formula}
S_n&=\frac{1}{2(n-1)}\sum_{j=1}^l \Big{(} 
\log(1-2\xi_j^2\cos (n\theta_j)+\xi^{2n}_j)\nonumber \\
& -n\log(1-2\xi_j\cos\theta_j+\xi^2_j) \Big{)}~.
\end{eqnarray} 
Note that, the findings presented in 
Eqs. (\ref{entanglment entropy formula}) and (\ref{renyi entropy formula}) are
our main result of this section.
In the next sections we will numerically evaluate the time evolution of 
von Neumann and R\'enyi entropies using 
Eqs. (\ref{entanglment entropy formula}) and (\ref{renyi entropy formula})
 and
compare the numerical results with the analytical predictions.
 
\section{Harmonic systems with local couplings}\label{SRHO EEDynamics sec}
\begin{figure}
\centerline{
\includegraphics[scale=0.5]{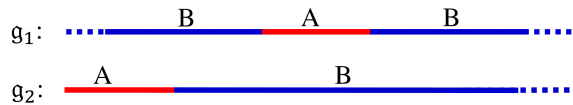}
}
\caption{(Color online) Different configurations of systems and
subsystems for entanglement dynamics with global quenches.
}
\label{configglobal}
\end{figure}

In this section we discuss numerical evaluation of the dynamics of entanglement entropy
 $S_A(t)$ and the R\'enyi entropy $S_n(t)$ for one dimensional 
 harmonic oscillators with short-range couplings. 
 This kind of spring-like coupling corresponds to
the kernel matrix 
${K} = \mathrm{tridiag}\lbrace -1,2+m^2 ,-1\rbrace$ 
in Eq. (\ref{harmonicOsc}),
 where it has nonzero elements only in the main diagonal and the first
diagonals below and above the main one. The parameter $m$ plays 
the role of the mass of the field theory. This kind of spring-like coupling 
can be considered as the discretization of a massive bosonic continuum
theory given by the Klein-Gordon hamiltonian. 
In the continuum limit the Eq. (\ref{harmonicOsc}) for the short-range 
harmonic oscillators has the following
form:
\begin{equation}\label{harmonicOscCont}
\mathcal{H}=\frac{1}{2}\int dx \left[ (\partial_t \phi)^2 + 
(\nabla \phi)^2 +m^2 \phi^2 \right]. 
\end{equation}
To determine the time evolution of the the von Neumann entropy $S_A(t)$ and
the R\'enyi entropy $S_n(t)$ for the short range harmonic oscillators we used 
Eqs. (\ref{entanglment entropy formula}) and (\ref{renyi entropy formula}). 
In this respect, we follow the method explained in the last section.
Let us first discuss different configurations
for the system and also subsystem that we have used
in our study. We consider two different
configurations of the system depicted in Fig. (\ref{configglobal}) 
among which we will calculate the entanglement:
\begin{itemize}
\item[$\mathfrak{g}_1$:]System is very large and $A$ is a small sub-system
with length $l$.
\item[$\mathfrak{g}_2$:]System is a semi-infinite line, 
$\left[0, \infty \right)$, and 
the subsystem $A$ is the finite interval $\left[0, l\right)$.
\end{itemize}

Let us now focus on the case where the subsystem is very small
with length $l$ and the system size $N$ is very large ($\mathfrak{g}_1$).
For this case
 we have used periodic boundary conditions so that $K(r) = K(r+N)$.

We are going to study the entanglement evolution after quantum quench.
 These are situations in which the system is initially 
prepared in the ground state of a hamiltonian
of a gapped field theory ($m_0\neq 0$ in Eq. \ref{harmonicOscCont}). 
At $t = 0$, the hamiltonian suddenly becomes massless, and
the system is then allowed to evolve undisturbed for $t > 0$. After 
the quench there is an energy excess in comparison with the ground state of the 
final hamiltonian $H$, which appear as quasiparticles
 and propagate to the entire system in time \cite{calabrese2009entanglement}. 
 One of the interesting feature of 
the entanglement dynamics is the behavior for very short time. 
In this limit, the evolution of entanglement entropy
is quadratic growth in time \cite{unanyan2010entanglement,liu2014entanglement}
 and follows
\begin{eqnarray}\label{quadratic entropy dynamics formula}
S_A(t\ll t^*) \equiv \kappa_2 t^2~,
\end{eqnarray}
where $\kappa_2$ is a function of the mass parameter $m_0$.

After that, conformal field
theory predicts the following formula for the dynamics of entanglement
entropy \cite{calabrese2009entanglement}:
\begin{eqnarray}\label{entropy case 1}
S_A(t) = - \frac{c}{3}\log m_0 + \Bigg{\lbrace}
  \begin{array}{l l}
    \frac{\pi m_0c }{6}t & \quad t<t^*\\
    \\
    \frac{\pi m_0c }{12}l & \quad t>t^*
  \end{array}
  ~,
\end{eqnarray}
where $m_0$ is the mass gap in the initial state and $t^*=l/2$ 
is the saturation time. This behavior has a simple
explanation in terms of quasiparticles excitations \cite{calabrese2009entanglement}.
The initially excited state acts as
a source for quasiparticle excitations. Highly entangled quasiparticle pairs
 created at a given point in space, travel with their maximum expected velocity $v$.
A quasiparticle of velocity $v$ produced at $x$ is therefore at $x+vt$ at time $t$. 
Now consider a pair of quasiparticles emitted at $t=0$ 
from the midpoint of the subsystem $A$ ($A$ is an interval of length $l$)
and travel with $v=1$. They arrive simultaneously
 at a point $x^\prime\in A$ and $x^{\prime\prime} \in B$ 
 at the same time $t=l/2$ \cite{calabrese2009entanglement}. 
So the entanglement entropy between $x$ and $x^\prime$, saturates if $t>l/2$. 
It is worth mentioning that quasiparticles emitted from far separated points
(with the distance much larger than the correlation length $\xi=m_0^{-1}$)
 are incoherent. 
    
\begin{figure}[t]
\centerline{
\includegraphics[scale=0.35]{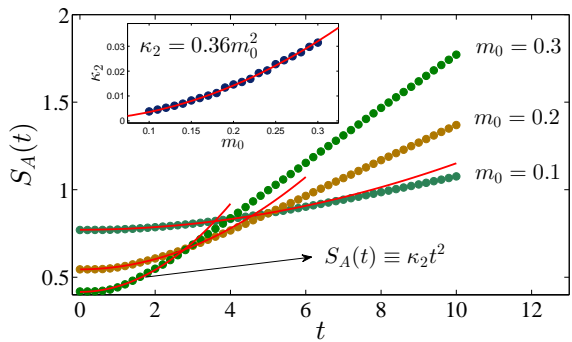}
}
\caption{(Color online) 
Quadratic growth in time for $S_A(t)$ in the region $t\ll t^*$ for the
 short-range harmonic oscillators with the configuration $\mathfrak{g}_1$. 
The total size of the system is $N = 500$, and the subsystem size $l = 60$. 
The solid red line corresponds to the $S_A(t) \equiv \kappa_2 t^2$. Inset: 
The value of the scaling parameter $\kappa_2$ as a function of mass
 parameter $m_0$.}
\label{dynamic case 1 small time}
\end{figure} 

 The dynamics of R\'enyi entropies is given
 by the following formula:
 \begin{eqnarray}\label{renyi entropy scaling case 1}
S_n(t) = - \frac{c_n}{3}\log m_0 + \Bigg{\lbrace}
  \begin{array}{l l}
    \frac{\pi m_0c_n }{6}t & \quad t<t^*\\
    \\
    \frac{\pi m_0c^\prime_n }{12}l & \quad t>t^*
  \end{array}
  ~,
\end{eqnarray} 
where $c_n =c^\prime _n= \frac{c}{2}(1+1/n)$ and $c=1$.

Now we numerically evaluate the dynamics of von Neumann and R\'enyi
entropies for the case $\mathfrak{g}_1$, where the system
is a chain of $N$ coupled harmonic oscillators, and the subsystem is a very small 
region with length $l\ll N$ 
(see Fig. (\ref{configglobal})).
Here we consider a one-dimensional 
system of $N$ bosonic oscillators and we will measure the
eigenvalues of the matrix $\mathcal{E}$ numerically. In order to calculate 
$\xi$ and $\theta$, we first need to construct the matrices 
$\Gamma$ and ${K}$. The matrix ${K}$ is the 
hamiltonian matrix after quench and the matrix $\Gamma = {K_0}^{\frac{1}{2}}$ 
corresponds to the hamiltonian matrix which is prepared at time $t = 0$. 
The matrix $A(t)$ can then easily be 
calculated from Eq. (\ref{A matrix with time}). Having the matrices
$A(t)$ and $A^{-1}(t)$ we can now calculate the matrix 
$\mathcal{E}=-2 (\tilde{\mathcal{C}})^{-1}\mathcal{D}$ from 
 $\mathcal{C}$ and $\mathcal{D}$ (see Eq. (\ref{C and D})).
 The numerical results are summarized in the
following.

Let us first address the behavior of the von Neumann entropy at the very short time $t\ll l/2$. 
 Our results show that $S_A(t)$ quadratically depends on $t$. 
 The von Neumann entropy $S_A(t\ll t^*)$ for different values of $m_0$ is plotted in
Fig. (\ref{dynamic case 1 small time}). The same behavior 
already noticed in Ref. \cite{unanyan2010entanglement}.
In Fig. (\ref{dynamic case 1 small time}) we also present the result for
the coefficient $\kappa_2$ as a function of mass parameter $m_0$ that is 
quadratic in $m_0$.
\begin{figure}[t]
\centerline{
\includegraphics[scale=0.35]{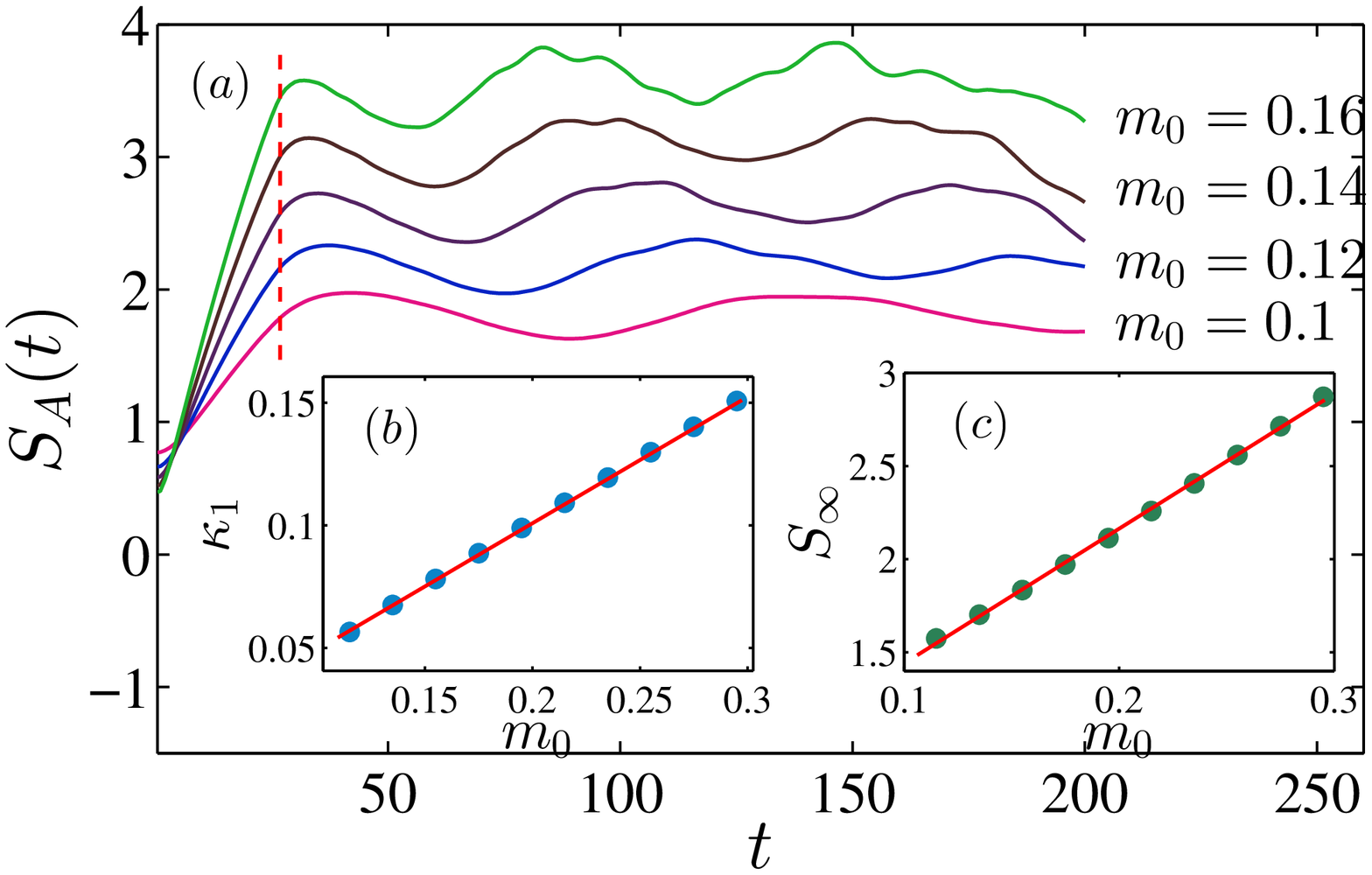}
}
\centerline{
\includegraphics[scale=0.35]{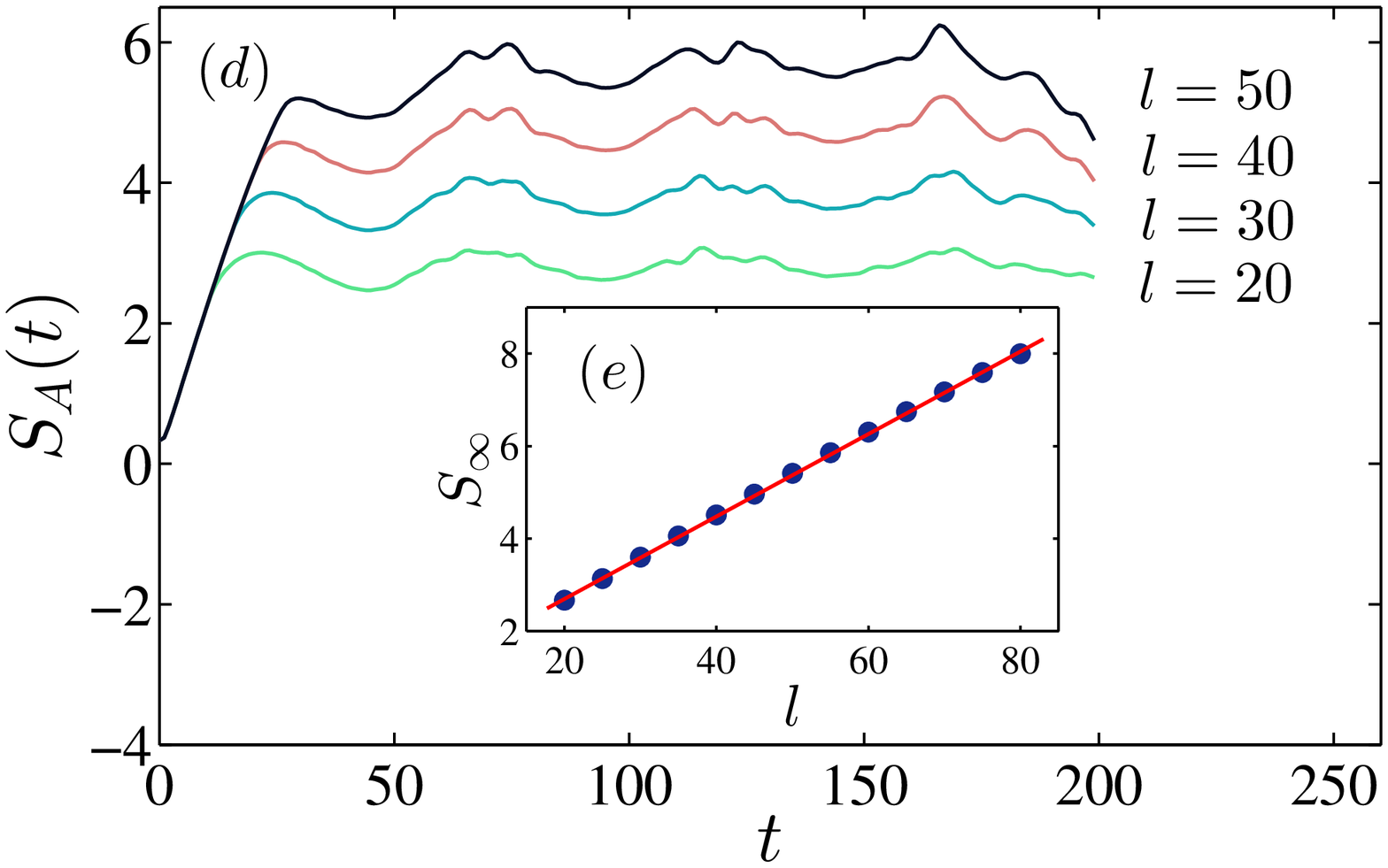}
}
\caption{(Color online) 
Top: (a) Entanglement entropy dynamics $S_A(t)$ for short-range harmonic
 oscillator with the configuration $\mathfrak{g}_1$ and different mass 
 parameter $m_0$. 
The total size of the system is $N = 400$, and the subsystem size $l = 50$. 
The dashed line corresponds to the saturation time $t^* = l/2$. (b) 
The von Neumann entropy $S_A(t)$ in the $t<l/2$, limit obeys
 the $S_A(t)\sim \kappa_1 t$, where 
$\kappa_1$ is a linear function of mass $m_0$. (c) In the $t>l/2$ limit $S_A(t)$ saturates to the 
$S_{\infty}$. The value of $S_{\infty}$ for the fixed value of $l=30$ 
is linear function of $m_0$. 
Bottom: (d)The von Neumann entropy $S_A(t)$ for short range harmonic oscillators
 with the configuration 
$\mathfrak{g}_1$ and fixed value of the mass parameter $m_0=0.4$. 
The total size of the system is $N = 400$, and the subsystem size 
$l \in \lbrace 20,30,40,50\rbrace $. 
(e) In the $t>l/2$ limit, $S(t)$ saturates to the $S_{\infty}$. 
The value of $S_{\infty}$ is a linear function of $l$.}
\label{dynamic case 1}
\end{figure}
\begin{figure}[t]
\centerline{
\includegraphics[scale=0.35]{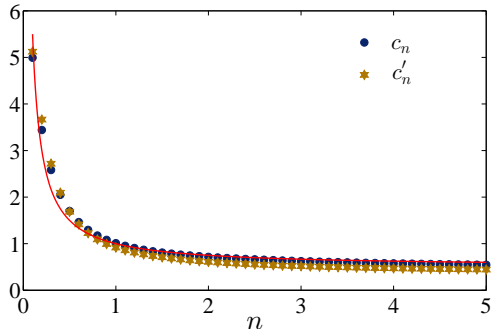}
}
\caption{(Color online) Prefactors $c_n$ and $c_n^{\prime}$ for dynamics of R\'enyi entropy 
for the short range harmonic oscillator system
with configuration $\mathfrak{g}_1$ and the system size $N=500$. The solid red line
 represents $\frac{1}{2}(1+1/n)$  (see Eq. (\ref{entropy case 2})).
}
\label{dynamic case 2 cn and cprimen for renyi}
\end{figure}
Then until 
 the saturation time $t^*\sim l/2$, entanglement dynamics $S_A(t)$ grows 
 linearly with time ($S_A(t)\sim \kappa_1 t$) and finally it saturates to the $S_{\infty}$ 
 (see Fig. (\ref{dynamic case 1})-a).
  The coefficient $\kappa_1$, is a linear function of 
mass parameter $m_0$ (see Fig. (\ref{dynamic case 1})-b) with the same
 slop as the predicted value $\pi/6$. 
 We also obtained the behavior of the saturation value $S_{\infty}$
 with respect to the mass term $m_0$. In Fig. (\ref{dynamic case 1}-c) we 
 show numerical results for the linear dependence between $S_{\infty}$ and $m_0$ which
  the slope matches perfectly with the theoretical value $\pi l/12$. 
  One can do the same calculation for different values of $l$ 
 which the result will be the same.
 In Fig. (\ref{dynamic case 1}-d) we also sketched $S_A(t)$ versus $t$ for 
 the fix value of $m_0 = 0.4$ and different
 values of $l$. The entanglement entropy for
such a system gets saturated in which $S_{\infty}$ increases linearly with $l$
(see Fig. (\ref{dynamic case 1})-e). 
A good agreement has been found between our numerical results and the 
theoretical prediction $\pi m_0/12$. 

Using Eq. (\ref{renyi entropy formula}), we can also evaluate the time 
evolution of R\'enyi entropy $S_n(t)$. 
In Fig. (\ref{dynamic case 2 cn and cprimen for renyi})
 we have plotted the numerical estimates of
the scaling factors $c_n$ and $c^\prime_n$. The agreement between the numerical 
results and the theoretical prediction 
Eq. (\ref{renyi entropy scaling case 1}) is fairly good.     
  
\begin{figure}[t]
\centerline{
\includegraphics[scale=0.35]{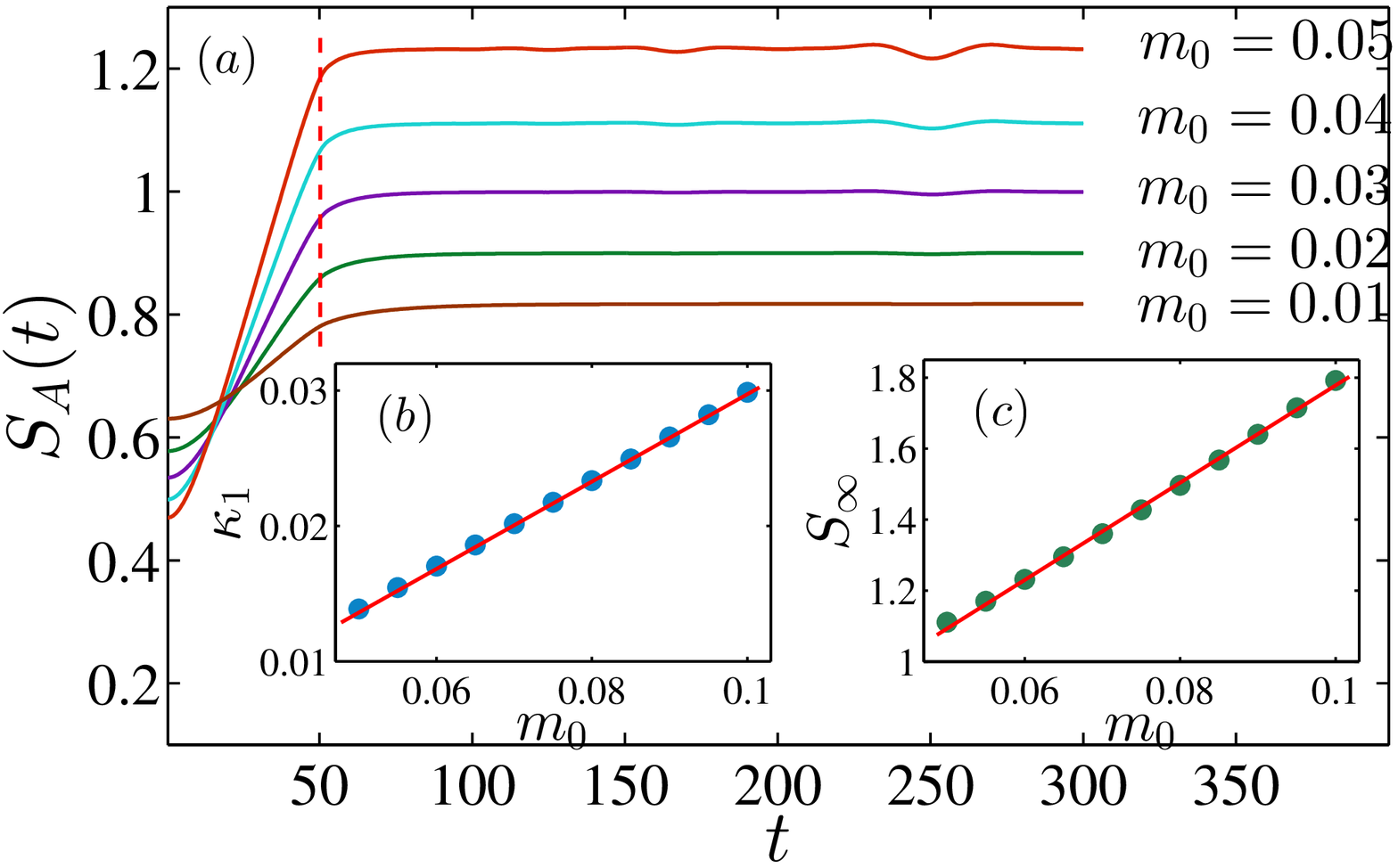}
}
\centerline{
\includegraphics[scale=0.35]{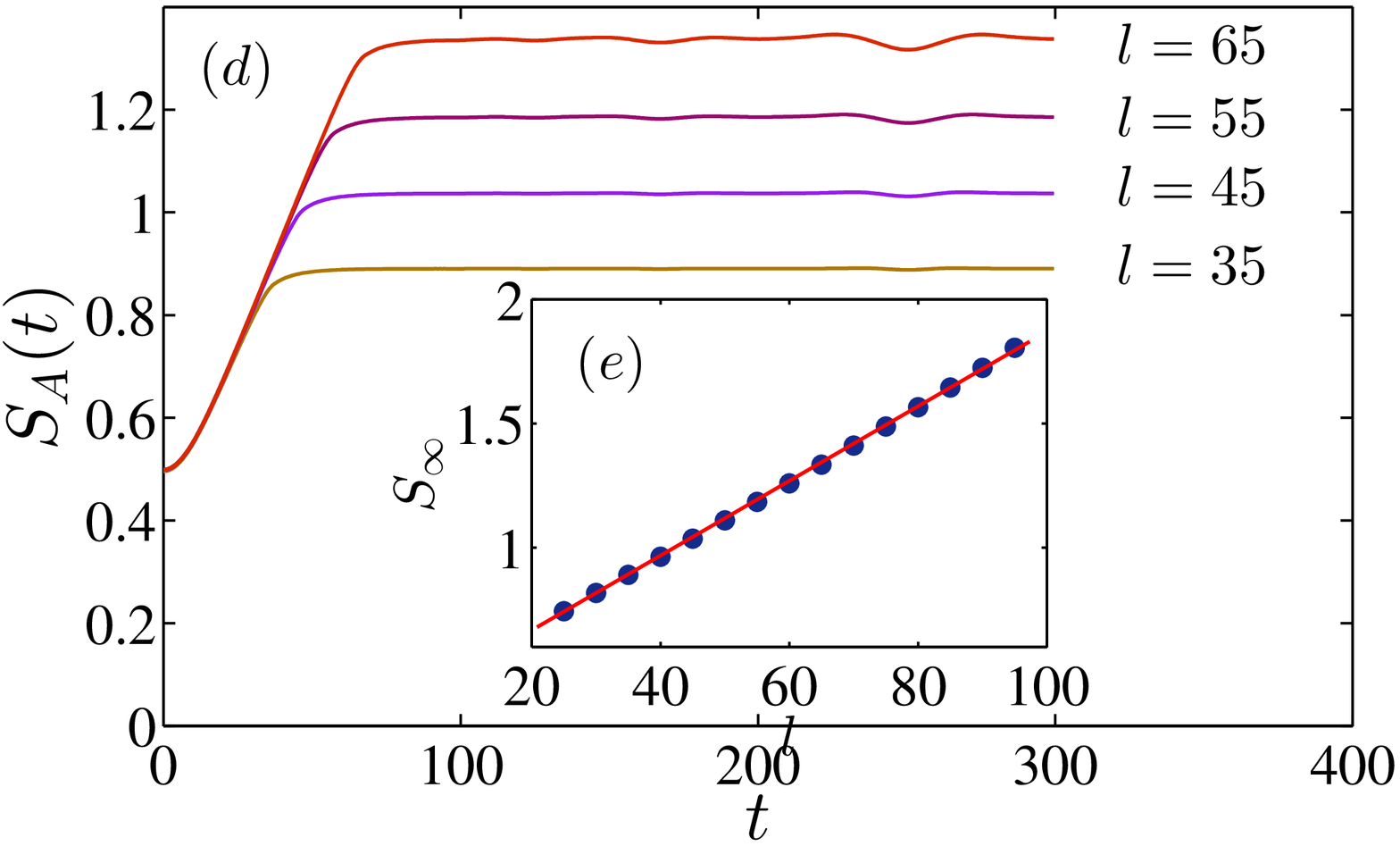}
}
\caption{(Color online) Top: (a) Entanglement entropy dynamics $S_A(t)$ 
for short-range harmonic
 oscillator with the configuration $\mathfrak{g}_2$ and different mass 
 parameter $m_0$. 
The total size of the system is $N = 500$, and the subsystem size $l = 50$. 
The dashed line corresponds to the saturation time $t^* = l$. (b) 
$S_A(t)$ in the $t<l$, limit obeys the $S(t)= \kappa_1 t$, where 
$\kappa_1$ is a linear function of mass $m_0$.
 (c) $S_A(t)$ in the $t>l$ limit saturates to the 
$S_{\infty}$. The value of $S_{\infty}$ 
is linear function of $m_0$.
Bottom: (d) Entanglement entropy dynamics $S_A(t)$ for short range
harmonic oscillators with the 
configuration $\mathfrak{g}_2$ and fixed value of the
  mass parameter $m_0=0.05$. 
The total size of the system is $N = 500$, and the subsystem size 
$l \in \lbrace 35,45,55,65\rbrace $. 
(e)The value of $S_A(t)$ in the $t>l$ limit saturates to the $S_{\infty}$ which is a 
linear function of $l$.
}
\label{dynamic case 2}
\end{figure}
The next step is to analyze the dynamics of von Neumann and R\'enyi
entropies for the case $\mathfrak{g}_2$, where the system
is semi-infinite line, and the subsystem is finite interval $\left[0,l\right)$
 (see Fig. \ref{configglobal}).
 It is clear that to understand the qualitative behavior of 
 entanglement dynamics, a very similar explanation that is based on 
 quasi-particle picture, should apply also for this special case of
 system and subsystem configuration.
 The only difference 
 here is that the quasiparticles emitted from $x=0$ in the region $A$
 need more time to be observed by the first point $x = l$ in the region $B$. 
 Therefore one can expect the saturation time is $t^* = l$. In this case
 we have the following formula
  \begin{eqnarray}\label{entropy case 2}
  S_n(t) = - \frac{c_n}{6}\log m_0 + \Bigg{\lbrace}
  \begin{array}{l l}
    \frac{\pi m_0c_n }{12}t & \quad t<t^*\\
    \\
    \frac{\pi m_0c^\prime_n }{12}l & \quad t>t^*
  \end{array}
  ~,
\end{eqnarray}
  where $c_n = c_n^\prime = \frac{1}{2}\left(1 +1/n \right)$.

To determine the time evolution of the von Neumann
entropy for short-range harmonic oscillator with configuration 
$\mathfrak{g}_2$, we computed 
the eigenvalues of the matrix $\mathcal{E}$ to evaluate $S_A$ and $S_n$ by
Eqs. (\ref{entanglment entropy formula}) and (\ref{renyi entropy formula}).
In Fig. (\ref{dynamic case 2}) 
we show $S_A(t)$ for different values of $m_0$ and $l$ respectively.
Our numerical results show that $S_A(t)$ in very short time is quadratic in time. 
Then the evolution of $S_A(t)$ has been studied for intermediate times that is 
$S_A(t) = \kappa_1 t$. 
The prefactor $\kappa_1$ as a linear function of $m_0$ is depicted in 
Fig. (\ref{dynamic case 2}-b). We also calculated the prefactor of the
$\kappa_1$ with respect to $m_0$. There is a very good agreement with
 the predicted value $\pi/12$.
  
 Next we examine the scaling properties of 
 the entanglement entropy $S_{\infty}$ in saturation regime. 
 Figures (\ref{dynamic case 2}-c) and (\ref{dynamic case 2}-e) 
 show a linear plot of $S_{\infty}$ versus of $m_0$ and $l$. 
 The straight lines, with slopes $\pi l/12$ and $\pi m_0/12$, respectively, 
 are drown to demonstrate the good agreement with the prediction
  Eq. (\ref{entropy case 2}). Finally we characterized the time evolution of the R\'enyi entropy and
 our numerical observation was in perfect agreement with the Eq. (\ref{entropy case 2}).

In the next section, our main results on the time evolution of the entanglement entropy
 for long-range harmonic oscillator are presented. To avoid any finite 
 size effect we will concentrate on the case where the system is very large and
the subsystem has small size $l$ (case $\mathfrak{g}_1$ in Fig. \ref{configglobal}).

\section{Harmonic systems with long-range couplings}\label{LRHO EEDynamics sec}
In this section, we study the time evolution of the entanglement after 
global quench for the system of harmonic oscillators with long-range couplings. 
Following Eq. (\ref{harmonicOsc}), let us introduce 
the $K$ matrix for coupled harmonic oscillators with long-range couplings. 
There are many ways
 to write a long range $K$ matrix for non-local scalar field theories \textit{i. e.} 
 see Ref. \cite{shiba2013volume}, however, 
we are interested in those that are defined by the fractional laplacian operator.
 In the continuum limit
the fractional laplacian has simple Fourier transform $|q|^{\alpha}$ 
or $(q^2)^{\frac{\alpha}{2}}$ which $q^2$ is just the Fourier transform
of a simple laplacian. Since the discrete 
laplacian in the fourier representation 
 is $2-2\cos q$, so that one may use the appropriate power
 of this to define the
discrete fractional laplacian. Then the elements 
of the interaction kernel $K$, representing the discretized 
fractional laplacian are then given by
\begin{eqnarray}\label{interaction kernel long range}
K_{i,j} &= -{\int_0^{2\pi} \frac{dq}{2\pi} e^{iq(i-j)} \lbrace \left[ 2(1-\cos(q))\right]^{\frac{\alpha}{2}}+m^{\alpha}\rbrace}\\ \nonumber
&=\frac{\Gamma(-\frac{\alpha}{2}+i-j)\Gamma(\alpha +1)}{\pi \Gamma(1+\frac{\alpha}{2}+i-j)} \sin(\frac{\alpha}{2}\pi)+{m^{\alpha}\delta_{i,j}}~.
\end{eqnarray}
The above equation for large distances behaves like  
$K_{i,j} \sim 1/|i-j|^{1+\alpha}$. This power-law interaction
for $\alpha\geq 2$ is effectively local. 
In the continuum limit the Eq. (\ref{interaction kernel long range}) can be written as 
\begin{equation} \label{fractional free field theory}
\frac{1}{2} \sum_{i,j=1}^{N}\phi_{i} K_{ij}\phi_{j}\rightarrow \int [-\frac{1}{2}\phi(x)(-\bigtriangledown)^{\alpha/2}\phi(x)+\frac{1}{2}m^{\alpha}\phi^2]dx.
\end{equation}
In the above formula $(-\bigtriangledown)^{\alpha/2}$ corresponds to the
 fractional laplacian operator.  
 
 The most important characteristics of the oscillator
system is the spatial correlation $\langle \phi_l \phi_m \rangle$, where
 for the system with periodic
boundary condition, one can find the spatial correlation length $\xi_s$ as:
\begin{eqnarray}\label{correlation general 1}
 \xi_s ^{-1}&\equiv -\lim_{r\rightarrow\infty}\frac{1}{r}\log |\langle \phi_l\phi_{l+r} \rangle| \\ \nonumber
 &= -\lim_{r\rightarrow\infty}\frac{1}{r}\log |K^{- 1/2}(r)|~.
 \end{eqnarray}
The elements of $K^{- 1/2}$ for a sufficiently large one-dimensional
 system can be expressed as a Fourier series
\begin{eqnarray} \label{HOLRcorr} 
  K^{- 1/2}(r,m)&=\hspace{5.5cm}\nonumber \\  -\int_0^{2\pi} \frac{dq}{2\pi} e^{iqr}
 &\lbrace \left[ 2(1-\cos(q))\right]^{\frac{\alpha}{2}}+{m}^{\alpha}\rbrace ^{- 1/2}~.
\end{eqnarray}
In the scaling limit Eq. (\ref{HOLRcorr}) behaves like 
$K^{- 1/2} = -\int dq e^{iq r}
 \lbrace |q|^\alpha +{m}^{\alpha}\rbrace ^{- 1/2}$ and  
one can evaluate this integral for $m = 0$ which is given by: 
\begin{eqnarray}\label{powerlawcorr case 1}
K^{-1/2}(r,0) \sim 1/r^{1-\alpha/2}~.
\end{eqnarray}
We now want to evaluate the integral Eq. (\ref{HOLRcorr}) 
for the massive case $m\neq0$ and large values of the distanse $r$. 
The integral can not be done explicitly in the most general case, 
therefore we studied this case just numerically (see Appendix. \ref{appendix sec 0}).
We find that the best fit to our numerical data is:
\begin{eqnarray}\label{powerlawcorr case 1 massive}
K^{-1/2}(r,m) \propto 1/r^{1+\alpha}~.
\end{eqnarray} 
One can observe that
 the above result leads to $\xi_s^{-1} = 0$ for
all values of the mass parameter $m$.  
Note that for the special case  $\alpha=2$, 
correlation function 
  decays exponentially and the correlation length is proportional to $\xi_s \sim 1/m$.
It is useful to note that $\alpha >2$ corresponds to a 
 a system of harmonic oscillators with long-range ($\alpha <2$) 
 plus shot-range ($\alpha =2$)
interactions where we have $\alpha=2$ and $\alpha<2$ both laplacians 
in the hamiltonian. Therefore, all the massive correlations are exponential 
because everything is dominated with $\alpha=2$.

\subsection{Weakly coupled long-range harmonic oscillators $1<\alpha<2$}
To understand the entanglement entropy growth behavior for harmonic oscillators with
weak couplings
 we consider the system with very large size and a subsystem with length $l$
 (configuration $\mathfrak{g}_1$ in the Fig. (\ref{configglobal})).
First one should construct the interaction kernel $K$ from Eq. (\ref{interaction kernel long range}).
The next step is to analyze the eigenvalues of the matrix $\mathcal{E}$ and
 measure the entanglement entropy $S_A(t)$ using Eq. (\ref{entanglment entropy formula}).
 We display the resulting quantity for different values
of $\alpha$,  
in Fig. (\ref{entropy dynamics long range different  alpha larger one}).
\begin{figure}
\centerline{
\includegraphics[scale=0.35]{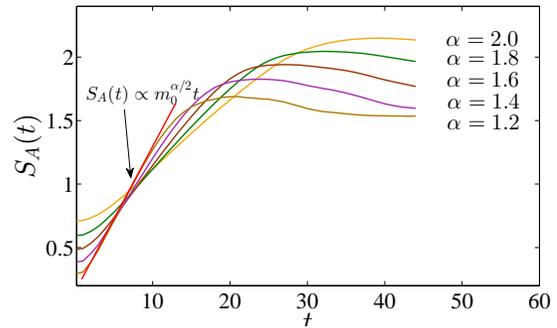}
}
\caption{(Color online) Entanglement entropy dynamics $S_A(t)$ in the long-range harmonic
 oscillators with the configuration $\mathfrak{g}_1$ for different values of
 $\alpha$. The curves correspond to the linear
 change of the $S_A(t)$ with time $t$ before the saturation time 
 $t^*$ for the $1<\alpha \leq 2$. The total size of the system is $N = 400$, the mass 
 parameter $m_0=0.1$ 
 and the subsystem size $l = 50$. 
}
\label{entropy dynamics long range different alpha larger one}
\end{figure}

\begin{figure}[t]
\centerline{
\includegraphics[scale=0.35]{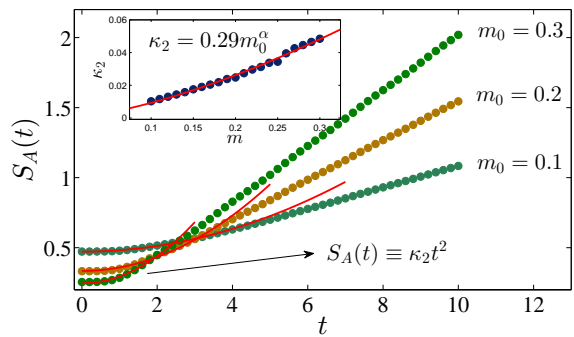}
}
\caption{(Color online)
Quadratic growth in time for $S_A(t)$ where $t\ll t^*$. The harmonic system is 
long-range ($\alpha=1.5$) with the configuration $\mathfrak{g}_1$. 
The total size of the system is $N = 500$, and the subsystem size $l = 60$. 
The solid red line corresponds to the $S_A(t) \equiv \kappa_2 t^2$. Inset: 
The value of the scaling parameter $\kappa_2$ as a function of mass
 parameter $m_0$.}
\label{dynamic case 1 small time long rane}
\end{figure}

An interesting behavior is 
the quadratic entropy growth at the very short-time $t\ll t^*$. In Fig. 
(\ref{dynamic case 1 small time long rane}) we plot the entanglement entropy
dynamics $S_A(t)$ shortly after the quench. This is the same
qualitative quadratic behavior $S_A \equiv \kappa_2 t^2$ 
as seen in the short range case that we mentioned
in the Eq. (\ref{quadratic entropy dynamics formula}). 
Our results indicate that the prefactor $\kappa_2$ obeys the 
power law formula $\kappa_2 \sim m_0^\alpha$. 

\begin{figure}
\centerline{
\includegraphics[scale=0.35]{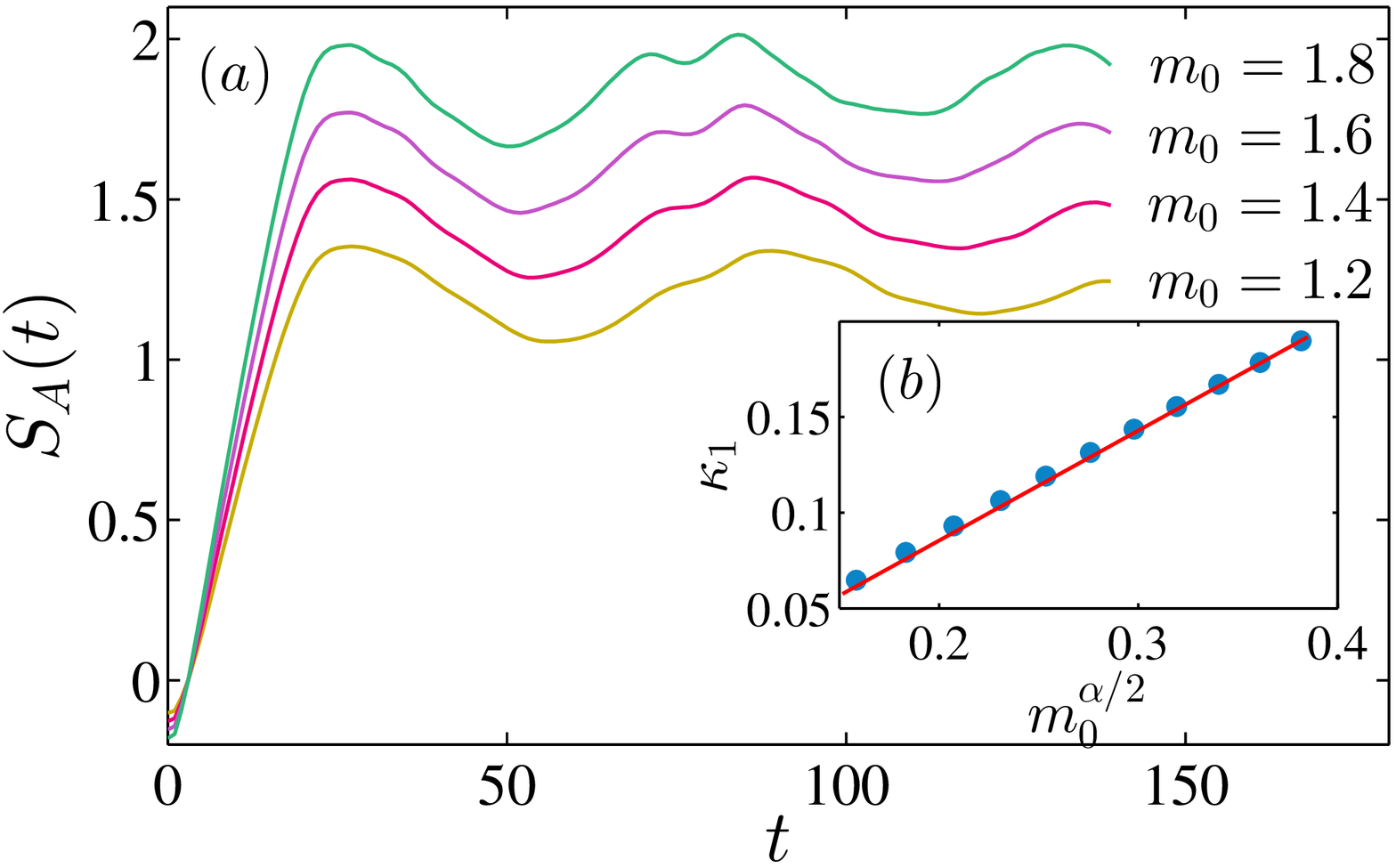}
}
\centerline{
\includegraphics[scale=0.35]{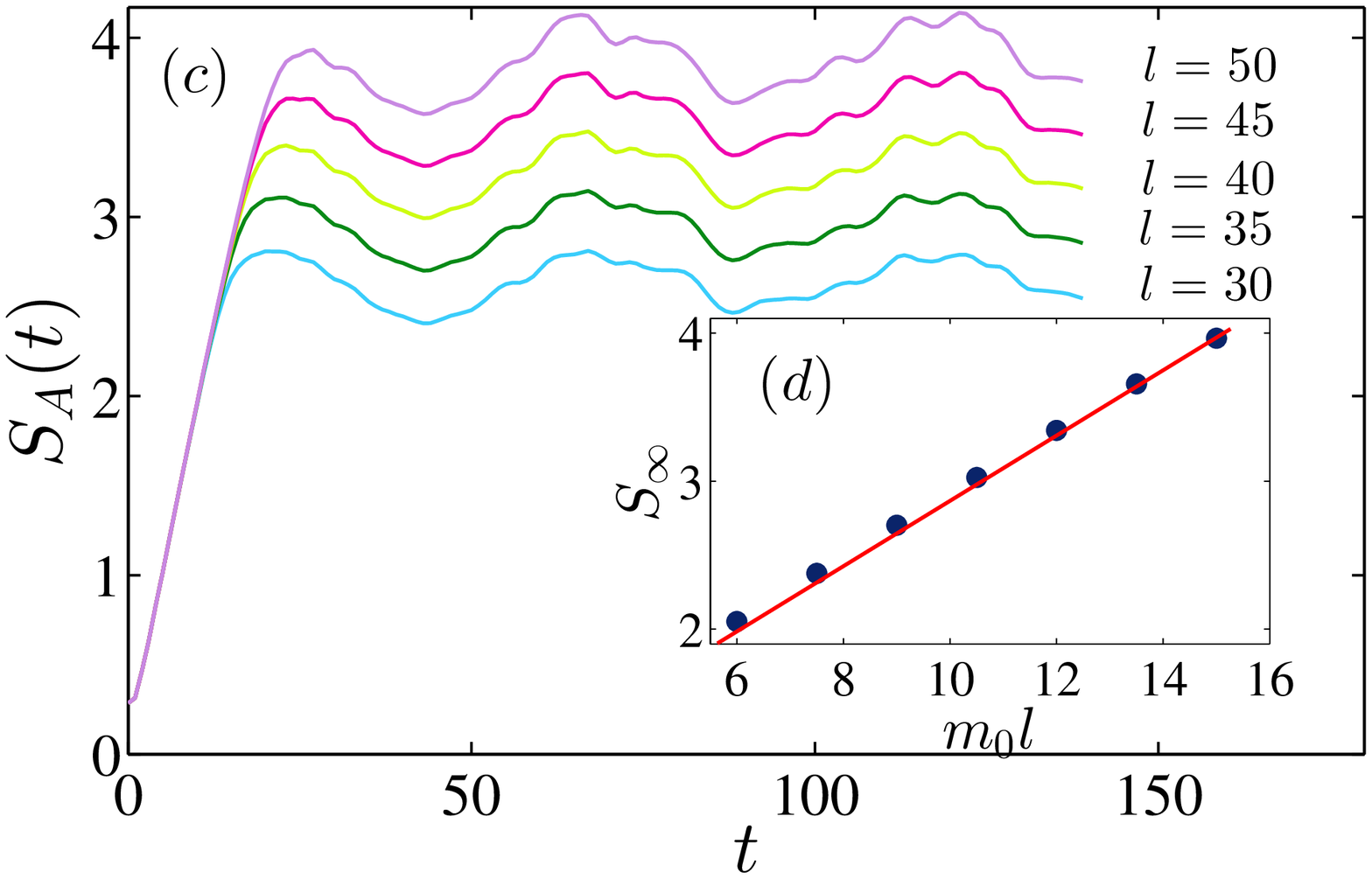}
}
\caption{(Color online) Top: (a) Entanglement entropy dynamics $S_A(t)$ for long-range harmonic
 oscillators ($\alpha=1.6$) with the configuration $\mathfrak{g}_1$ and 
 different mass parameter $m_0$. 
 The total size of the system is $N = 500$, and the subsystem size $l = 50$. (b) 
  Before the saturation time $t<t^*$, entanglement entropy dynamics
   obeys the $S(t)\sim \kappa_1 t$, where 
$\kappa_1$ is a linear function of $m^{\alpha/2}$. 
Bottom: (c) The entanglement entropy dynamics $S_A(t)$ for long-range harmonic
 oscillator with the configuration $\mathfrak{g}_1$ and fixed value of
 mass parameter $m_0=0.3$. The total size of the system is $N = 300$, and 
 the subsystem size $l \in \lbrace 30,35,40,45,50  \rbrace$. 
(d) $S_A(t)$ in $t\gg t^*$, saturates to the $S_{\infty}$ where it changes linearly 
with $m_0l$. 
}
\label{entropy dynamics long range different alpha 1.6}
\end{figure}

\begin{figure}
\centerline{
\includegraphics[scale=0.35]{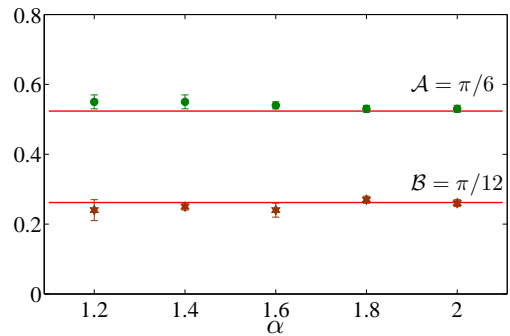}
}
\caption{(Color online) 
Numerical results for the coefficients $\mathcal{A}$ and $\mathcal{B}$ 
as a function of $\alpha$ (see 
Eq. (\ref{EE dynamic for LRHO})).
}
\label{A and B coefficient}
\end{figure}

As we remarked before, if we consider the configuration $\mathfrak{g}_1$ for
  harmonic oscillators with $\alpha = 2$, according to Eq. (\ref{entropy case 2})
 we expect a linear growth of the entropy as a function of time. 
 Notice that this linear growth is only true before saturation time $t^*$
 and after that the $S_A(t)$ will become time independent. Figure 
 (\ref{entropy dynamics long range different alpha larger one})
clearly shows that $S_A(t)$ grows as $\sim t$ before saturation begins. 
We see that this linear form for the entanglement entropy 
is hold in all cases with $1<\alpha \leq 2$. Figure
(\ref{entropy dynamics long range different alpha 1.6})
 is the plot of $S_A(t)$ for long-range harmonic oscillator with $\alpha=1.6$
 and different values of the initial mass parameter $m_0$. 
 Our numerical simulations
 show that the slope of linear 
 entanglement growth in time is $ \kappa_1 = \mathcal{A}m_0^{\alpha/2}$.
 In Fig. (\ref{entropy dynamics long range different alpha 1.6}),
  we plot a comparison of our estimate with the numerical results. 
  We also see that the prefactor $\mathcal{A}$ as we demonstrated in
Fig (\ref{A and B coefficient}) is $\alpha$-independent and it is equal to 
the value $\pi/6$. 

We will
now ask how the entanglement dynamics $S_A(t)$ for system with long-range 
couplings depends on the sub-system size $l$. 
In Fig. (\ref{entropy dynamics long range different alpha 1.6})
 we show the numerical results for the time evolution of the entanglement 
 dynamics in the long-range harmonic oscillators with $\alpha = 1.6$ for 
 different values of the subsystem size $l$.
  It is clear from Fig. (\ref{entropy dynamics long range different alpha 1.6})
 that at long time $S_A(t)$ saturates to $S_\infty$. 
 We find that the saturation
value, $S_\infty$, grows linearly as $S_{\infty} = \mathcal{B}m_0l$. 
As shown in the Fig. (\ref{A and B coefficient}), 
the prefactor $\mathcal{B}$ is surprisingly $\alpha$-independent and 
equal to the value $\pi/12$. 
 
In summary,  
the time evolution of the entanglement entropy
 for long-range harmonic oscillators follows 
\begin{eqnarray}\label{EE dynamic for LRHO}
S_A(t) = - \frac{c^g(\alpha)}{3}\log m_0 + \left\lbrace
  \begin{array}{l l }
    \mathcal{A} m_0^{\alpha/2}t & \quad t<t^*\\
    \\
    \mathcal{B} m_0l & \quad t \gg t^*
  \end{array}\right.
  ~.
\end{eqnarray}
where the prefactor $c^g(\alpha)$ is introduced in 
Refs. \cite{Nezhadhaghighi2012,Nezhadhaghighi2013}. 
In all cases with $1<\alpha <2$, we find that the prefactors $\mathcal{A}$ 
and $\mathcal{B}$, as shown in Fig. (\ref{A and B coefficient}), 
are  independent of $\alpha$.

It is instructive to compare the scaling parameters 
$\kappa_2 \sim m_0^{\alpha}$ and $\kappa_1 \sim m_0^{\alpha/2}$ 
the respective coefficient of the quadratic and linear part of the $S_A(t)$.
 This gives us an insight into the special behavior of the entanglement
 dynamics of the harmonic systems with long-range couplings. 
 As we will soon see, this behavior is closely related to the
 form of the hamiltonian Eq. (\ref{fractional free field theory}).

\begin{figure}[t]
\centerline{
\includegraphics[scale=0.35]{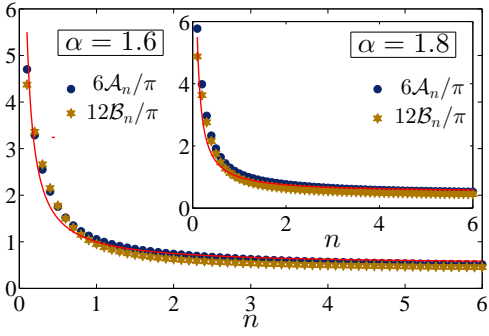}
}
\caption{(Color online) 
Scaling prefactors $\mathcal{A}_n$ and $\mathcal{B}_n$ vs $n$ for different 
$\alpha$'s ($\alpha=1.6,1.8$).    
The solid red line
 represents $\frac{1}{2}(1+1/n)$.
}
\label{dynamic case 2 An and Bn for renyi long range }
\end{figure}

Let us consider now
the time evolution of the R\'enyi entropy for the harmonic system with long-range 
couplings. Figure (\ref{dynamic case 2 An and Bn for renyi long range }) 
is devoted to these analysis. Interestingly, we observe that
  
\begin{eqnarray}\label{renyi dynamic for LRHO}
S_n(t) = - \frac{c_n^g(\alpha)}{3}\log m_0 + \left\lbrace
  \begin{array}{l l }
    \mathcal{A}_n m_0^{\alpha/2}t & \quad t<t^*\\
    \\
    \mathcal{B}_n m_0l & \quad t \gg t^*
  \end{array}\right.
  ~,
\end{eqnarray}
where the prefactor $c_n^g(\alpha)$ is introduced in 
Refs. \cite{Nezhadhaghighi2012,Nezhadhaghighi2013}. The scaling parameters 
$\mathcal{A}_n$ and $\mathcal{B}_n$ are depicted in the 
Fig. (\ref{dynamic case 2 An and Bn for renyi long range }) for
different values of $\alpha$. It is surprising
that $6\mathcal{A}_n/\pi$ and $12\mathcal{B}_n/\pi$ are $\alpha$-independent
and follow the relation $\frac{1}{2}(1+1/n)$. The specific reason for 
this $\alpha$-independent behavior remains an open problem.

\begin{figure}
\centerline{
\includegraphics[scale=0.35]{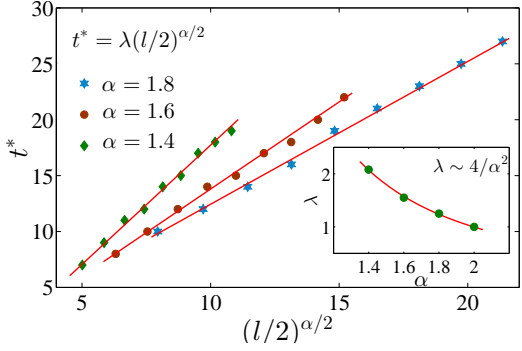}
}
\caption{(Color online)
(a) The saturation time $t^*$ versus $(l/2)^{\alpha/2}$ for different values 
of $\alpha$. (b) Scaling parameter $\lambda$ with respect to $\alpha$. 
}
\label{saturation time for LRHO}
\end{figure}

Since we were able to calculate the entanglement entropy dynamics $S_A(t)$, 
we can now numerically evaluate another aspect to consider 
which is the saturation time $t^*$. 
It is easy to see that 
(see Fig. (\ref{entropy dynamics long range different alpha larger one})) 
the entanglement 
  entropy saturates at time $t = t^*$. We measured this value
  from our numerical data. It is clear from Fig. (\ref{saturation time for LRHO})
  that $t^*$ scales with the subsystem size as 
\begin{eqnarray}\label{saturation time eq for LRHO}
t^* = \lambda(l/2)^{\alpha/2}~.
\end{eqnarray}  
  The prefactor $\lambda$ is $\alpha$-dependent quantity where the best fitting function
  to our numerical data is $\lambda = 4/\alpha^2$. 
  It is worth mentioning that this finding nicely matches the special case $t^* = l/2$ for those harmonic
  systems with short range couplings. 
  
  Let us now discuss an important consequence regarding the above results. 
  The differences in the behavior of the entanglement entropy dynamics
  for short and long-range harmonic oscillators
  come from the difference between length and time scales. It
is well known that in the long-range systems the dynamical
exponent is $z = \alpha/2$. The dynamical exponent controls the relative
scaling of time and space leading to the invariant form $t/l^z$. 
Therefore this argument hints that for the massive system there are two 
correlation lengths $\xi_t = 1/m^{z}$ and $\xi_s = 1/m$ where they will be
equal for harmonic system with short-range couplings ($\alpha=2$). 
Returning to our entanglement dynamics problem, we can conclude that the saturation time
$t^*$ should scale as Eq. (\ref{saturation time eq for LRHO}). In a similar way
one can guess the behavior in the Eq. (\ref{EE dynamic for LRHO}).
 
\subsection{Strongly coupled long-range harmonic oscillators $0<\alpha <1$}

\begin{figure}
\centerline{
\includegraphics[scale=0.35]{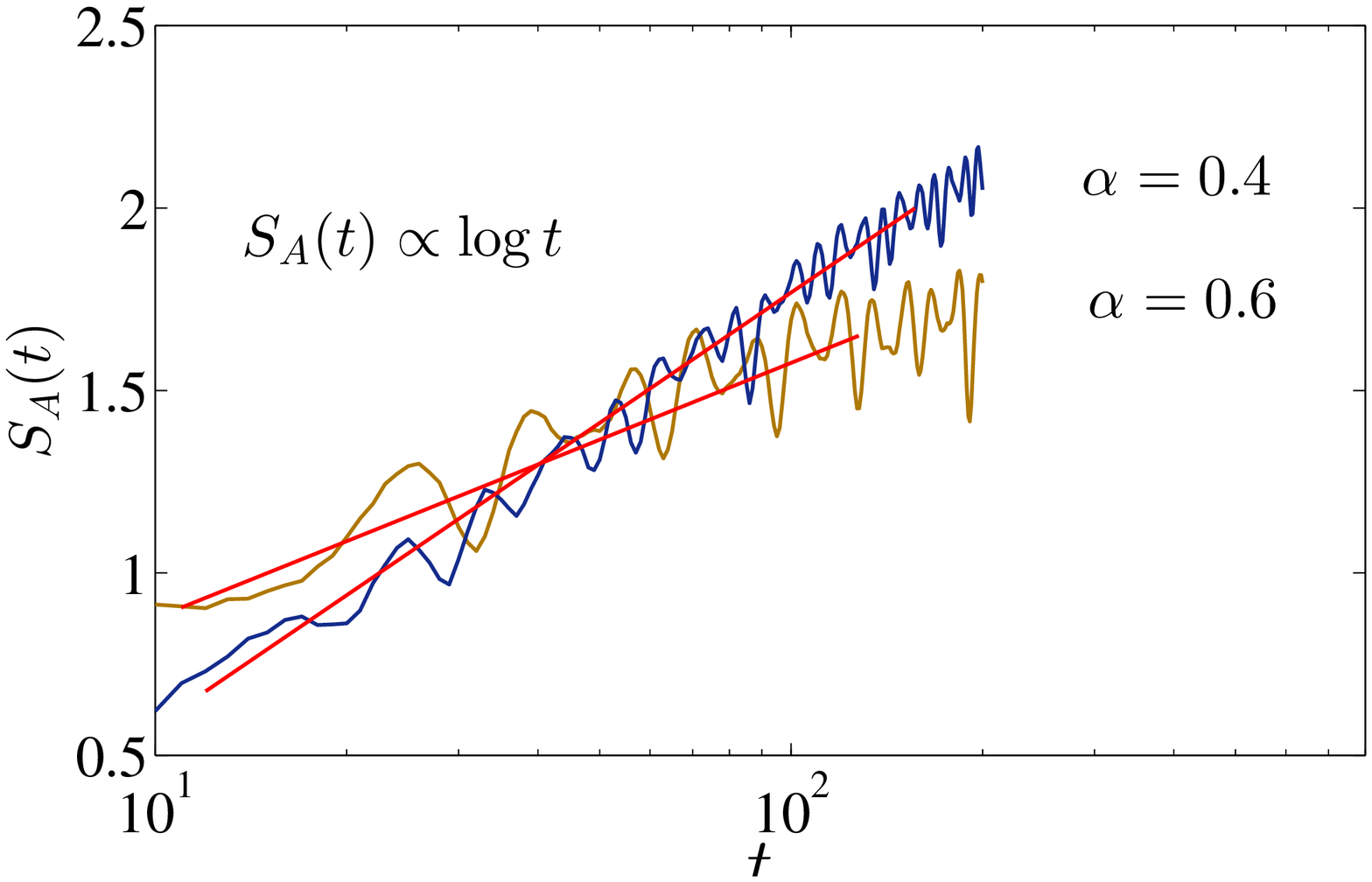}
}
\centerline{
\includegraphics[scale=0.35]{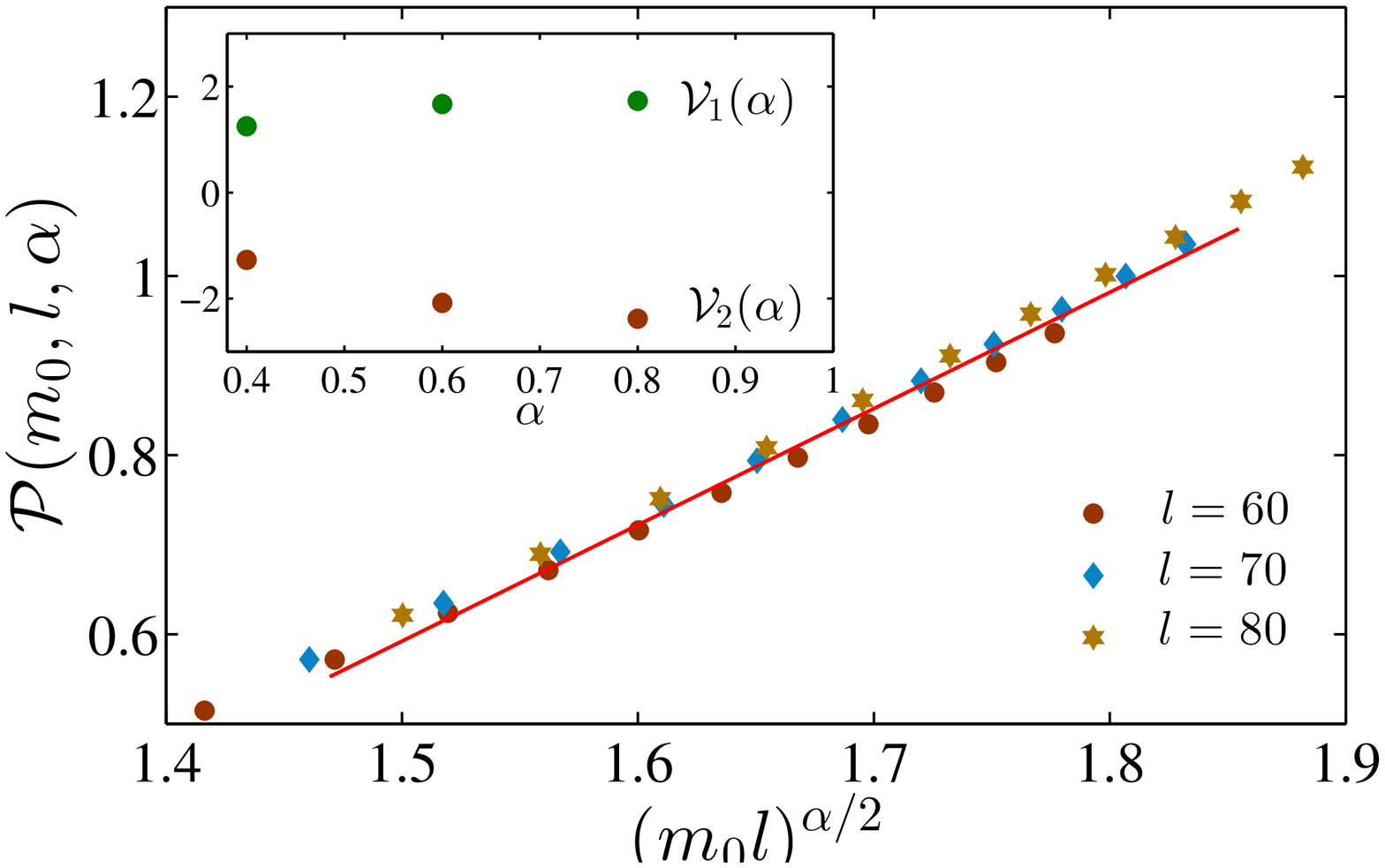}
}
\caption{(Color online) Top: Entanglement entropy dynamics $S_A(t)$ in long-range harmonic
 oscillators ($\alpha <1$) with the configuration $\mathfrak{g}_1$. 
 The total size of the system is $N = 500$, the mass 
 parameter $m_0=0.1$ 
 and the subsystem size $l = 50$. The red lines corresponds to 
 logarithmic change of $S_A(t)\sim \mathcal{P}(m_0,l,\alpha) \log t$ 
 where the coefficient $\mathcal{P}(m_0,l,\alpha)$ is
 a function of $m_0$, $l$ and $\alpha$. Bottom: The coefficient 
 $\mathcal{P}(m_0,l,\alpha)$ for different values of $m_0$ and $l$ is a linear
 function of $(m_0l)^{\alpha/2}$, i.e. 
 $\mathcal{P}(m_0,l,\alpha)= \mathcal{V}_1 (\alpha) (m_0l)^{\alpha/2} +\mathcal{V}_1 (\alpha)$ where
  the coefficients $\mathcal{V}_1$ ($\mathcal{V}_2$) as shown in 
  the inset, is an increasing (decreasing)
  function of $\alpha$. 
  }
\label{entropy dynamics long range different alpha smaller one}
\end{figure}

\begin{figure}
\centerline{
\includegraphics[scale=0.35]{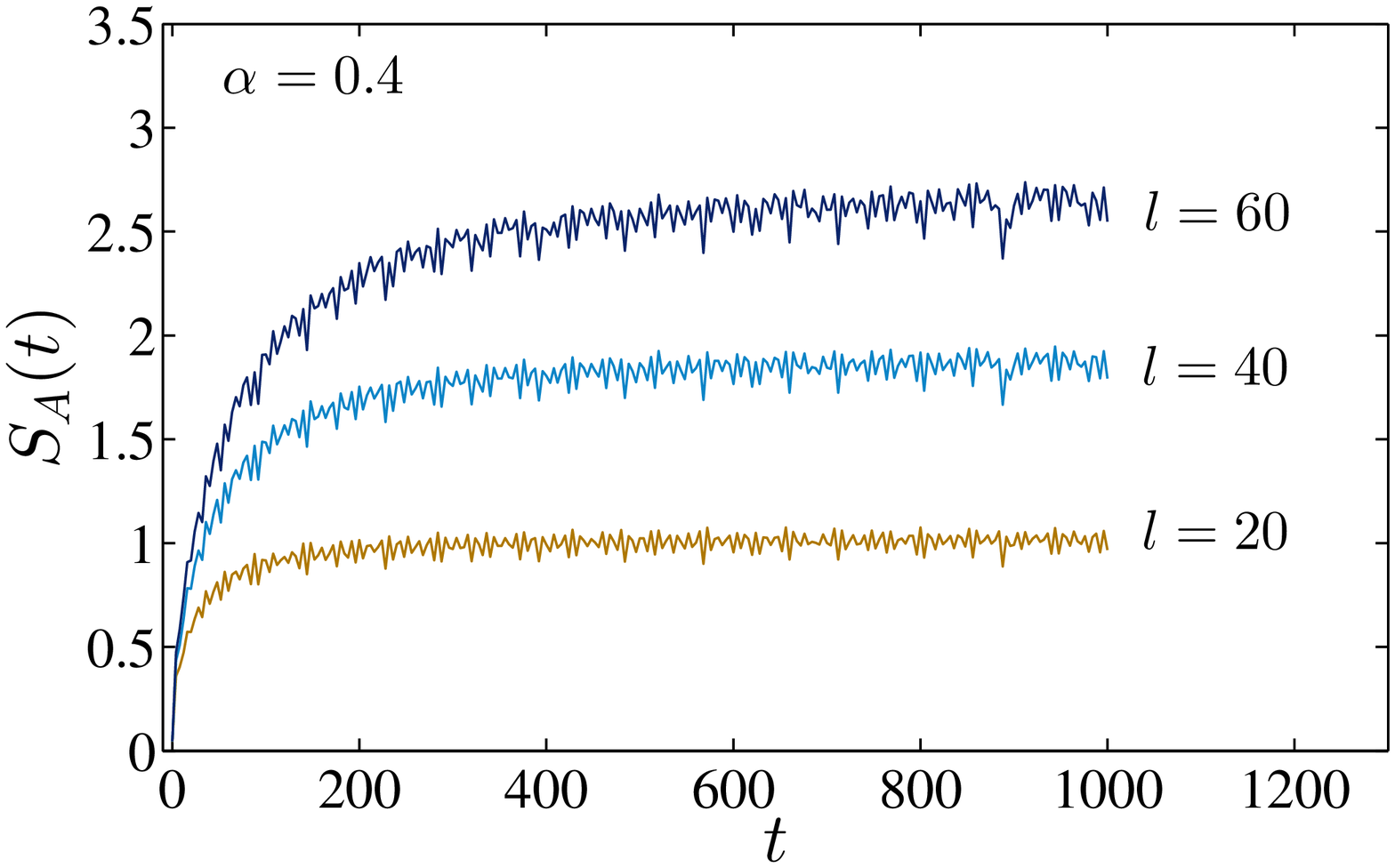}
}
\centerline{
\includegraphics[scale=0.35]{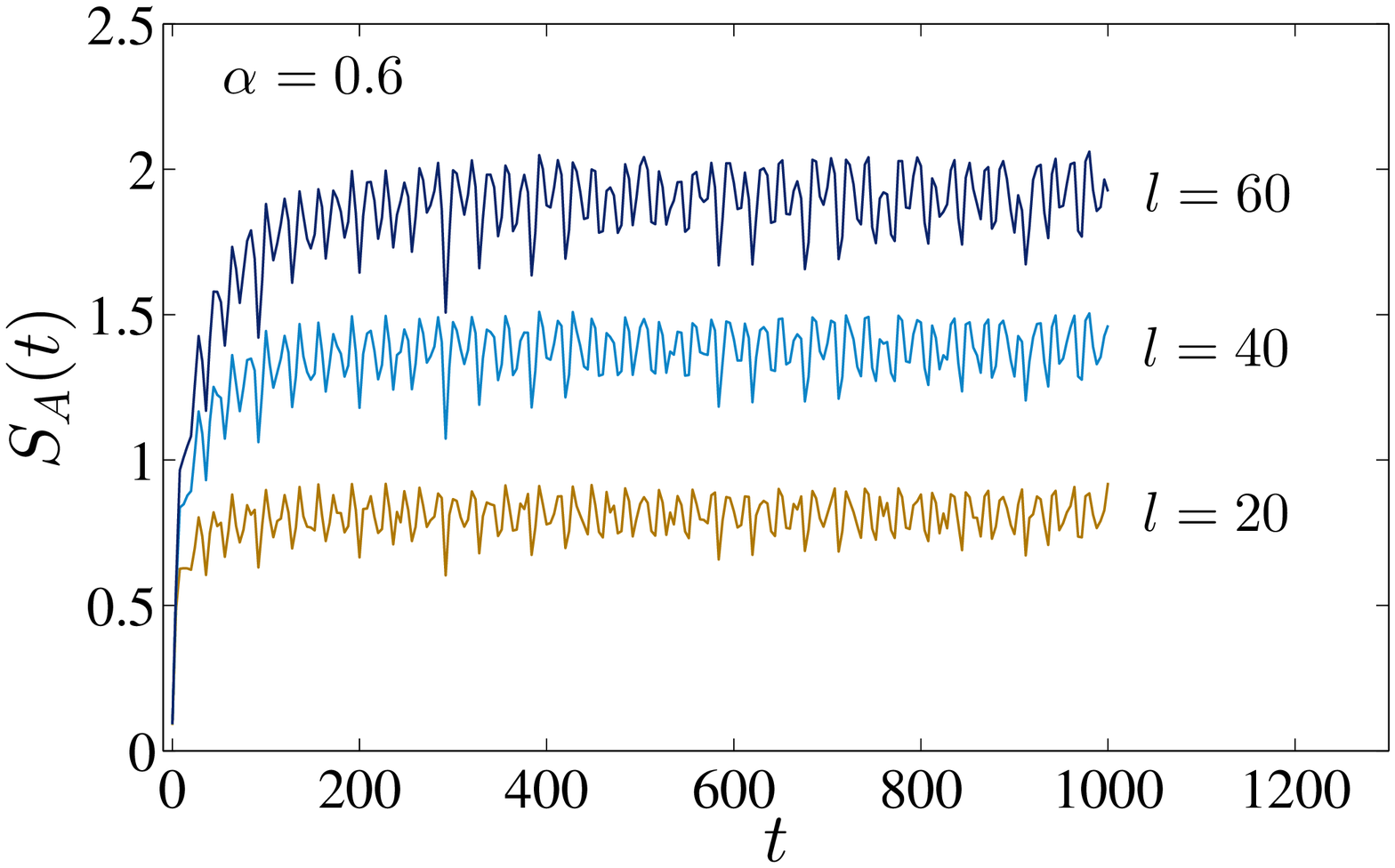}
}
\caption{(Color online) The role of finite size effects on the 
entanglement entropy dynamics $S_A(t)$ in long-range harmonic
 oscillators ($\alpha <1$) with the configuration $\mathfrak{g}_1$. 
 The total size of the system is $N = 500$, the mass 
 parameter $m_0=0.1$. Top: $\alpha = 0.4$, Bottom: $\alpha = 0.6$. 
  }
\label{FSE long range different alpha smaller one}
\end{figure}

So far, we have only considered long-range harmonic chains with $1<\alpha \leq 2$. 
Here, we explore entanglement dynamics after a quantum quench in
 long-range harmonic oscillators with strong couplings ($\alpha <1$).
 We observe that, $S_A(t)$ does not grow linearly and also it does not saturate in the 
 finite time interval that we considered. It is interesting to note that
the entanglement dynamics for this case has the form,
\begin{eqnarray}\label{logarithmic EED stronge interaction}
S_A(t) = - \frac{c^g(\alpha)}{3}\log m_0 + 
\left\lbrace
  \begin{array}{l l l}
    \kappa_2 t^2 & \quad t \ll 1\\
    \\
    \mathcal{P}(m_0,l,\alpha) \log t   \\
    \\
    \mathcal{B} m_0 l &\quad t \gg 1
  \end{array}\right.
  ~,
\end{eqnarray} 
where $c^g(\alpha)$ is introduced in 
Refs. \cite{Nezhadhaghighi2012,Nezhadhaghighi2013}, and 
$\kappa_2 \propto m_0^\alpha$. In our numerical analysis, we found that, 
$\mathcal{B} \sim \pi/12$ is independent of $\alpha$. 
In Fig. (\ref{entropy dynamics long range different alpha smaller one}) we
 report the logarithmic behavior of the entanglement entropy dynamics for the
 strongly long range harmonic oscillator. 
  As we show in Fig. (\ref{entropy dynamics long range different alpha smaller one}),
   $\mathcal{P}(m_0,l,\alpha)$ is a 
function of mass parameter $m_0$ and size of the sub-system $l$, as
\begin{eqnarray}
\mathcal{P}(m_0,l,\alpha)= \left[ \mathcal{V}_1 (\alpha) (m_0l)^{\alpha/2} +
\mathcal{V}_2(\alpha)\right]~,
\end{eqnarray} 
 where $\mathcal{V}_1$ and $\mathcal{V}_2$ as shown in the 
 Fig. (\ref{entropy dynamics long range different alpha smaller one}) 
 are functions of $\alpha$.
 In Fig. (\ref{FSE long range different alpha smaller one})
 we depicted the entanglement dynamics for strongly long-range 
 harmonic oscillators in the saturation regime. 
 In this regime the entanglement entropy  for all 
 values of $\alpha$ saturates to the
 $S_{\infty}\propto m_0 l$ (see Eq. (\ref{logarithmic EED stronge interaction})). 
 
\vspace{1cm}
As pointed out above, in this study we considered a system of 
long-range harmonic oscillators with the interaction kernel 
Eq. (\ref{interaction kernel long range}). To study the time evolution of
von Neumann and R\'enyi entropies after quantum quench we assume that
 the mass parameter in the massive hamiltonian Eq. (\ref{interaction kernel long range})
is suddenly changes from $m_0$ to a different value $m \sim 0$. 
Hence after the quench there is an energy excess which 
acts as a source of quasiparticles moving with maximum group velocity. 
 The maximum group velocity for $1<\alpha <2$ generally depends on the mass parameter $m$ as 
\begin{eqnarray}
v_g^{max}\propto m^{\alpha/2-1}~,
\end{eqnarray}  
which obviously diverges in the limit $m\to 0$  
(see Appendix. \ref{appendix sec}). Therefore, for those long-range systems
 that have been quenched to the critical point of the system ($m=0$),
  there is no maximum group velocity for quasiparticles.
As already reported in the previous sections,
a very small mass parameter ($m\ll m_0$) has been chosen to 
avoid rapid oscillations of $S_A(t)$ in saturation regime.
It is worth mentioning that 
our results have not been affected by non-zero parameter $m$ until
$m\gg 1/N$ which $N$ is the size of the system. 
In the Appendix we explain all of the above in details.
 This means that interestingly
we found the linear behavior for the time evolution of
von Neumann entropy $S_A(t)$ even for those systems with infinite maximum 
group velocity. To compare such result with the theoretical predictions for
free field theory, we have taken into account
 that maximum group velocity for short-range harmonic oscillator is finite. 

Following Ref. \cite{schachenmayer2013entanglement} 
it has been shown that the time
 evolution of the entanglement entropy in the transverse field Ising model 
 with long-range interactions ($J_{i,j}\propto |i-j|^{-\sigma}$) 
 shows three different regimes as a function of $\sigma$.  
 For relatively short-range interactions $\sigma \geq 1$ a linear
growth of the entanglement entropy as a function of time has been found.
 For those long-range interactions with $\sigma \in (0.8,1.0)$
 a regime of logarithmic growth of entropy has been proposed. 
 Finally for strongly long-range interactions with $\sigma <0.2$ 
 rapid oscillations of the entanglement entropy around small values are found.
 In light of our entanglement dynamics results in the harmonic oscillators 
 with long-range couplings, similar regimes 
 possibly with similar signatures, as in, Ref. \cite{schachenmayer2013entanglement}
  have been found. Note that in this study we have $\sigma = 1+\alpha$.  
Our findings show that the linear
growth of entanglement entropy breaks down at $\alpha =1$ ($\sigma = 2$).
The behavior of $S_A(t)$ for $\alpha>1$ ($\sigma > 2$) is a linear function 
of time and on the other hand
 it follows the logarithmic growth with time for $\alpha <1$ ($\sigma < 2$). In principle following
 \cite{gong2014persistence} one should be able to find some useful linear bounds on the growth of entanglement entropy for all the 
 values of $\alpha$. However, it is not clear that why in our system for $\alpha>1$ the bound is saturated and for $\alpha<1$ it is not.

\begin{figure} 
 \centerline{
\includegraphics[scale=0.35]{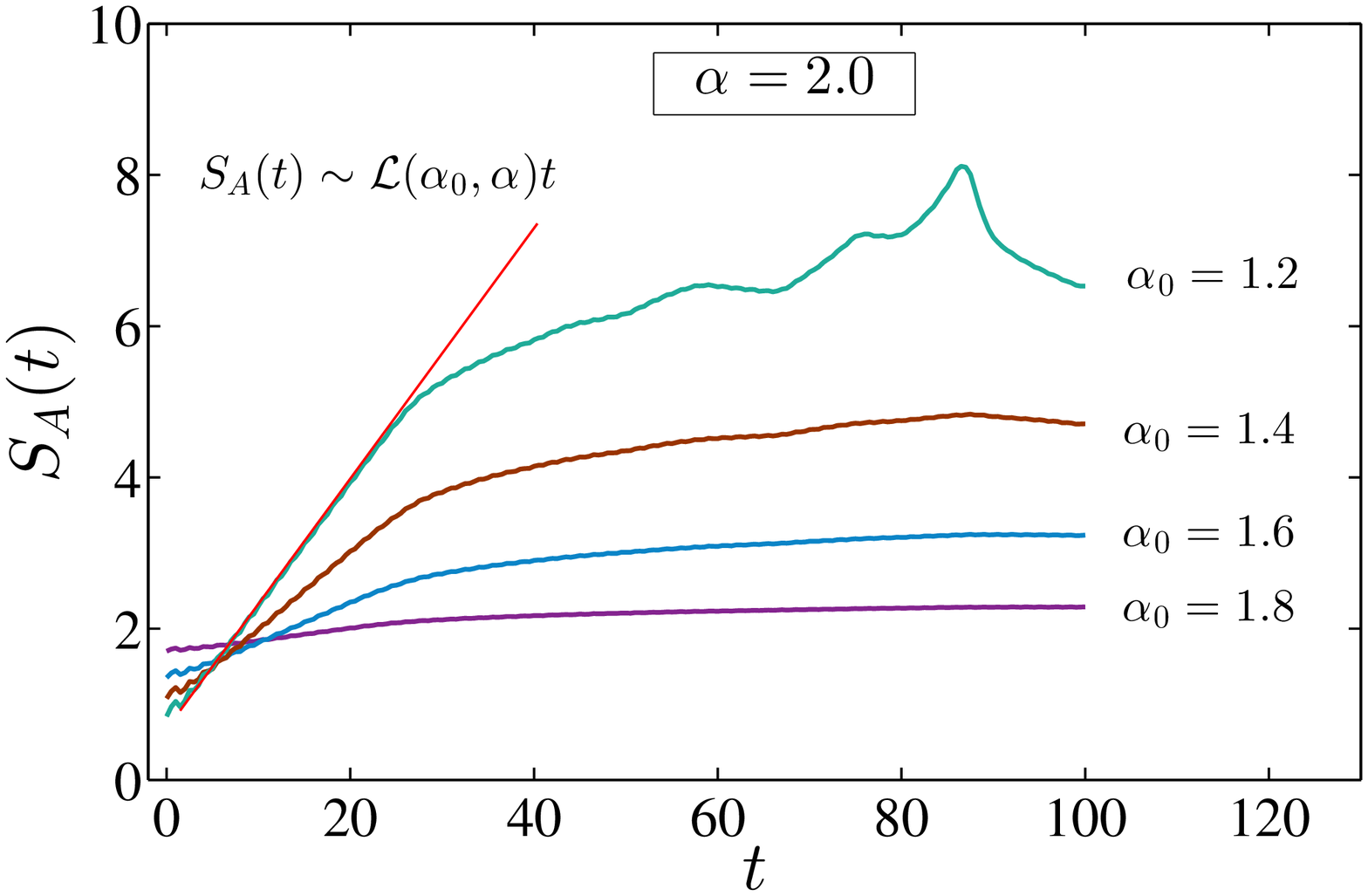}}
\centerline{
\includegraphics[scale=0.35]{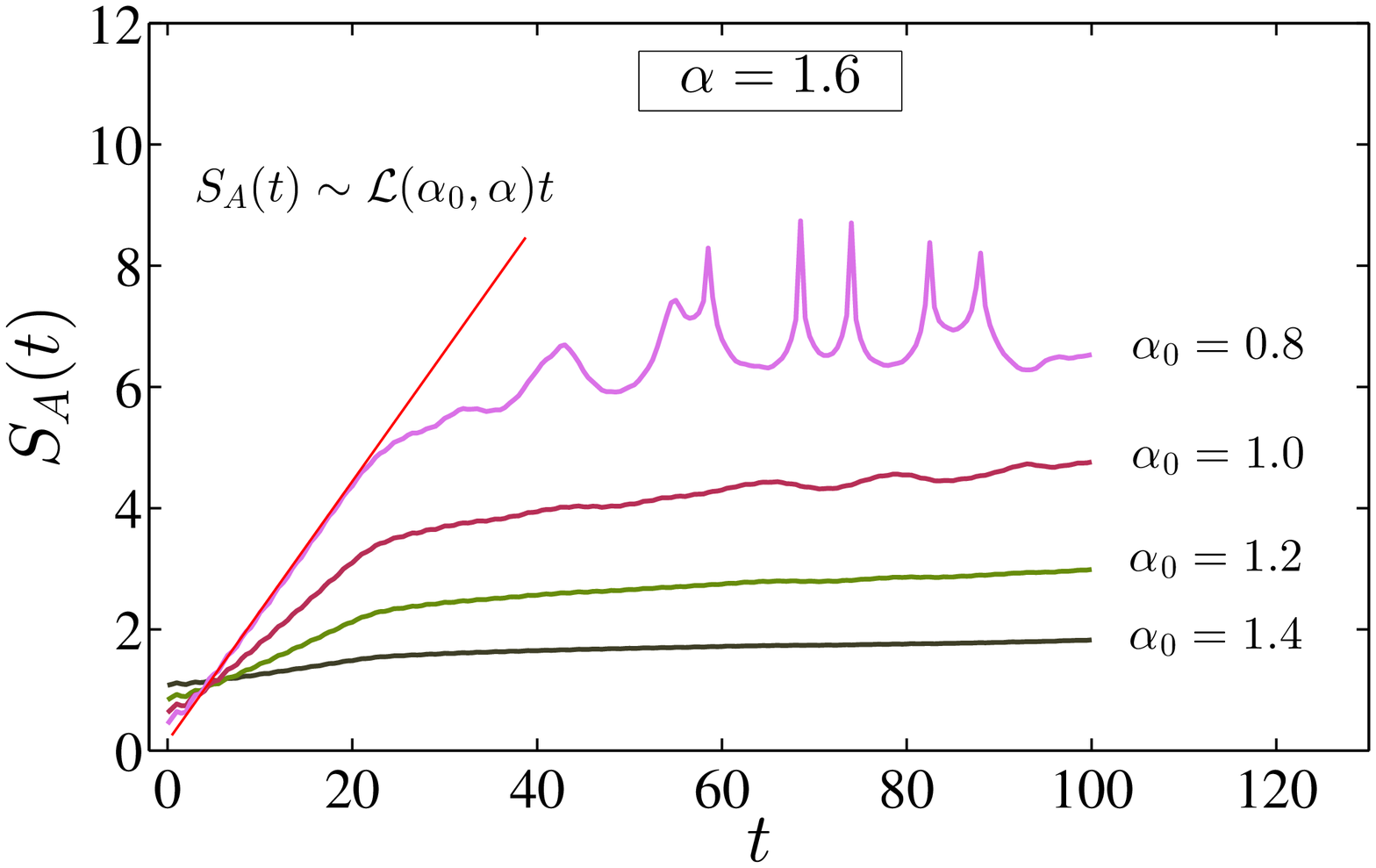}}
\centerline{
\includegraphics[scale=0.35]{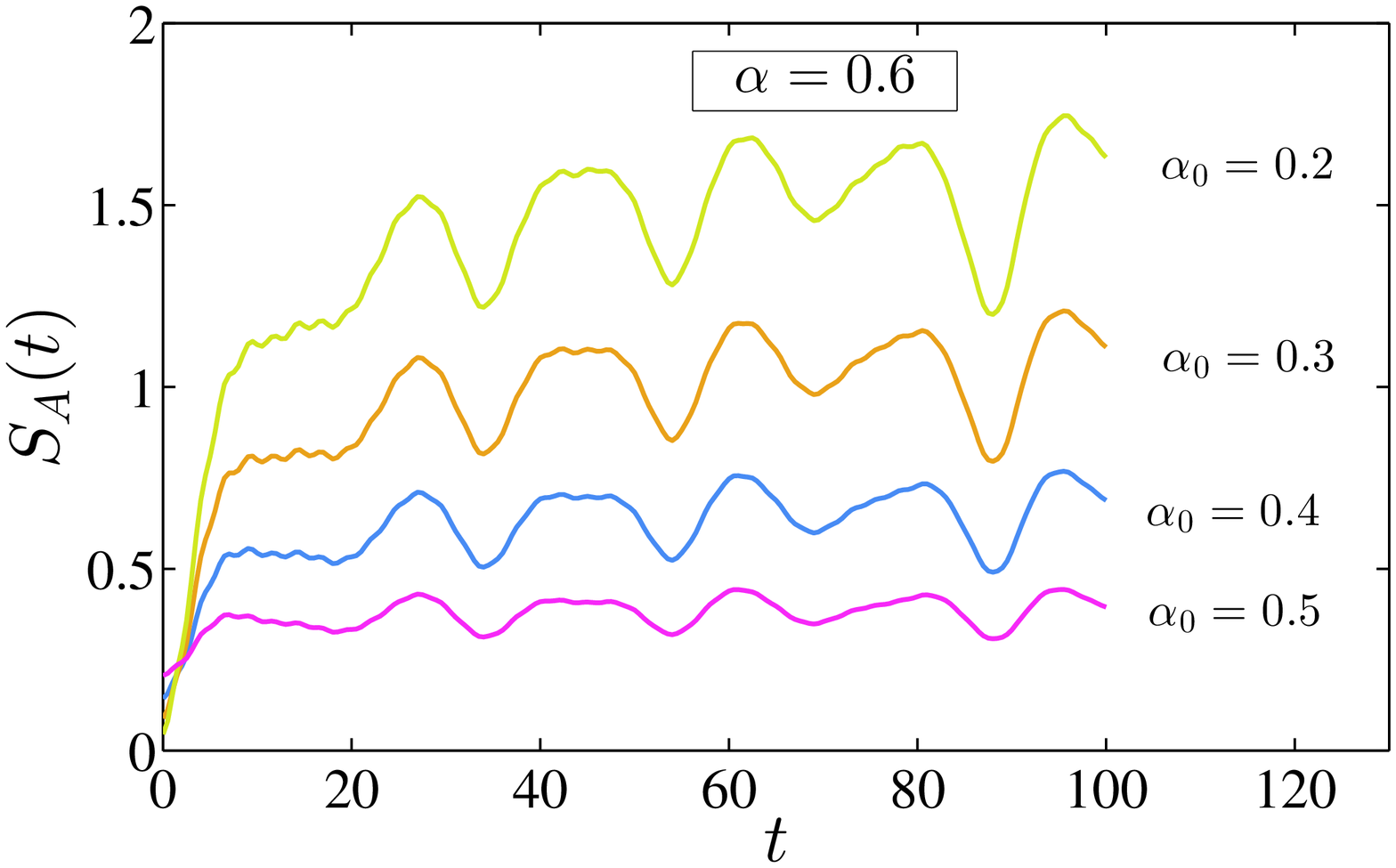}}
\caption{(Color online) Top: The time evolution of the von Neumann entropy $S_A(t)$ 
in the short-range harmonic oscillators
with the total size $N = 400$ when the initial state is gapless power-law 
with long-range scaling exponent $\alpha_0$. 
$S_A(t)$ starts from non-zero value and grows linearly with time then saturates. 
Middle: The time evolution of the von Neumann entropy $S_A(t)$ 
in the harmonic oscillators with weak long-range coupling ($\alpha = 1.6$).   
Bottom: The same result for the harmonic oscillators with strong long-range couplings ($\alpha = 0.6$).
}
\label{dynamics for quench from alpha0 to alpha}
\end{figure}

\section{Initial state effects}\label{initial state sec}
In the previous sections, we studied the time evolution of the entanglement 
entropy after global quantum quench. We have considered the harmonic oscillator
systems with short and long-range couplings.
The most crucial point about the long-range hamiltonian 
Eq. (\ref{interaction kernel long range}) is that
 the gapfull initial state has the power
 law correlation function (see Eq. (\ref{powerlawcorr case 1 massive})).
One of the fundamental questions that has to be addressed is how 
the entanglement dynamics is related to the initial state of the system. 
 Here, we would like to study
  the time evolution of entanglement entropy 
 for the long-range systems with the 
 gapless power-law, exponential decaying and unentangled initial states.

\subsection{Gapless power-law initial state}
\begin{figure} 
 \centerline{
\includegraphics[scale=0.35]{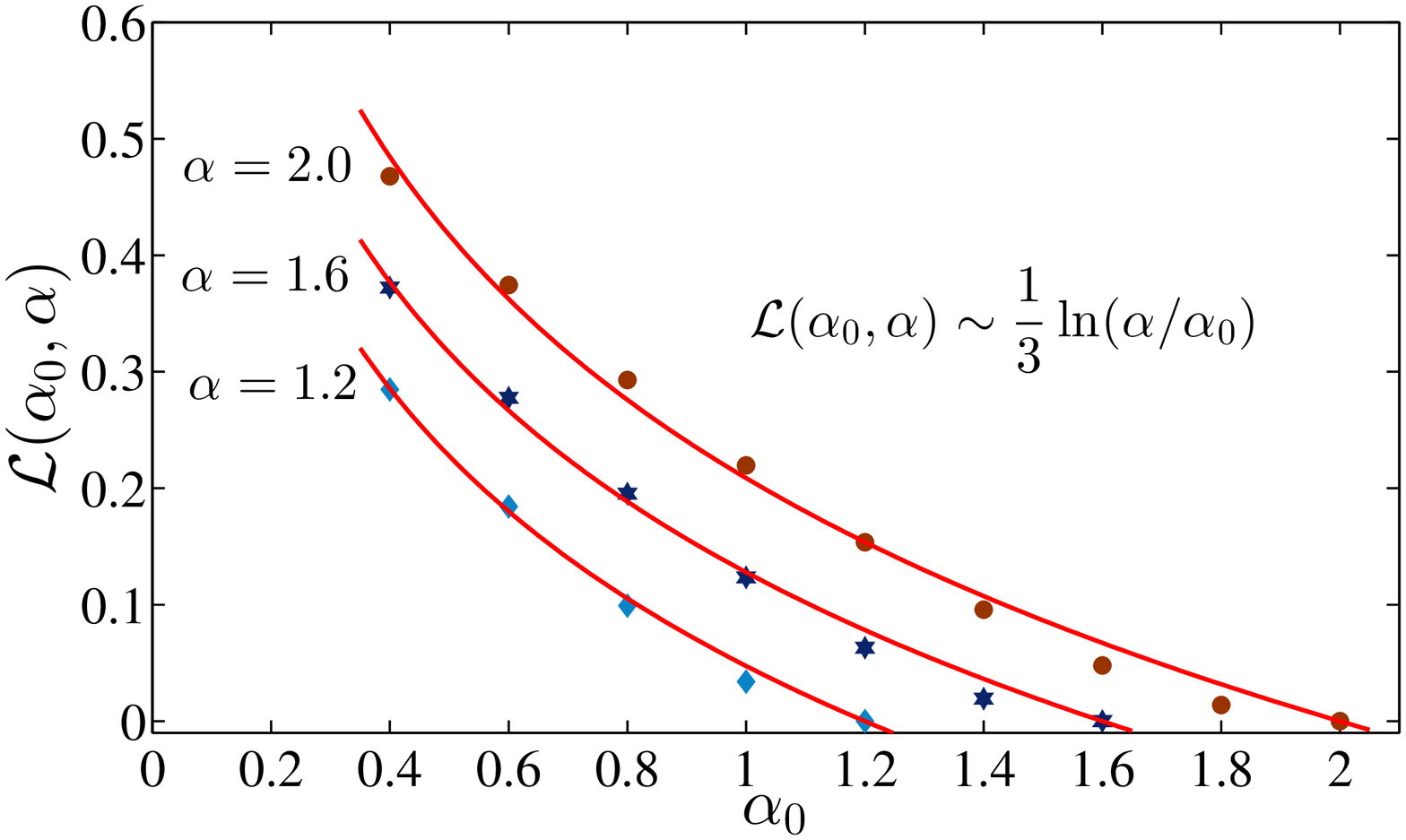}}
\centerline{
\includegraphics[scale=0.35]{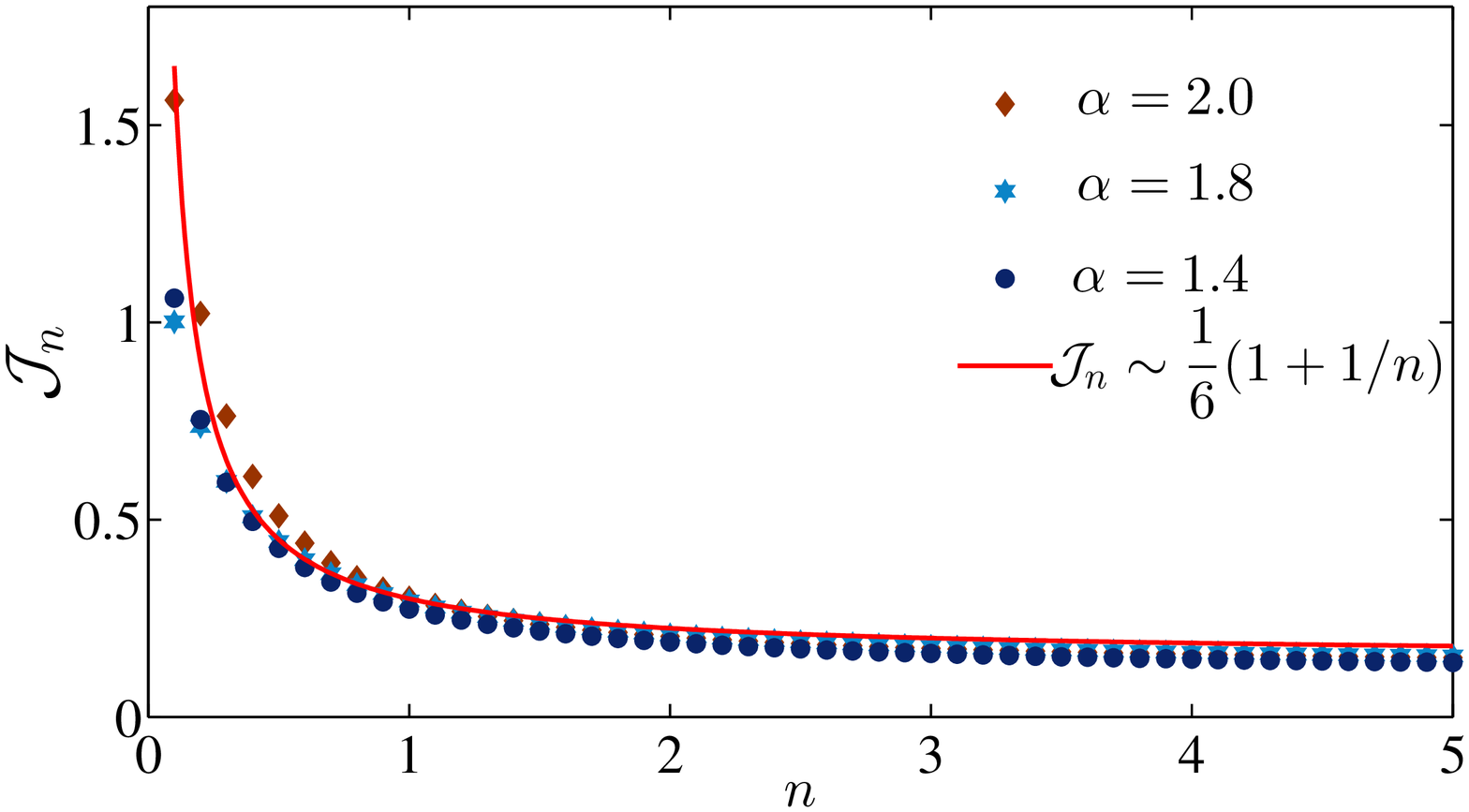}}
\caption{(Color online) Top: The numerical estimates of the prefactor $\mathcal{L}(\alpha_0,\alpha)$ for
different values $\alpha = 2.0, 1.6, 1.2$ as a function of $\alpha_0$.
 The solid red lines correspond to $\mathcal{L}(\alpha_0,\alpha) = \frac{1}{3}\ln (\alpha/\alpha_0)$.
Bottom: The prefactor $\mathcal{J}_n$ for the dynamics of the R\'enyi entropy
after quench from gapless power-law initial state 
(see Eq. (\ref{renyi quench from alpha0 to alpha}) and (\ref{prefactor renyi quench from alpha0 to alpha})). 
The solid red line corresponds to $\frac{1}{6}(1+1/n)$.   
}
\label{coefficient for quench from alpha0 to alpha}
\end{figure}

To extend our analysis to gapless initial state with
power-law correlation function, we first need to modify the previous
definition of the global quench. As mentioned earlier,
 for the gapless ($m=0$) long-range harmonic oscillators
 given by Eq. (\ref{fractional free field theory}), 
 the correlation function is of the form 
 $K^{-1/2}(r) \propto 1/r^{1-\alpha/2}$ (see Sec. (\ref{LRHO EEDynamics sec})).
Let us now consider the quench between gapless power law initial state
with the scaling exponent $\alpha_0$ to another gapless
 long-range hamiltonian with the scaling factor $\alpha$. We are 
 mainly interested in the general time and subsystem size dependence of
the entanglement entropy in the case $\alpha_0 \neq \alpha$. 
Here we assume that $\alpha_0 < \alpha$. Shown in
Fig. (\ref{dynamics for quench from alpha0 to alpha}) 
are plots of the time evolution of the von-Neumann 
entanglement entropy for various values of $\alpha_0$ and $\alpha$. 
We observe that the von-Neumann and R\'enyi entropies for the different 
values of the $\alpha_0$ and $\alpha$ grow linearly in time, 
then saturate at long time. We see that the numerical data are fit very well by
the following formula for the R\'enyi entropi
follows
\begin{eqnarray}\label{renyi quench from alpha0 to alpha}
S_n(t) = \frac{\tilde{c}_n(\alpha_0)}{3}\log l +
\left\lbrace
  \begin{array}{l l }
    \mathcal{L}_n(\alpha_0,\alpha) t & \quad t < t^*\\
    \\
    \mathcal{L}_n(\alpha_0,\alpha)b(\alpha)~l  &\quad t > t^*
  \end{array}\right.
  ~,
\end{eqnarray} 
where $l$ is the size of the subsystem and 
the prefactor $\tilde{c}_n(\alpha_0)$  is introduced in 
Ref. \cite{Nezhadhaghighi2013} and $t^* = b(\alpha)~l$. 
Since the coefficient $\mathcal{L}_n(\alpha_0,\alpha)$ only
depends on $n$ and the initial and final scaling parameters $\alpha_0$ and $\alpha$,
we obtain the asymptotic relation
\begin{eqnarray}\label{prefactor renyi quench from alpha0 to alpha}
\mathcal{L}_n(\alpha_0,\alpha) = \mathcal{J}_n \ln (\alpha/\alpha_0)~,
\end{eqnarray} 
where our numerical results suggest that $\mathcal{J}_n = \frac{1}{6}(1+1/n)$. 
In Fig. (\ref{coefficient for quench from alpha0 to alpha})
the prefactor $\mathcal{L}(\alpha_0,\alpha)$ (for $n=1$)
as a function of $\alpha_0$ and $\alpha$ is reported. 
The numerical results for the prefactor $\mathcal{J}_n$ is shown in 
the bottom panel of Fig. (\ref{coefficient for quench from alpha0 to alpha}),
pointing at a good agreement between $\mathcal{J}_n = \frac{1}{6}(1+1/n)$ and
our numerical results.  

\begin{figure} 
\centerline{
\includegraphics[scale=0.35]{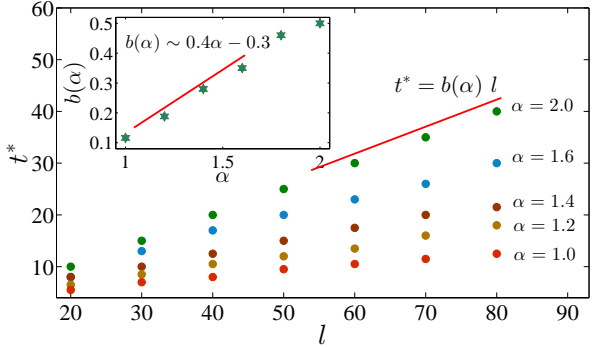}}
\caption{(Color online) The relation between the saturation time $t^*$ 
and the subsystem size $l$ for different values $\alpha$. 
The solid red line represents $t^* = b(\alpha)~l$.
Inset: The prefactor $b(\alpha)$ as a function of $\alpha$ 
shows the linear behavior $b(\alpha) \simeq 0.4\alpha -0.3$.   
}
\label{tstar for quench from alpha0 to alpha}
\end{figure}

We should note that our numerical calculations show that 
for the fixed values of the scaling parameters $\alpha_0$ and $\alpha$,
the saturation regime begins earlier for smaller value of the subsystem size $l$. 
In Fig. (\ref{tstar for quench from alpha0 to alpha}) we show numerical results for the saturation time $t^*$ as function
of the subsystem size $l$ for different values of $\alpha$ (the results are $\alpha_0$ independent). 
The figure shows that $t^*$ is a linear function of $l$ as $t^* = b(\alpha)~l$. 
As mentioned previously, the saturation time for the gapfull power-law 
initial state scales like Eq. (\ref{saturation time eq for LRHO}).
The specific reason for the linear behavior in the case of 
gapless power-law initial state remains unsolved.
The parameter $b(\alpha)$ is shown in the Fig. (\ref{tstar for quench from alpha0 to alpha})
and best fit to our numerical data was $b(\alpha) \simeq 0.4\alpha -0.3$.
The value of $b(\alpha)$ is zero at $\alpha \sim 0.75$.  
It is interesting to note that for those long-range harmonic oscillators
with strong long-range couplings ($\alpha<1.0$) the typical behavior of
the entanglement entropy changes qualitatively (see Fig. (\ref{dynamics for quench from alpha0 to alpha})). 
In other words, there is no linear behavior in the entanglement growth. 

\subsection{Exponential decaying initial state} 
 In order to analyze the 
 initial state effects, we define the
hamiltonian of the long-range harmonic oscillator with the following $K$
matrix:
\begin{eqnarray}\label{LRHOexpcorr}
K_{i,j} &= {\int_0^{2\pi} \frac{dq}{2\pi} e^{iq(i-j)}  
\left[ \left(2-2\cos(q)\right) + m^2\right]^{\frac{\alpha}{2}}}~,
\end{eqnarray}
where $m$ is a positive real number. 
The interesting point about Eq. (\ref{LRHOexpcorr}) is that
 one can easily find the correlation function
 $\langle \phi_l \phi_{l+n} \rangle \equiv K^{-1/2 }(n)$ 
 explicitly and the result is
 \begin{eqnarray}
 K^{-1/2 }(r) &= \int_{-\infty}^{\infty} e^{iqr}\left(|q|^2 + m^2 \right)^{-\alpha/4} dq\\ \nonumber
 &= \frac{\sqrt{\pi} 2^{(\alpha-6)/4}}{\Gamma(\alpha/4)}(\frac{r}{m})^{(\alpha-2)/4}\mathbf{K}\left[(\alpha - 2)/4,mr\right]~, 
 \end{eqnarray}
 where $\mathbf{K}$ is the modified Bessel function. 
 The large distance asymptotic behavior of the correlation function is
 \begin{eqnarray}
 K^{- 1/2}(r) \propto e^{-mr}/r^{(\alpha-4)/4}~.
 \end{eqnarray}
Above result leads to a
finite spatial correlation length $\xi^{-1}_s \propto m$. 

\begin{figure}
\centerline{
\includegraphics[scale=0.35]{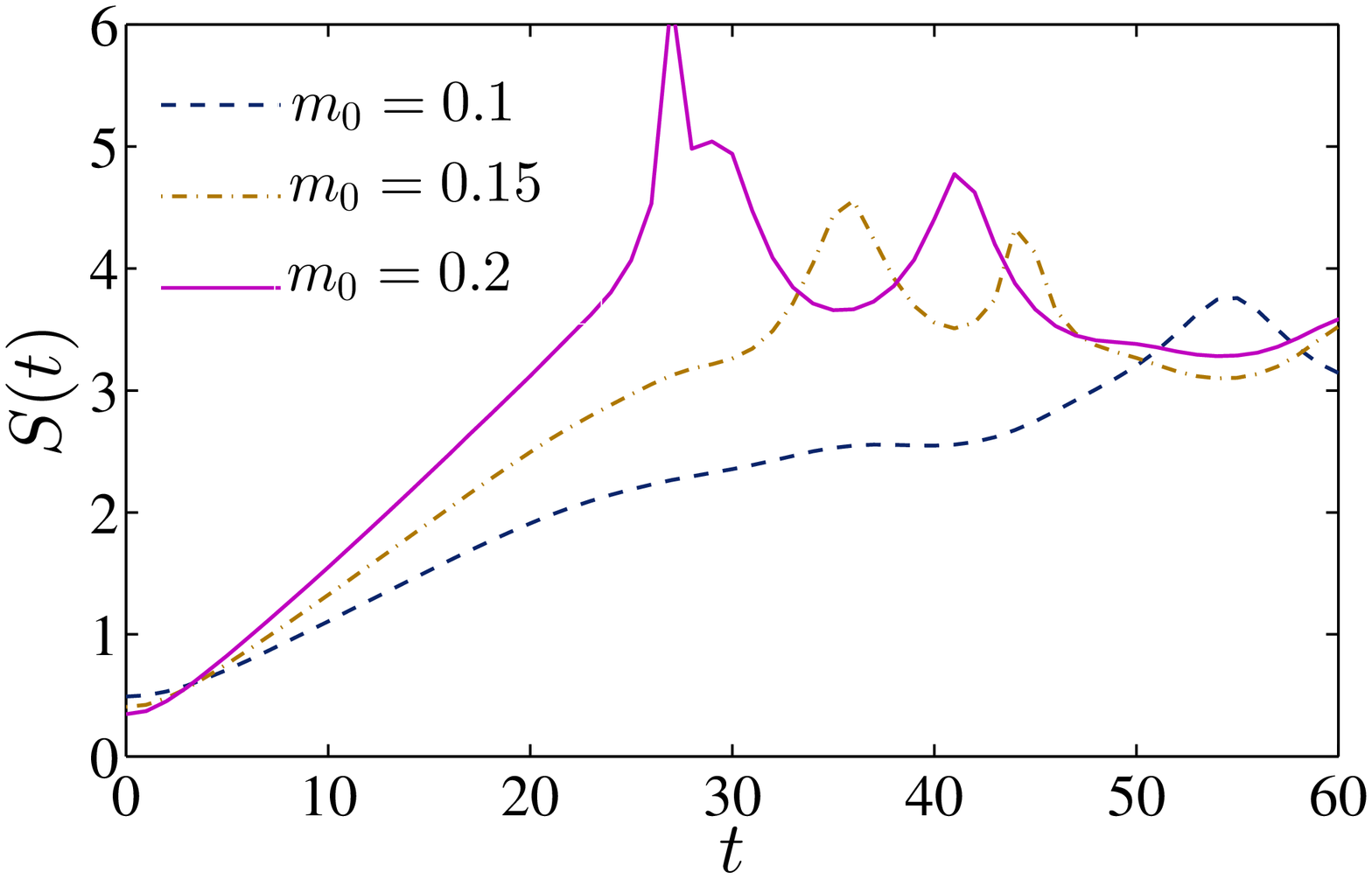}
}
\centerline{
\includegraphics[scale=0.35]{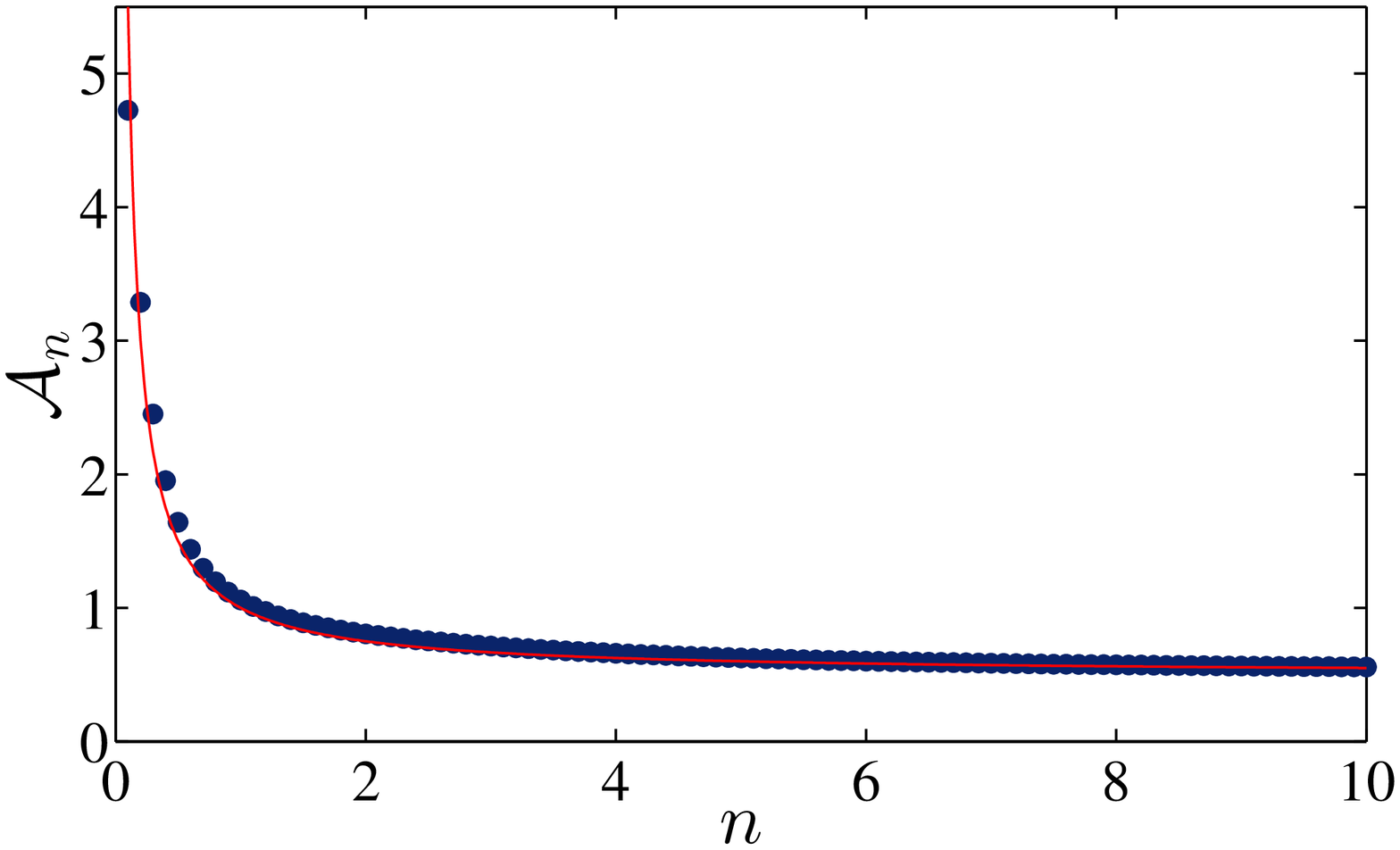}
}
\caption{(Color online) Top: Entanglement entropy dynamics $S_A(t)$ for long-range harmonic
 oscillator Eq. (\ref{LRHOexpcorr}) with an exponential decaying correlation function. 
 The total size of the system is $N = 400$, the subsystem size $l = 50$ and 
 $\alpha =1.5$. Different lines corresponds to 
 different values of the mass parameter $m_0$. 
 Bottom: Prefactor $\mathcal{A}_n$ 
for dynamics of the R\'enyi entropy 
for the long-range harmonic oscillator system ($\alpha=1.5$).
The solid red line
 represents $\frac{1}{2}(1+1/n)$.
  }
\label{EE LRHO15 exp init condition}
\end{figure}

Using the technique of the previous sections,
 we provided the numerical test for the entanglement evolution $S_A(t)$ of
the long-range harmonic oscillators ($\alpha\neq 2$ in Eq. (\ref{LRHOexpcorr})). 
In our study, initially ($t=0$) the harmonic chain is massive $m_0>0$, 
and after the quench for $t>0$ 
the hamiltonian (\ref{LRHOexpcorr}) is critical with $m=0$.

Our results for $\alpha >1$ clearly shows that the qualitative
and quantitative behavior of $S_A(t)$ for $t<t^*$ are similar to those 
long-range
systems with power-law correlation functions. 
For example we find numerically
that the entanglement dynamics $S_A(t\ll t^*)$ in very short time
have the same quadratic growth as the previous sections. 
A more careful analysis also
shows that the $S_A(t)$ grows linearly in time 
until the saturation time $t^*$. Our numerical estimate of the linear
coefficient $\mathcal{A}\sim \pi/6$ was in agreement with 
the Eq. (\ref{EE dynamic for LRHO}). 
In summary for the long-range harmonic oscillators with $\alpha>1$ 
we get the following results: 
\begin{eqnarray}
S_A(t) \equiv  - \frac{c^g(\alpha)}{3}\log m_0 +
\left\lbrace
  \begin{array}{l l }
    \kappa_2 t^2 & \quad t \ll 1\\
    \\
    \mathcal{A} m_0^{\alpha/2} t  &\quad t < t^*
  \end{array}\right.
  ~.
\end{eqnarray} 
In the above equation we checked that the prefactors $c^g(\alpha)$ and $\kappa_2$
are same as that in Eq. (\ref{EE dynamic for LRHO}).
 
Now we can ask, what is the time-dependent behavior of R\'enyi entropy?
 We have seen that the results for $t<t^*$ is exactly the same
as for the previous sections. As shown in 
Fig. (\ref{EE LRHO15 exp init condition}) the 
 agreement between the previous results given
by Eq. (\ref{renyi dynamic for LRHO}) and 
Fig. (\ref{dynamic case 2 An and Bn for renyi long range })
are fairly good.

Unfortunately, in the saturation regime it was difficult to study the 
scaling behavior of the $S_A(t\to \infty)$ and $S_n(t)$ with 
the mass parameter $m_0$ and the subsystem size $l$
 because of the rapid oscillations
in $S_A(t)$ and $S_n(t)$.

\begin{figure}
\centerline{
\includegraphics[scale=0.35]{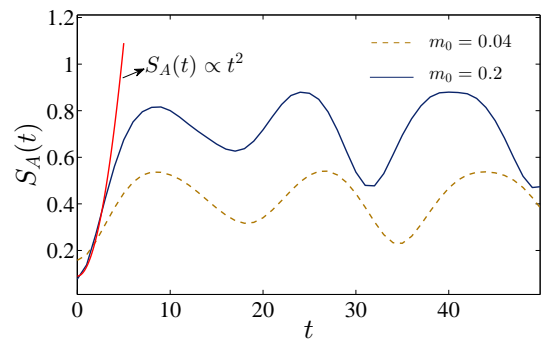}
}
\caption{(Color online) Entanglement entropy dynamics $S_A(t)$ for long-range harmonic
 oscillator Eq. (\ref{LRHOexpcorr}) with strong coupling ($\alpha <1$)
  which the initial state's correlation function decays exponentially. 
 The total size of the system is $N = 400$, the subsystem size $l = 50$ and 
 $\alpha =0.6$. Different lines corresponds to 
 different values of the initial mass parameter $m_0$. 
  }
\label{EE strongLRHO exp init condition}
\end{figure}

It is also straightforward to change the initial state for the harmonic system
with strongly long-range couplings. 
In Fig .(\ref{EE strongLRHO exp init condition}),
we plot the time evolution of the entanglement
entropy for the harmonic oscillators with 
very long-range interactions ($\alpha <1$). Here we only find rapid oscillations
 around a constant value which depends to the subsystem size $l$ 
 and mass parameter $m_0$. For short time as we expect,
  the $S_A(t)$ grows quadratically in time. The key observation here 
  is that the initial state of the system
  completely changes the behavior of the entanglement dynamics for strongly 
  coupled harmonic oscillators. 
 
Finally let us now consider the maximum group velocity for the quasiparticles 
which produced after quantum quench from exponential decaying initial state 
to the critical point of the hamiltonian Eq. (\ref{LRHOexpcorr}).  
When the system size $N$ is large enough, 
it is easy to verify that Eq. (\ref{LRHOexpcorr}) leads to 
$v_g(k) = \frac{\alpha}{2} |k| \left[ |k|^2 + m^2 \right]^{\alpha/4 -1}$
and one can obtain an exact result for the maximum group velocity
 $v_g^{max}\propto m ^{\alpha/2 -1}$. When the system is quenched to
  the critical point ($m=0$), $v_g^{max}$ diverges and
   there is no maximum group velocity for
 the quasiparticles created during the quench.  
 We emphasize that in the case where the long-range harmonic
system is weakly coupled, there is no maximum group velocity
but the $S_A(t)$ grows linearly with time. 
This means that in our long-range system, there is no direct signature of
 energy quasiparticles in the special behavior of the 
entanglement entropy growth.

\subsection{Unentangled initial state}
In this subsection we analyzed the case in which
the harmonic oscillator system is initially (at $t=0$) prepared in a 
given unentangled state. Therefore, we begin with a chain in the absence of
any couplings between two different sites of the chain and then 
quench the system to the critical point of 
the hamiltonian Eq. (\ref{interaction kernel long range}). 
In this case the initial entanglement entropy is zero ($S_A(t=0)=0$). 

\begin{figure} 
 \centerline{
\includegraphics[scale=0.35]{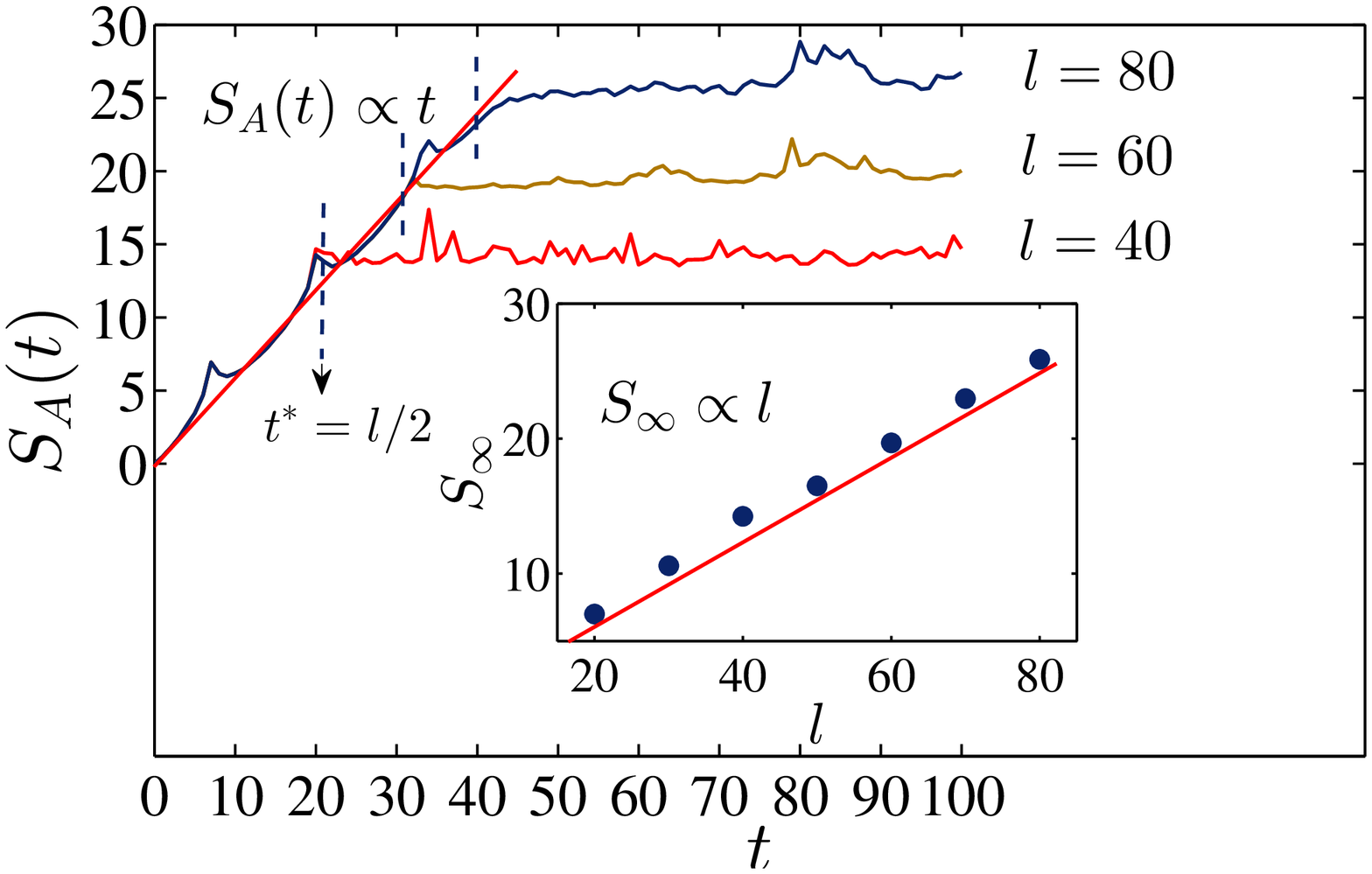}}
\centerline{
\includegraphics[scale=0.35]{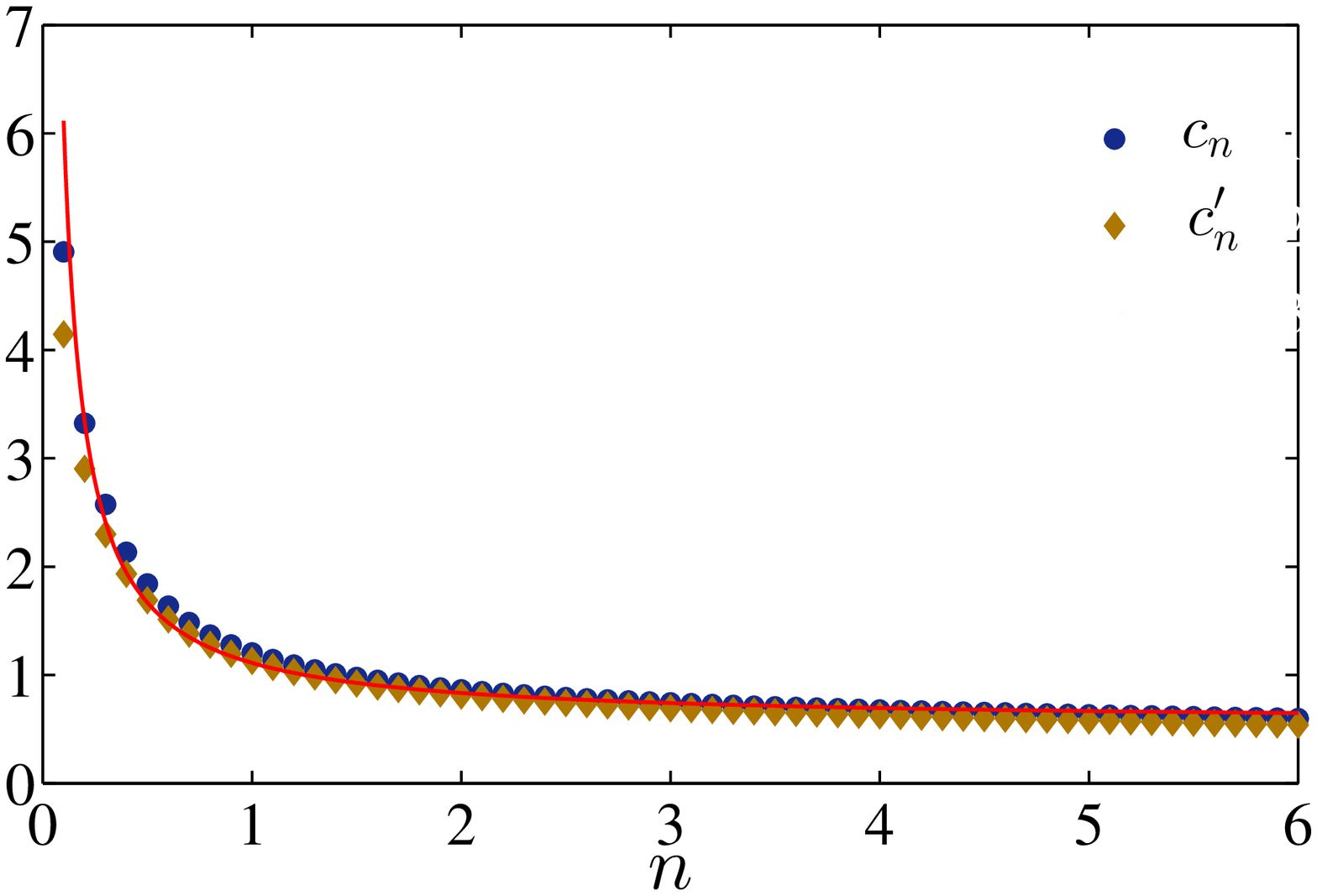}}
\caption{(Color online) Top: The time evolution of the von Neumann entropy $S_A(t)$ 
in the short-range harmonic oscillators
with the total size $N = 500$ when the initial state is uncoupled. 
$S_A(t)$ starts from zero and grows linearly with time then saturates to 
the value $S_{\infty}$ which is a linear function of subsystem size $l$ (see inset). 
Bottom: The numerical estimates of the prefactors $c_n$ and $c^\prime_n$. The solid
red line corresponds to $\frac{1}{2}(1+1/n)$.   
}
\label{dynamics for uncopupled state to SRHO}
\end{figure}

We first consider the time evolution of the entanglement entropy
 of short-range harmonic oscillators with configuration $\mathfrak{g}_1$.
  We present results from 
 numerical calculations of the von Neumann and R\'enyi entropies according to 
 Eqs. (\ref{entanglment entropy formula}) and (\ref{renyi entropy formula}), 
 respectively. The evolution of von Neumann entropy $S_A(t)$
  is shown in Fig. (\ref{dynamics for uncopupled state to SRHO}) which clearly showing 
that $S_A(t)$ exhibits linear behavior in
 time $t$ until it saturates at $t^* = l/2$. However, some peaks
 appear in the linear regime for short periods of time which we believe 
 that the peaks will disappear in $N\to \infty$ limit.

 All the data of the time evolution of von Neumann and
 R\'enyi entropies are well fitted by
 \begin{eqnarray}
S_n(t) =
\left\lbrace
  \begin{array}{l l}
    \frac{\pi c_n}{6} t & \quad t < l/2\\
    \\
    \frac{\pi c^\prime_n}{12} l  &\quad t>l/2
  \end{array}\right.
  ~,
 \end{eqnarray}
where in our numerical simulations, we found $c_n = \frac{1}{2}(1+1/n)$
and $c^\prime_n = \frac{1}{2}(1+1/n)$. 
In Fig. (\ref{dynamics for uncopupled state to SRHO}), we show our numerical results for the
prefactors $c_n$ and $c^\prime_n$. 

\begin{figure} 
 \centerline{
\includegraphics[scale=0.35]{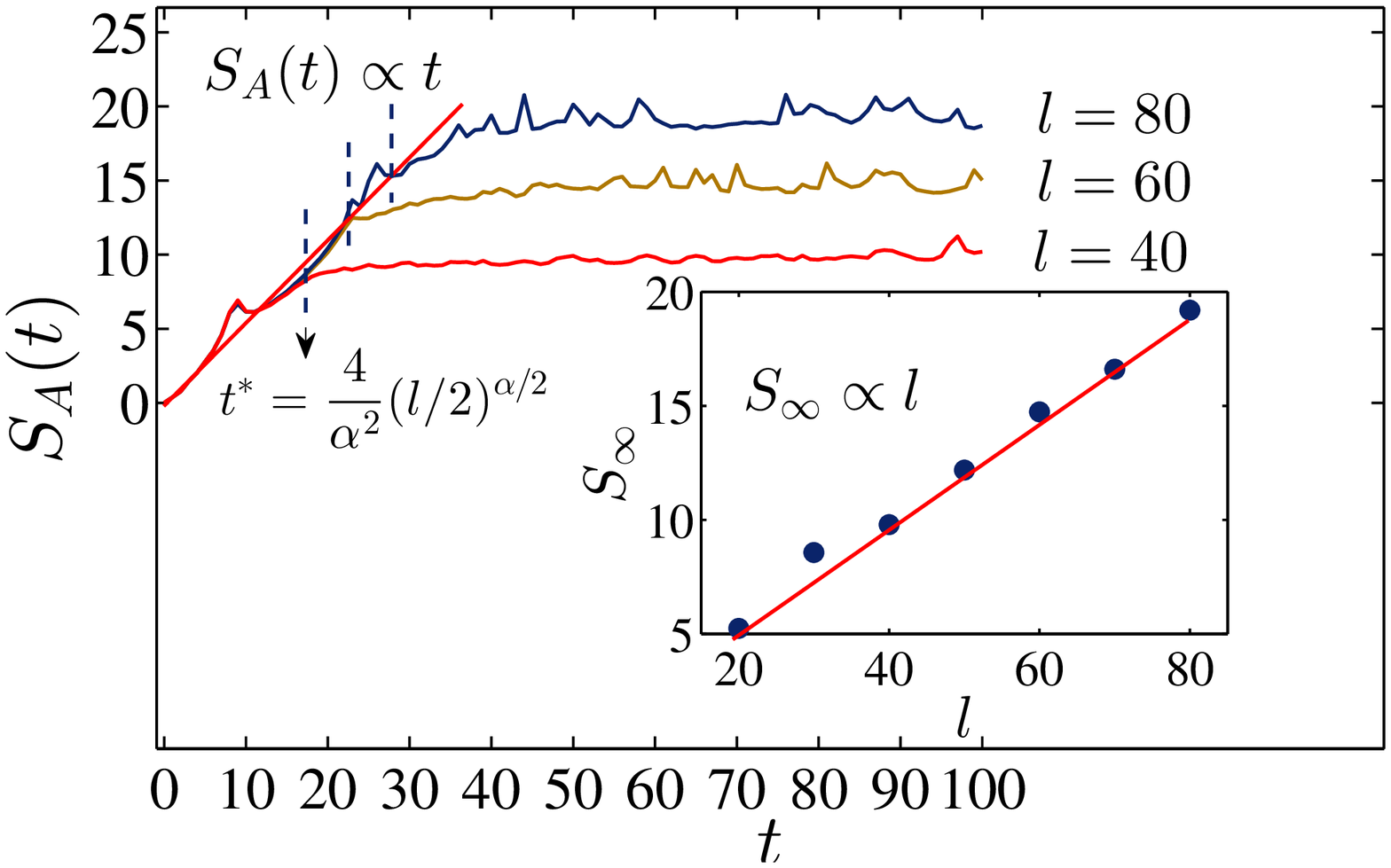}}
\centerline{
\includegraphics[scale=0.35]{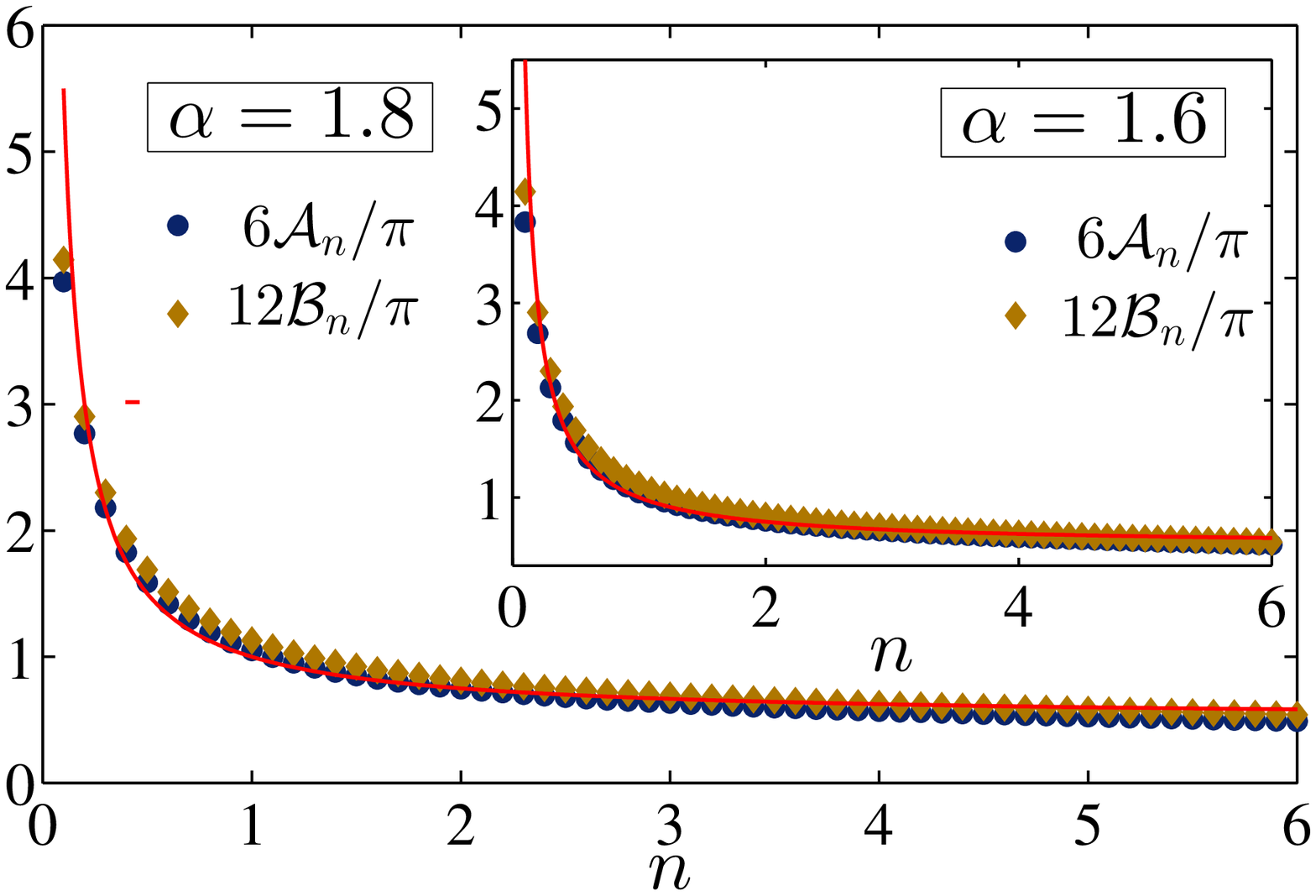}}
\caption{(Color online) Top: The time evolution of the von Neumann entropy $S_A(t)$ 
in the long-range harmonic oscillators ($\alpha=1.5$)
with the total size $N = 500$ when the initial state is uncoupled. 
$S_A(t)$ starts from zero and grows linearly with time then saturates to 
the values $S_{\infty}$ which is a linear function of subsystem size $l$ (see inset). 
Bottom: The numerical estimates of the prefactors $\mathcal{A}_n$ and $\mathcal{B}_n$ 
for different value $\alpha=1.6, 1.8$. The solid
red lines correspond to $\frac{1}{2}(1+1/n)$.   
}
\label{dynamics for uncopupled state to LRHO}
\end{figure}

We also studied time evolution of the R\'enyi entropy 
 in the long-range
harmonic oscillators with unentangled initial state. 
The resulting values of entanglement dynamics $S_A(t)$ for
 long-range harmonic chain with $\alpha=1.5$
 is shown in Fig. (\ref{dynamics for uncopupled state to LRHO}).
  For weakly coupled long-range harmonic oscillators $1<\alpha<2$ 
we found that the best fit for the data is
\begin{eqnarray}
S_n(t) =
\left\lbrace
  \begin{array}{l l}
     \mathcal{A}_n t & \quad t < t^*\\
    \\
    \mathcal{B}_n l  &\quad t> t^*
  \end{array}\right.
  ~.
 \end{eqnarray}
Interestingly, we find that the best fit to $\mathcal{A}_n$ and 
 $\mathcal{B}_n$ is
$\frac{6\mathcal{A}_n}{\pi} = \frac{12\mathcal{B}_n}{\pi} = \frac{1}{2}(1+1/n)$.  
The coefficients $\mathcal{A}_n$ and $\mathcal{B}_n$ 
as functions of $n$ are represented in Fig. (\ref{dynamics for uncopupled state to LRHO}).  
We then compute the saturation time $t^*$ which our numerical results 
clearly show that $t^* = \frac{4}{\alpha^2} (l/2)^{\alpha/2}$.  

Finally, we should mention that the time evolution of the von Neumann entropy 
which the initial state is uncoupled and the final state is
the critical state of the strongly long-range harmonic oscillator, follows 
a simple formula 
\begin{eqnarray}
S_A(t)=  \mathcal{P}(l,\alpha) \log t ~,
\end{eqnarray} 
where $\mathcal{P}(l,\alpha)= \mathcal{V}_1 (\alpha) (l)^{\alpha/2} +\mathcal{V}_1 (\alpha)$ 
and the prefactors $\mathcal{V}_1$ and $\mathcal{V}_2$ are the same as those
introduced in Sec. (\ref{LRHO EEDynamics sec}). 

 At this stage we should mention that our calculations show a universal
  behavior in the time evolution of the von Neumann and R\'enyi entropies
   for weakly coupled long-range harmonic oscillators ($1<\alpha<2$). 
  In other words initial state of the system does not change the
  behavior of the entanglement dynamics. It is worth mentioning that the 
  prefactors of the linear parts of $S_A(t)$ and $S_n(t)$
   in weak couplings regime ($1<\alpha<2$) 
   are not affected by the initial states of the system. 
   Another interesting aspect to consider
   is that the entanglement dynamics for strongly coupled harmonic oscillators,
   identifying qualitatively different behavior as the initial state of the 
   system varied.

\section{Mutual information dynamics}\label{mutual section}
\begin{figure*}[htb]
\begin{center}
\includegraphics[angle=0,scale=0.4]{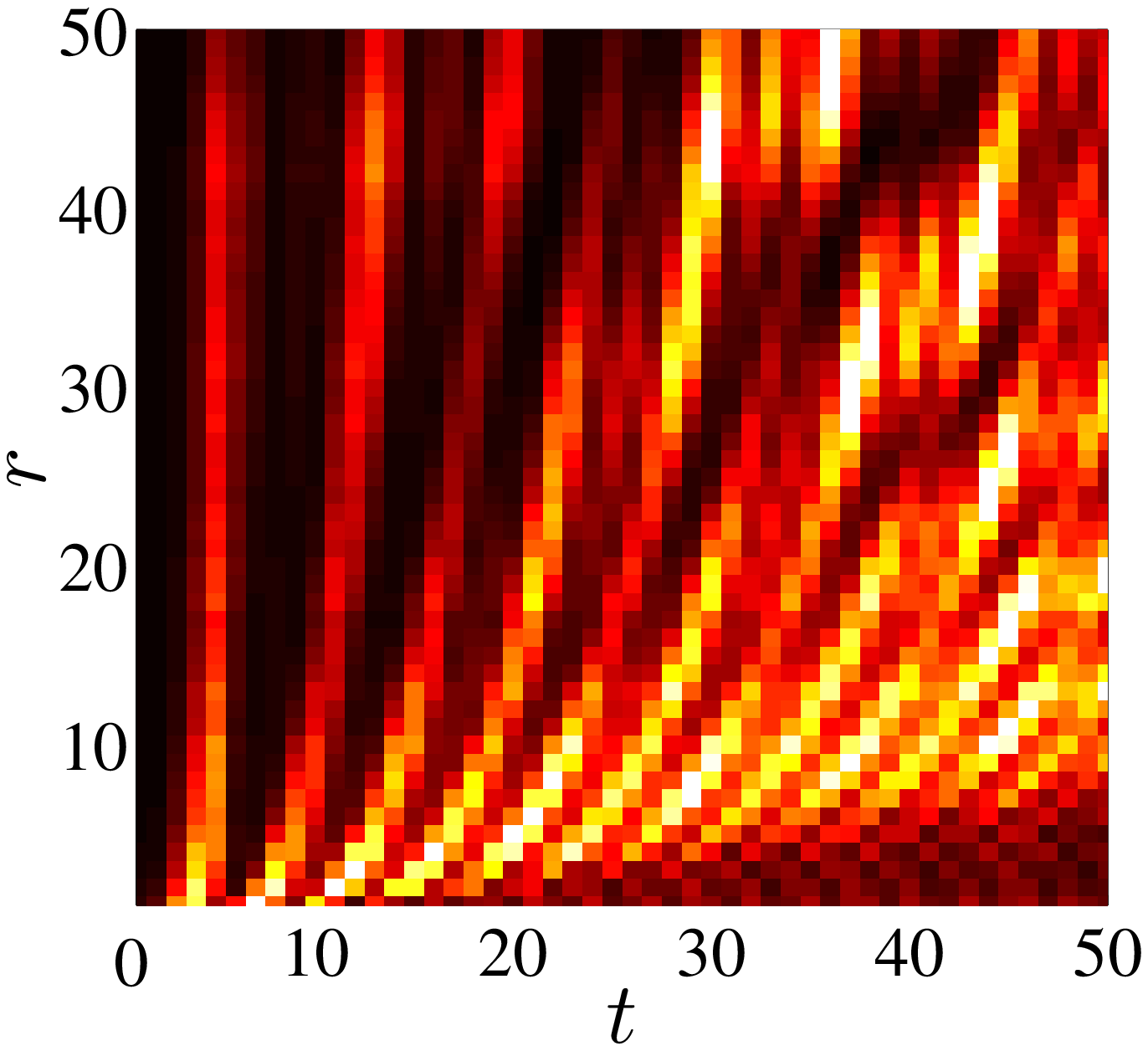}
\includegraphics[angle=0,scale=0.4]{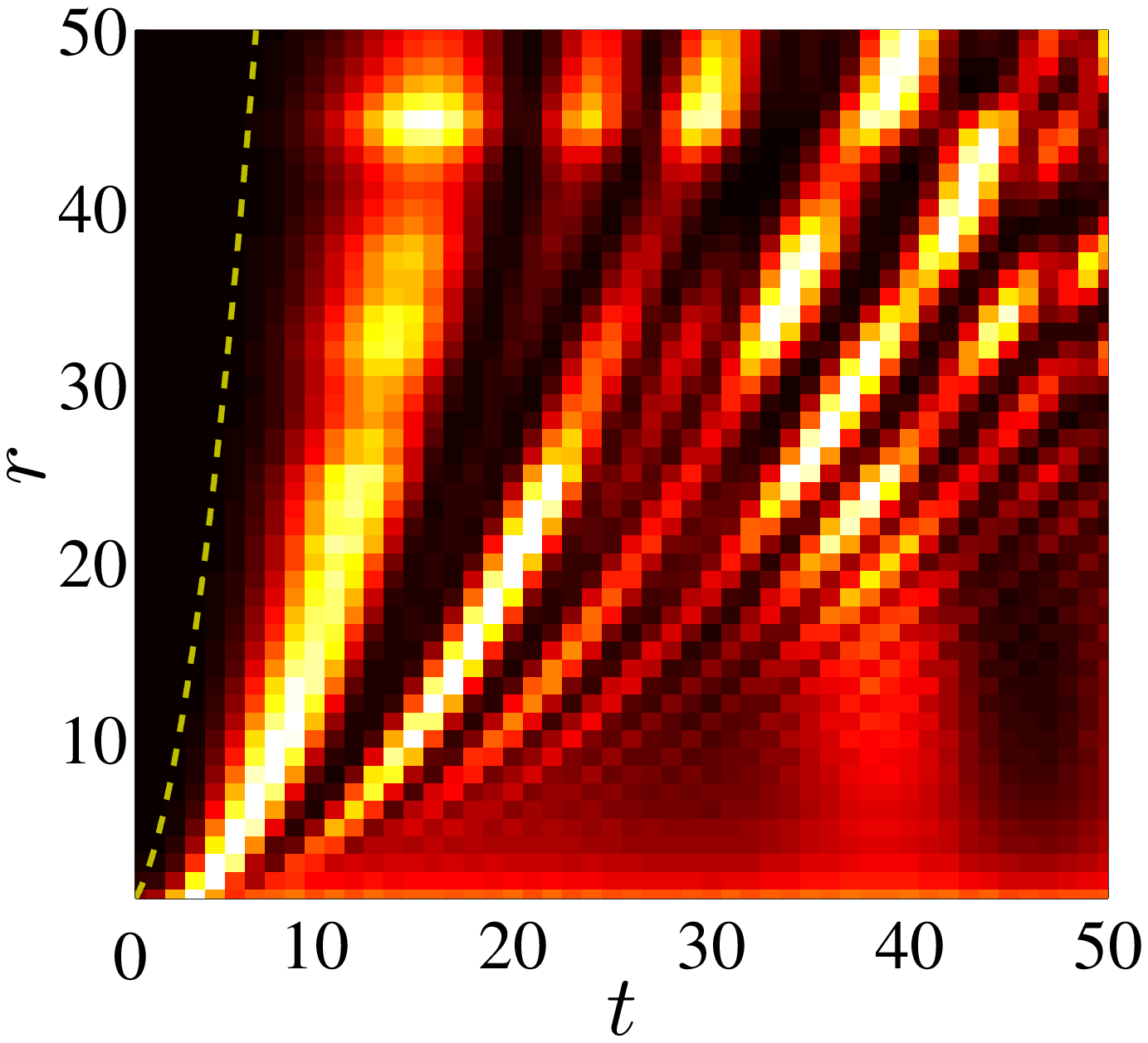}
\includegraphics[angle=0,scale=0.4]{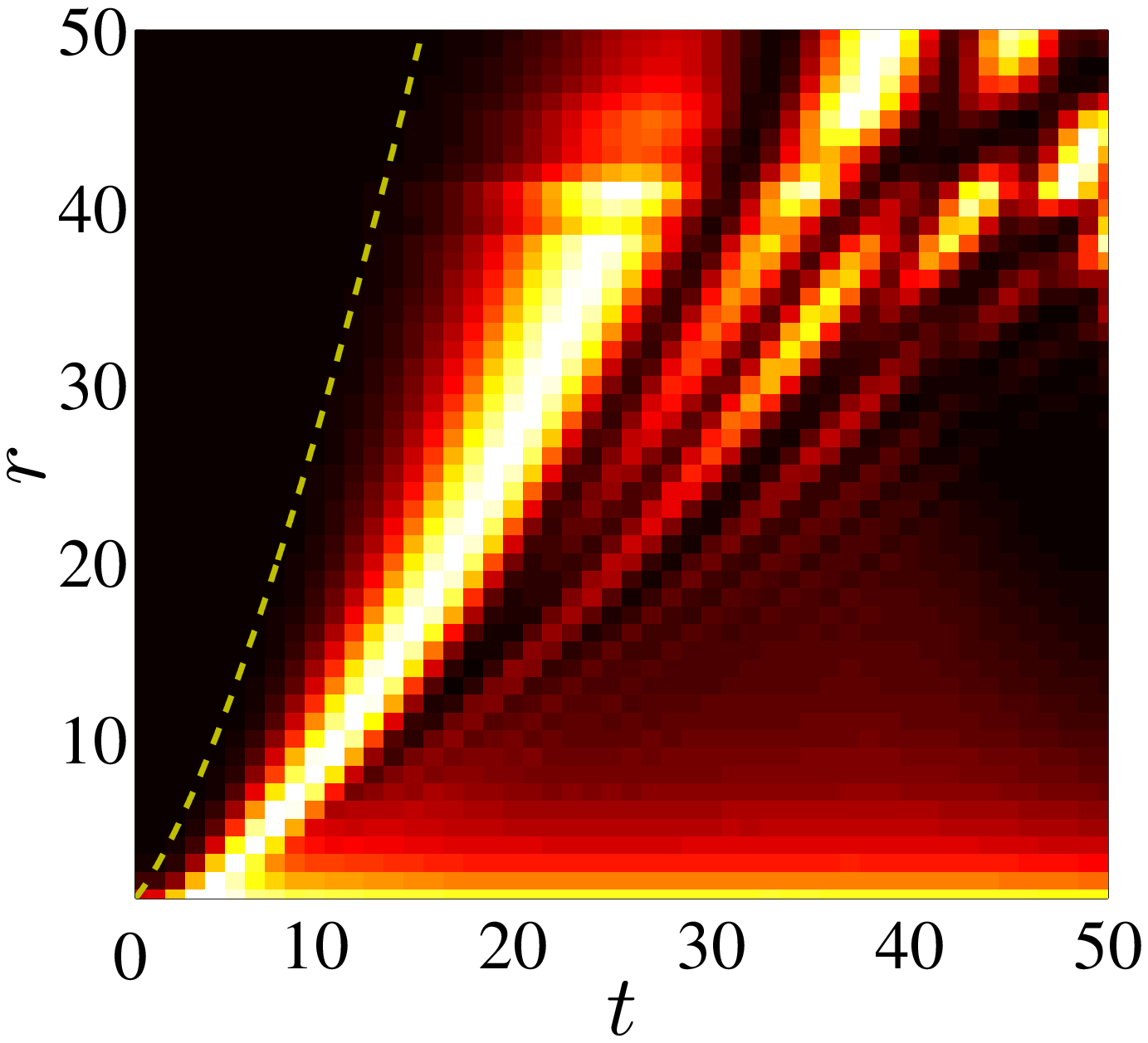}
\includegraphics[angle=0,scale=0.4]{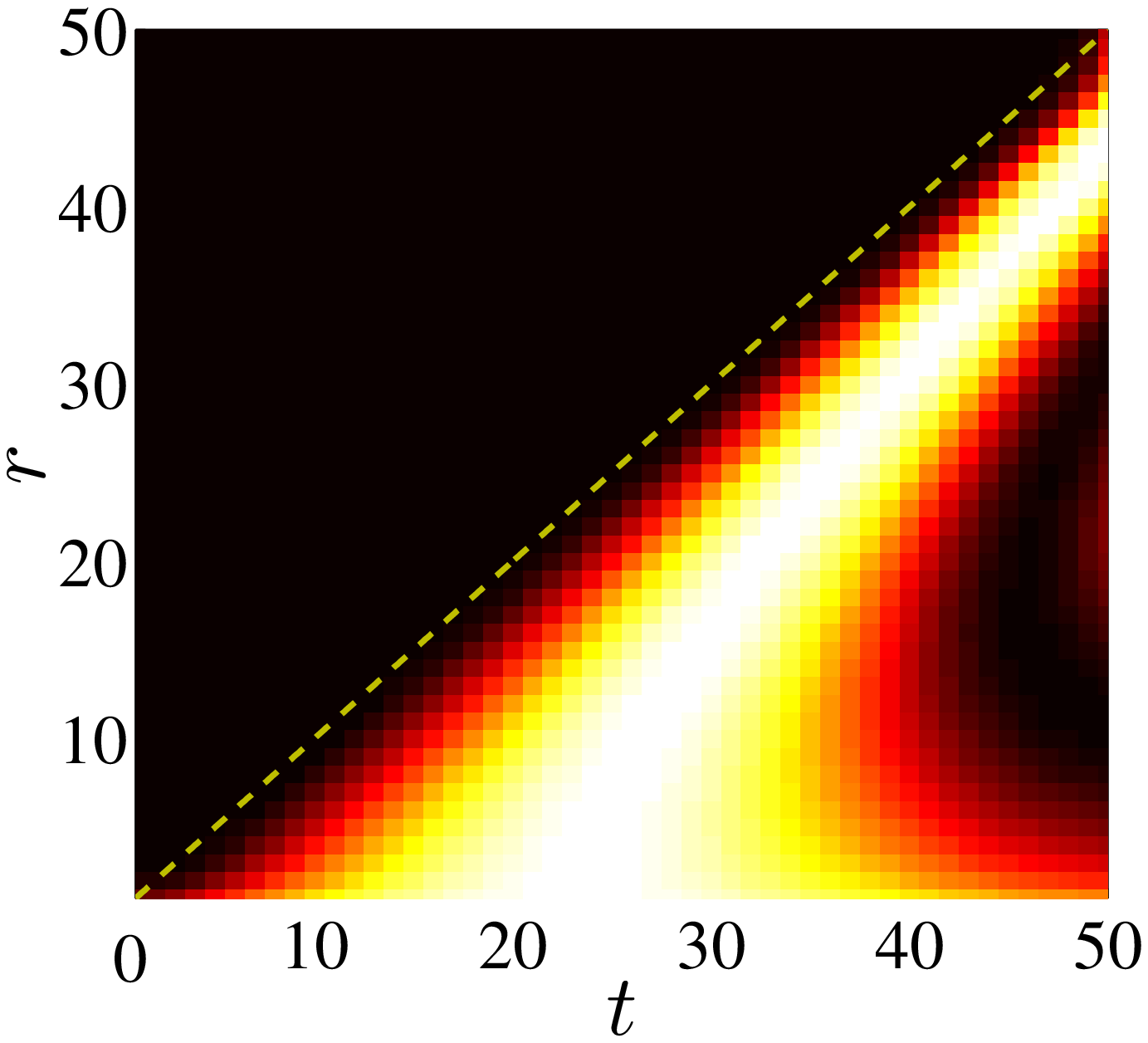}
\end{center}
\caption{(Color online) The density plot of mutual information 
$I_{i,j}$
between two sites with the separation distance $r = |i-j|/2$ in short and long-range harmonic 
oscillators. The total size and initial mass parameter are equal to  $N = 200$
and $m_0=0.3$, respectively. Top
left ($\alpha = 0.4$), right ($\alpha = 1.0$), and bottom left ($\alpha = 1.4$), right ($\alpha = 2.0$).
Yellow (light Gray) regions correspond to
larger value of $I_{i,j}$ and black regions correspond
to smaller value of $I_{i,j}$.
 Dashed lines correspond to the $t_I^* = r^{\alpha/2}$. 
 At time $t^*_I$, the mutual information $I_{i,j}$ 
suddenly changes from zero to nonzero value.  
For $t<t_I^*$ the mutual information $I_{i,j}$ is zero and for 
$t>t_I^*$ it remains nonzero and oscillates due to finite-size effects.
}
\label{mutual information plot}
\end{figure*}

In this section, we study the time evolution of the mutual information 
between two distant point $i$ and $j$ of the harmonic chain  
with long-range couplings. 
 The quantum mutual information is defined as follows
\begin{eqnarray}
I_{A_1,A_2}(t) = S_{A_1}(t) + S_{A_2}(t) - S_{A_1 \cup A_2}(t)~,
\end{eqnarray}
where $A_1$ and $A_2$ are the $i$-th and $j$-th points of the harmonic chain 
respectively. First, we consider the time evolution of the mutual information
$I_{i,j}$ between two sites $i$ and $j$ of the chain of harmonic oscillators with 
short range interaction. In the 
Fig. (\ref{mutual information plot})
one can find the mutual information between two points of 
the chain while increasing the distance. Obviously one can find that 
the mutual information of two particular points with distance 
$|i-j|$, remains nearly
zero for a period of time, until it suddenly changes to the non-zero
value. 

In the Fig. (\ref{mutual information plot}),
 we plot the time evolution of the mutual information between two points
  of the long-range harmonic oscillator with the separation distance
$|i-j| \in [10,60]$. It is interesting to note that the time evolution
 of the mutual information for this case is the same as the harmonic chain with short
 range interaction. This behavior is the key point to understand 
 the linear growth of the entanglement entropy for the long-range harmonic 
 systems with $1<\alpha \leq2$. 
    
In contrast with the previous cases, we observe that for $\alpha <1$ 
the qualitative behavior of the time evolution of the mutual information between 
two distant points of the harmonic chain is completely different from
 the reported one for $\alpha>1$. The Fig. 
 (\ref{mutual information plot}) shows 
 a plot of the mutual information dynamics in harmonic chain with
 strong long-range couplings. It is clear
  that the mutual information grows rapidly from zero
  value at the beginning in which the behavior is independent of the distance separation
  of the two points.

\begin{figure} 
 \centerline{
\includegraphics[scale=0.35]{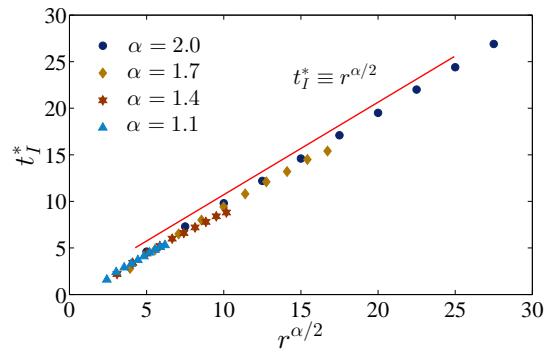}}
\caption{(Color online)
The numerical estimation for $t^*_I$ as a function of distance $r$.
 At time $t^*_I$, the mutual information $I_{i,j}$  of two particular points
 with distance $r = |i-j|/2$ 
suddenly changes from zero to nonzero value.
}
\label{tstar for mutual information for short and longe range oscillator}
\end{figure}

In order to clarify the role of long-range couplings in the mutual
 information dynamics, we measure the time $t^*_I$ which before that 
 time there is no mutual interaction between two points $i$ and $j$ of
  the harmonic chain. For later time $t>t^*_I$ the mutual information 
  $I_{i,j}(t)$ changes to the non-zero value. In our simulation, 
  as shown in Fig. (\ref{tstar for mutual information for short and longe range oscillator}) 
  we observe 
  \begin{eqnarray}\label{t star mutual}
  t^*_I = r^{\alpha/2}~,
  \end{eqnarray}
 where $r = |i-j|/2$ is the 
  separation distance. It is worth
mentioning that in Eq. (\ref{t star mutual})
we find the same scaling behavior as Eq. (\ref{saturation time eq for LRHO}).

It is worth mentioning that the similar
 regimes are reported in Ref. \cite{schachenmayer2013entanglement} for 
 spin chains with variable range interaction. As it was previously mentioned,
  in Ref. \cite{schachenmayer2013entanglement} they studied the time
 evolution of the mutual information between spins in the transverse field Ising model 
 with long-range interactions ($J_{i,j}\propto |i-j|^{-\sigma}$) which 
 shows different regimes as a function of $\sigma$. 
 It is interesting to note that they found that distant spins
become entangled instantaneously for spin chains with strong couplings for $\sigma <1$.
 In contrast, we found that in our long-range harmonic chains, two 
 different sites become entangled instantaneously when $\alpha<1$ ($\sigma <2$).

\section{Conclusion and remarks}\label{conc sec}
We have investigated a detailed study of the time evolution of
 entanglement entropy and mutual information that results after a global quench
in a chain of harmonic oscillators with short and long-range couplings. 
To do so first we have proposed an efficient method to numerically compute 
the time evolution of the von Neumann and R\'enyi entropies after a global quench in a general 
$d$-dimensional hamiltonian. 
For example we presented a detailed numerical evaluation for 
the entanglement dynamics in short-range harmonic oscillators.  
All our results are in good agreement with the theoretical predictions.
Another interesting question to study is the time evolution of
entanglement entropy in the harmonic chain with long-range couplings. 
We found that a regime of quadratic and linear entanglement entropy
growth is present even for long-range coupled harmonic oscillators with 
$\alpha >1$. For the strongly coupled long-range systems
 with $\alpha <1$ we found a regime of logarithmic 
 entanglement entropy growth. 
 However, it is important to mention that it does not look like possible 
 to explain this 
 behavior with the energy quasiparticle picture
  of a global quench. Note that there
 is no maximum group velocity for those long-range systems that have been
 quenched to critical point of the system. We have also determined the
 time evolution of the entanglement entropy for different initial states.
 It is remarkable that the logarithmic growth of
 entanglement for strongly
 long-range coupled harmonic oscillators affected by the initial state. 
 Focusing on the long-range harmonic oscillators we observed 
 that the mutual information dynamics after the global quench 
 exhibits numerous interesting
dynamical behaviors. In contrast to the weakly coupled 
long-range harmonic oscillators, different sites become entangled 
instantaneously when the system is strongly coupled with $\alpha<1$. 

For future studies, it could be of interest to study the
time evolution of entanglement entropy and mutual information
in  harmonic oscillators with short and long-range couplings 
on higher dimensional lattices.

\section*{Acknowledgments}
 MGN kindly acknowledges numerous discussions with
 P. Calabrese, B. Doyon and M. van den Worm. MAR thanks FAPESP for financial support.

\appendix
\section{Gapfull power-law initial state}\label{appendix sec 0}
\begin{figure} 
 \centerline{
\includegraphics[scale=0.35]{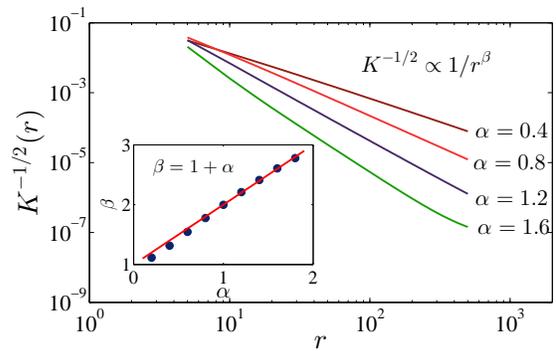}}
\caption{(Color online)
Numerical evaluation of $K^{-1/2}(r)$ vs $r$ for power-law initial state
shows the scaling behavior $K^{-1/2}(r)\propto 1/r^\beta$.
Inset: Scaling exponent $\beta$ as a function of $\alpha$. The
solid red line represents $\beta = 1+\alpha$. 
}
\label{power law correlation length}
\end{figure}

In this Appendix we discuss the numerical evaluation of the 
correlation length for the hamiltonian 
(\ref{interaction kernel long range}). 
   We performed numerical integration Eq. (\ref{HOLRcorr}) to 
 find the correlation length. 
 In the
Fig. (\ref{power law correlation length}), $K^{-1/2}(r)$ is plotted vs $r$ using logarithmic scales.
 The figure shows that $K^{-1/2}(r)$ scales
as $1/r^{\beta}$. 
In the inset of Fig. (\ref{power law correlation length}) we depict the
scaling parameter $\beta$ as a function of $\alpha$, 
which indicates $\beta =1+\alpha$. 
This power-law behavior specifies that the correlation length 
$\xi_s \equiv
 \left[\lim_{r\rightarrow\infty}\frac{\log(r)}{r}\right]^{-1}$ 
  remains infinite even at $m\neq 0$. 
  
\section{Details on numerical calculations}\label{appendix sec}
The entanglement dynamics for coupled harmonic oscillators 
can be efficiently studied by the algorithm proposed in section \ref{qq section}. 
In this study a discrete
scheme has been developed to simulate the entanglement dynamics of 
the short and long-range 
harmonic oscillator. 
 We hereby, require a chain of $N$ harmonic oscillators described by 
 Eq. (\ref{interaction kernel long range}).
It was pointed out in the previous sections that to study the time evolution 
of the von Neumann and R\'enyi entropies,
 the system is prepared in the 
ground state of a massive hamiltonian $H(m_0)$ and at time $t = 0$
the parameter $m_0$ is changed suddenly to a different value
$m$. It is important to note that the Eqs. (\ref{entropy case 1})
 and (\ref{renyi entropy scaling case 1}) have been
proved for the system with size $N\to \infty$ and $m= 0$.
Here we will discuss the dependence of our results on finite 
$N$ and $m\neq 0$. In other words, it will be
interesting to check whether and how the numerical results
will be affected by the changes in chain size $N$ and mass
 parameter $m$. 
 
Let us first consider the dispersion relation of the 
 hamiltonian Eq. (\ref{interaction kernel long range}) as: 
\begin{eqnarray}\label{dispersion finite system}
\epsilon _k = \sqrt{\left[ 2-2\cos(ka) \right]^{\alpha/2}+m^\alpha}~,
\end{eqnarray}
where $a$ is the lattice constant and $m$ is the mass parameter. 
In the previous sections the lattice constant had been set equal to one.

Note that, because of the finiteness of the number of oscillators $N$, 
the energy spectrum is quantised, which the resulting quantization is
$k = k_n = 2n\pi/Na$ and $n\in [0,N-1]$.

 It would also be instructive to extract the 
 group velocity of the quasiparticle excitations which is defined as
 \begin{eqnarray}\label{group velocity finite system}
 v_{g}(k_n) = d\epsilon_{k_n} /dk_n = \frac{\alpha |\sin(k_n)|^{\alpha -1}}{2\sqrt{|\sin(k_n)|^{\alpha}+m^\alpha}}\cos(k_n)~.
 \end{eqnarray}
\begin{figure} 
 \centerline{
\includegraphics[scale=0.35]{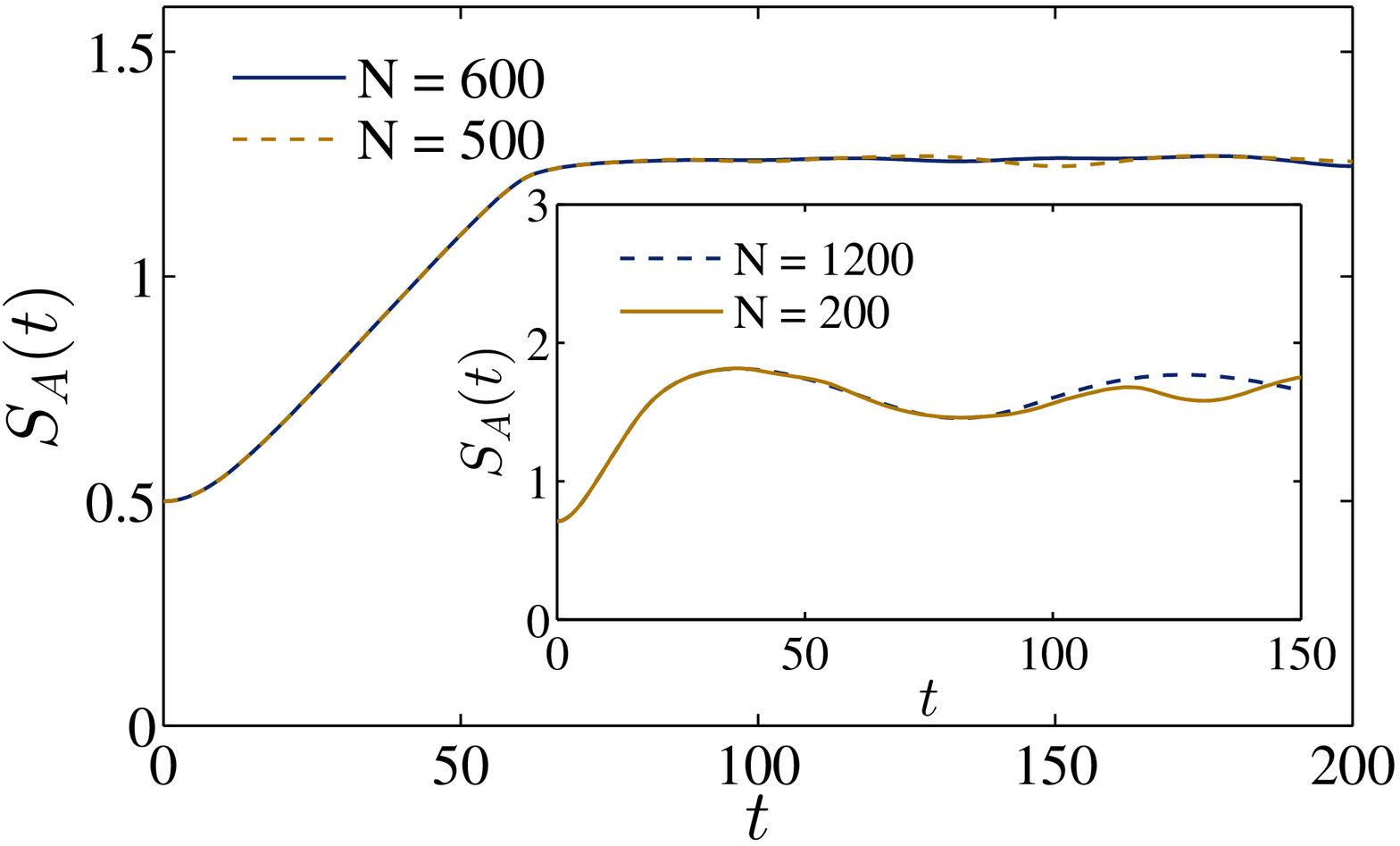}}
\centerline{
\includegraphics[scale=0.35]{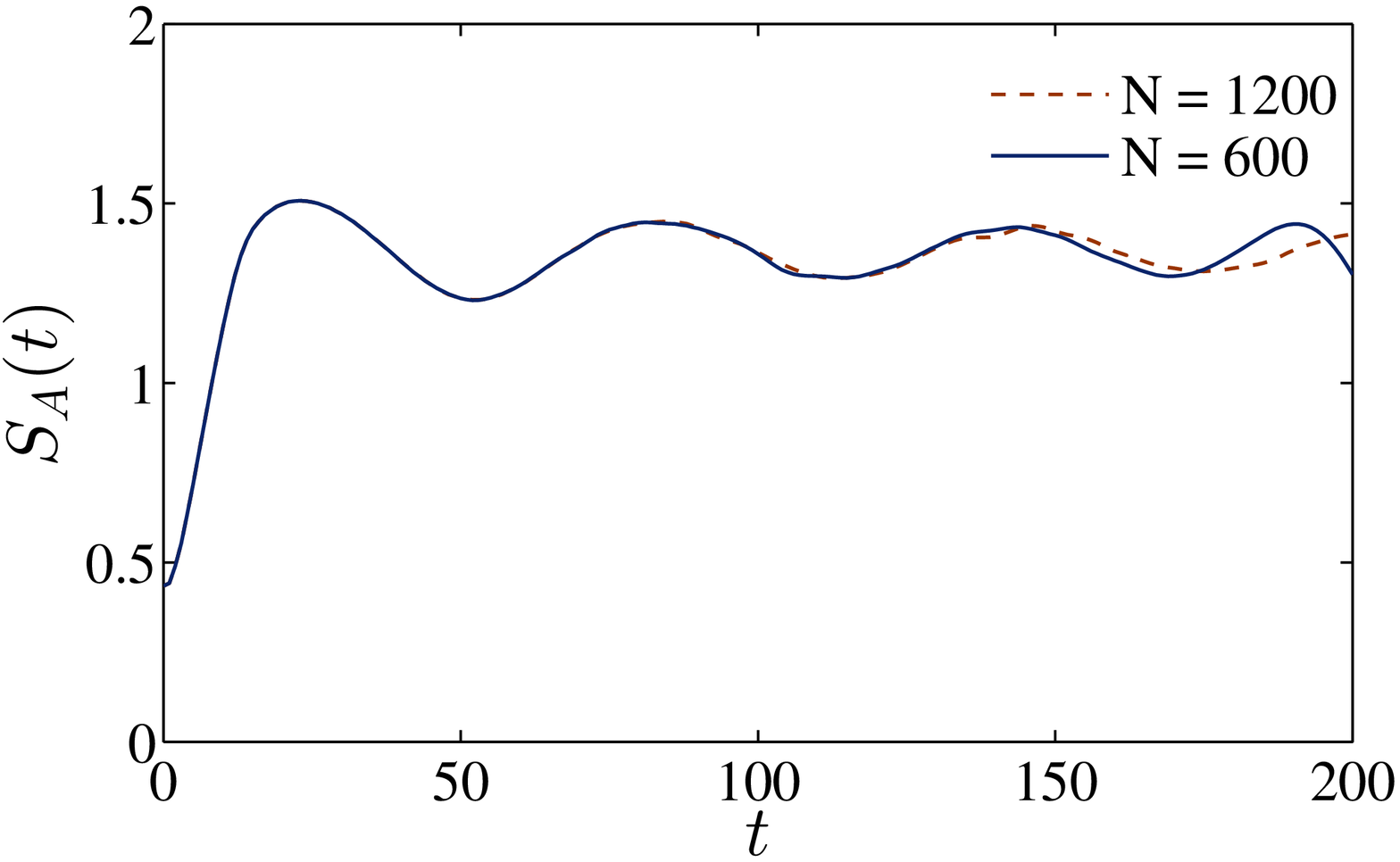}}
\caption{(Color online)
Top: Entanglement entropy dynamics $S_A(t)$ in short range harmonic
 oscillator with the configuration $\mathfrak{g}_2$ and different system size $N$ 
 ($l = 30$ and $m_0 = 0.05$). The same results for the configuration 
 $\mathfrak{g}_1$ ($l = 30$, $m_0 = 0.12$) shown in the inset. Bottom: 
 $S_A(t)$ for long-range harmonic
 oscillator ($\alpha = 1.5$) with different system size $N$. Notice that 
 the results do not depend on $N$.  
}
\label{pic for change of chain size}
\end{figure} 
 
For the massless case ($m=0$) the maximum group velocity
 can be found for $k\sim 0$. It is straightforward to calculate that 
  $\lim _{k_n \to 0}v_{g}(k_n)  \approx \frac{\alpha}{2} |k_n|^{\alpha/2-1}$
which means that there is no maximum allowed velocity of
the quasiparticles for $0<\alpha<2$. It is clear that for 
the short-range harmonic oscillators even for the 
system with finite size, the maximum group velocity of excitations 
is equal to unity ($\max \left[v_{g}(k)\right] = 1$). 
 
\begin{figure}
\centerline{
\includegraphics[scale=0.35]{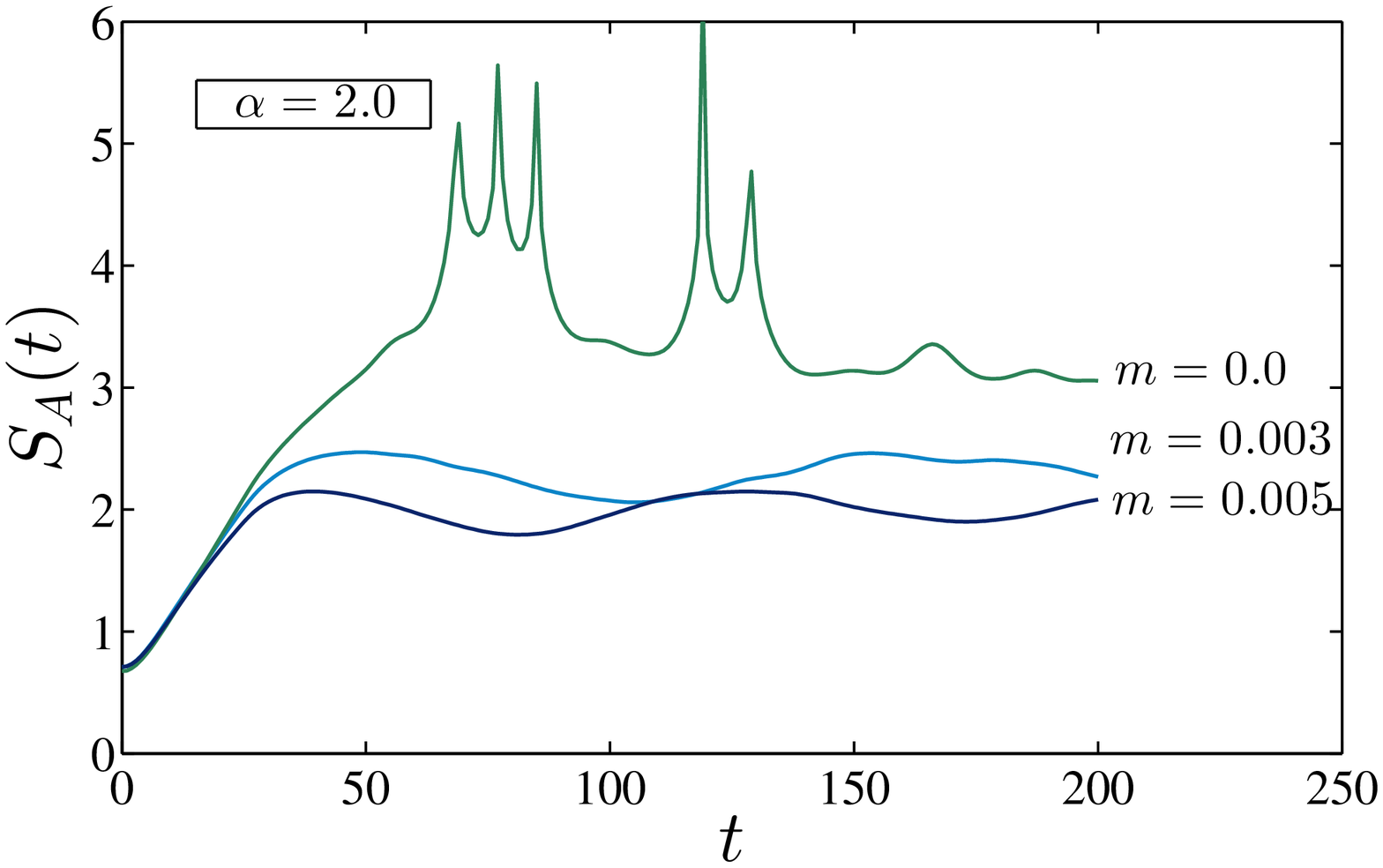}
}
\centerline{
\includegraphics[scale=0.35]{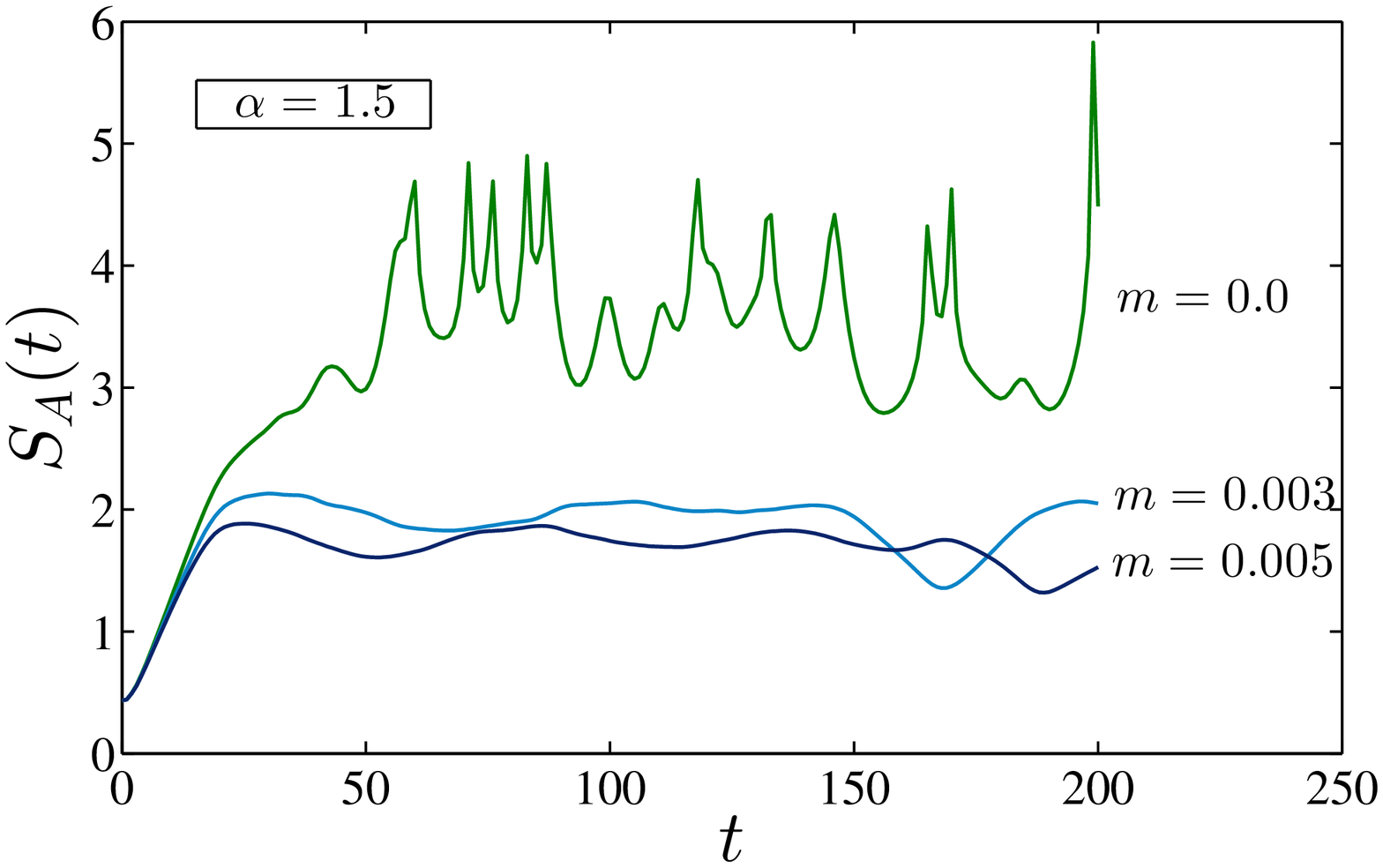}
}
\centerline{
\includegraphics[scale=0.35]{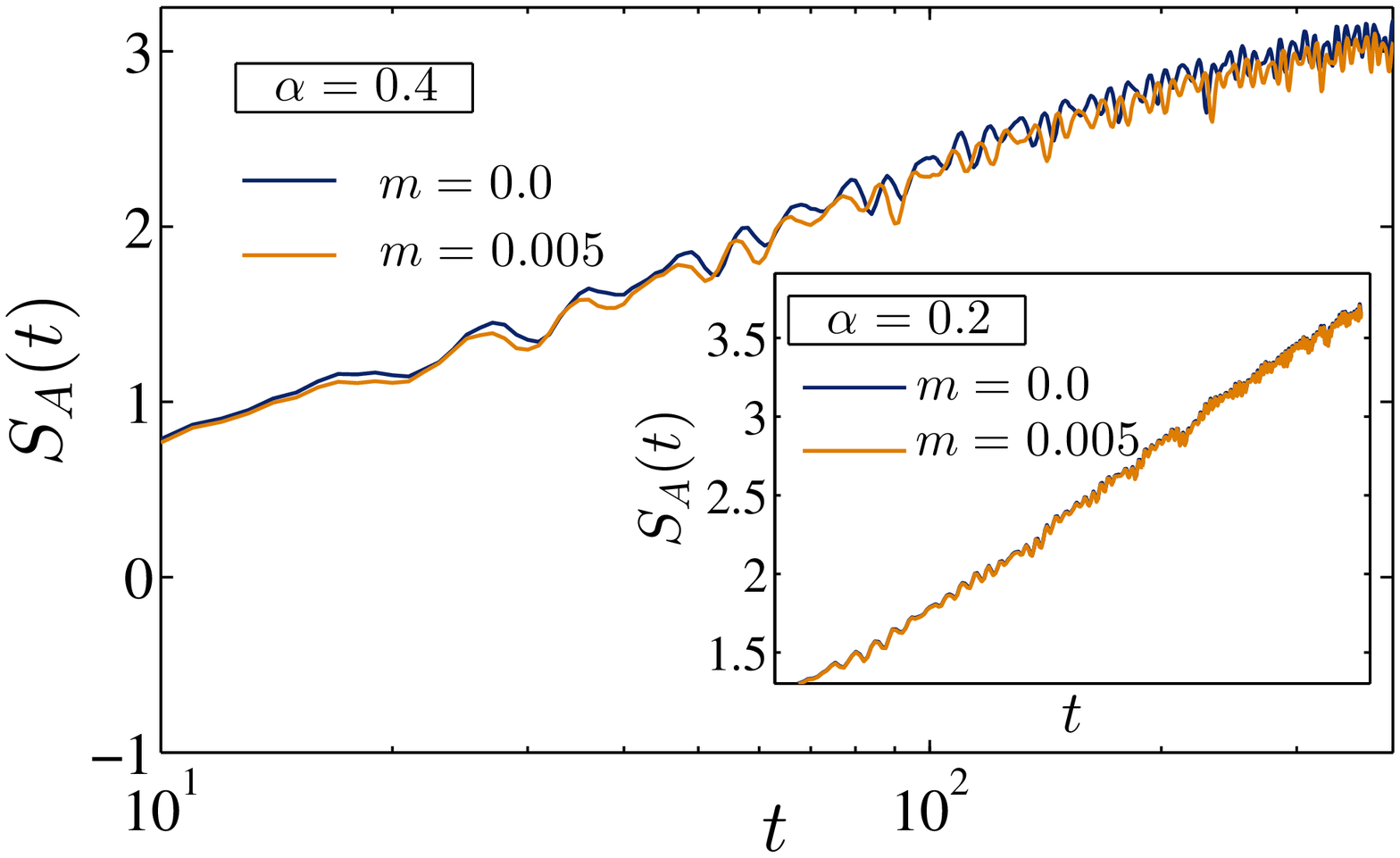}
}
\caption{(Color online) Top: Entanglement entropy dynamics $S_A(t)$ in short-range harmonic
 oscillator with the configuration $\mathfrak{g}_1$. 
 The total size of the system is $N = 500$, the mass 
 parameter $m_0=0.12$ 
 and the subsystem size $l = 50$. Different lines corresponds to 
 different values of the small mass parameter $m$. 
 Middle: The same results for long-range harmonic
 oscillator with $\alpha=1.5$ with the configuration $\mathfrak{g}_1$, 
 the system is $N = 500$, the mass 
 parameter $m_0=0.12$ 
 and the subsystem size $l = 50$. 
 Bottom: Entanglement entropy dynamics $S_A(t)$ for long-range harmonic
 oscillators with strong coupling $\alpha = 0.4$ and $\alpha = 0.2$. 
  }
\label{EE change mprime}
\end{figure}
  
If the system is gapped, based on the Eq. (\ref{group velocity finite system})
 there is no maximum group velocity also for the harmonic chain with 
 strong long-range couplings ($\alpha<1$). 
We are mainly interested in the specific
dependence of the maximum group velocity of excitations for those harmonic oscillators
 with $1 \leq \alpha \leq 2$, on the mass parameter $m$ and lattice constant $a$. 
 The numerical results show that the maximum group velocity does not change
  with $N$ for large chain sizes. We checked numerically $N > 500$ is large enough
   to find the $N$-independent value for the maximum group velocity. Then 
without changing the results
one can safely analyses the maximum value of the Eq. (\ref{group velocity finite system}) 
 in the limit $N\to \infty$. The Eq. (\ref{group velocity finite system})
   becomes especially simple in the scaling limit:
\begin{eqnarray}
v_{g} = \frac{\alpha |k|^{\alpha-1}}{2\sqrt{|k|^\alpha +m^\alpha}}~.
\end{eqnarray} 
Then by simple algebra one can show that 
 the maximum group velocity for the
dispersion relation Eq. (\ref{dispersion finite system}) 
has the following form:
\begin{eqnarray}\label{max group velocity finite massive system}
v_{g}^{max} = \frac{\alpha}{2}(2/\alpha-1)^{1/2}\left(\frac{\alpha -1}{1-\alpha/2}\right)^{\frac{\alpha-1}{\alpha}} m^{\alpha/2 -1}~.
\end{eqnarray}

Let us now consider the $m$ and $N$ dependence in time evolution of the 
entanglement entropy. In Fig. (\ref{pic for change of chain size}) 
we provide an examples of typical behavior of 
entanglement dynamics $S_A(t)$ for short range harmonic oscillator
with configurations $\mathfrak{g}_1$ and $\mathfrak{g}_2$ and different values of 
the system size $N$. It is not surprising that our results are independent
 of $N$. As shown in the same figure the entanglement dynamic for harmonic
 oscillators with long-range couplings is $N$ independent.

It is important to note that when the the system is very
 large and the sub system is a small portion with length $l$ ($\mathfrak{g}_1$) 
 our
numerical results for $m =0$ show many oscillations in the saturation
regime (see Fig. (\ref{EE change mprime})). Therefore it is very hard to analysis the entanglement dynamics 
after saturation regime. The parameter $m \ll m_0$ has been chosen to 
decrease these oscillations. Figure (\ref{EE change mprime}) 
shows the entanglement dynamics
 $S_A(t)$ in the case which $m_0=0.12$ 
and $m\neq 0$.
\begin{figure}
\centerline{
\includegraphics[scale=0.35]{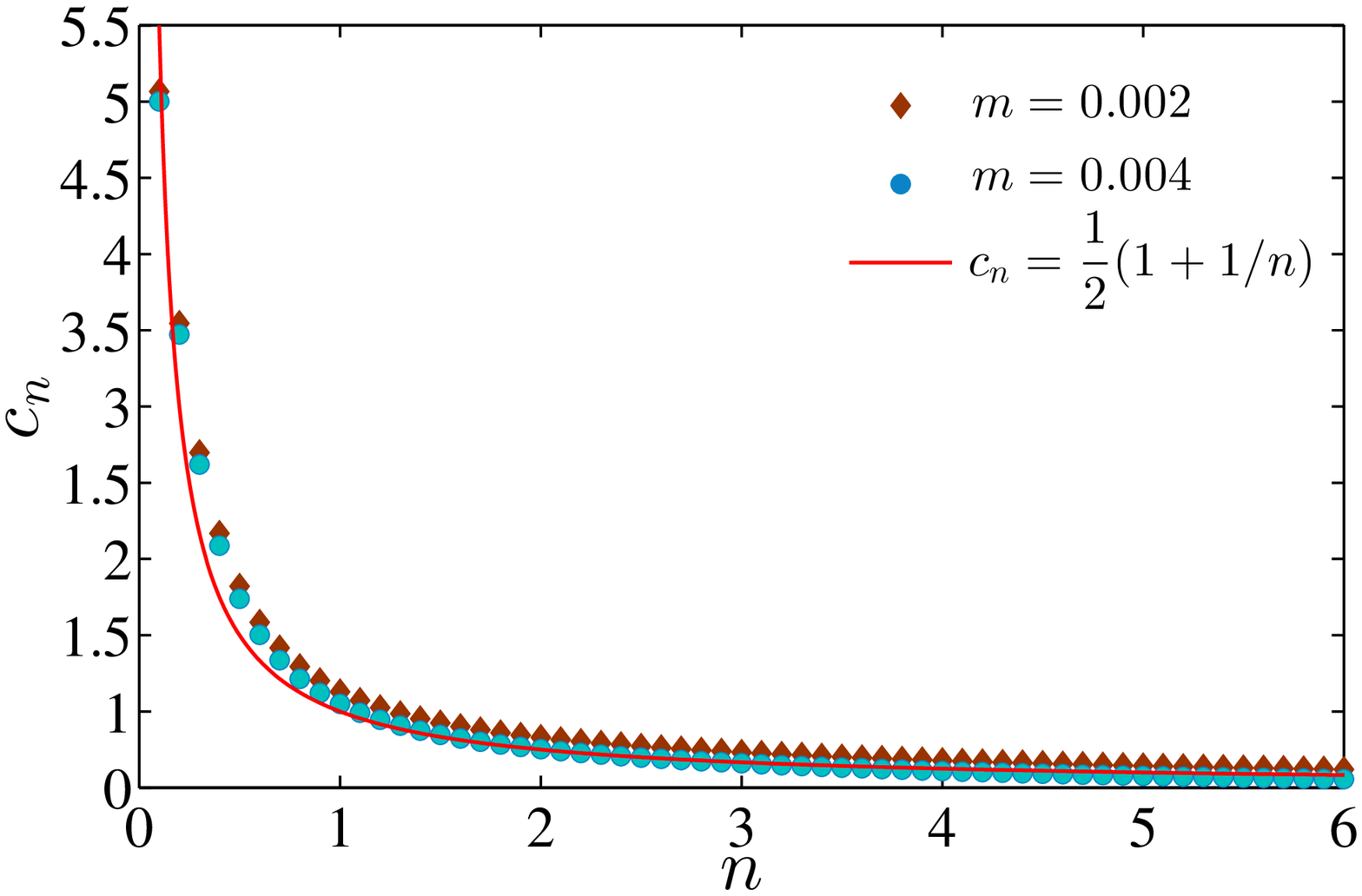}
}
\centerline{
\includegraphics[scale=0.35]{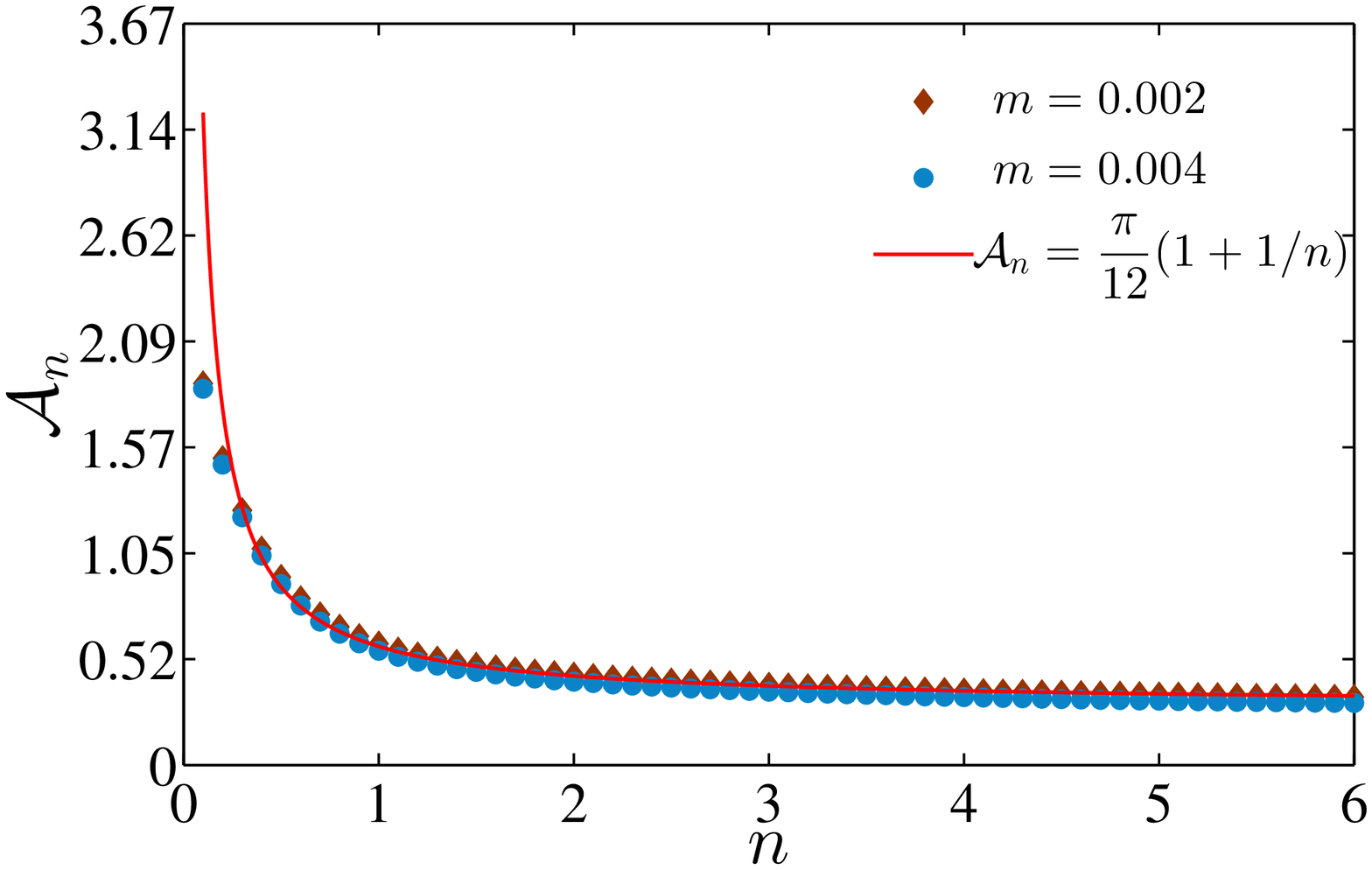}
}
\caption{(Color online) 
Top: The scaling prefactor $c_n$ for the
entanglement dynamics in short-range harmonic
 oscillator before saturation regime (see Eq. (\ref{renyi entropy scaling case 1})). Different
 symbols correspond to different values of the small mass parameter $m$.  
Bottom: The scaling prefactor $\mathcal{A}_n$ for the
entanglement dynamics in long-range harmonic
 oscillator ($\alpha = 1.5$) before saturation regime (see Eq. (\ref{renyi dynamic for LRHO})). Different
 symbols correspond to different values of the small mass parameter $m$.  
  }
\label{prefactors for diffrent m}
\end{figure} 
\begin{figure}[htp]
\centerline{
\includegraphics[scale=0.35]{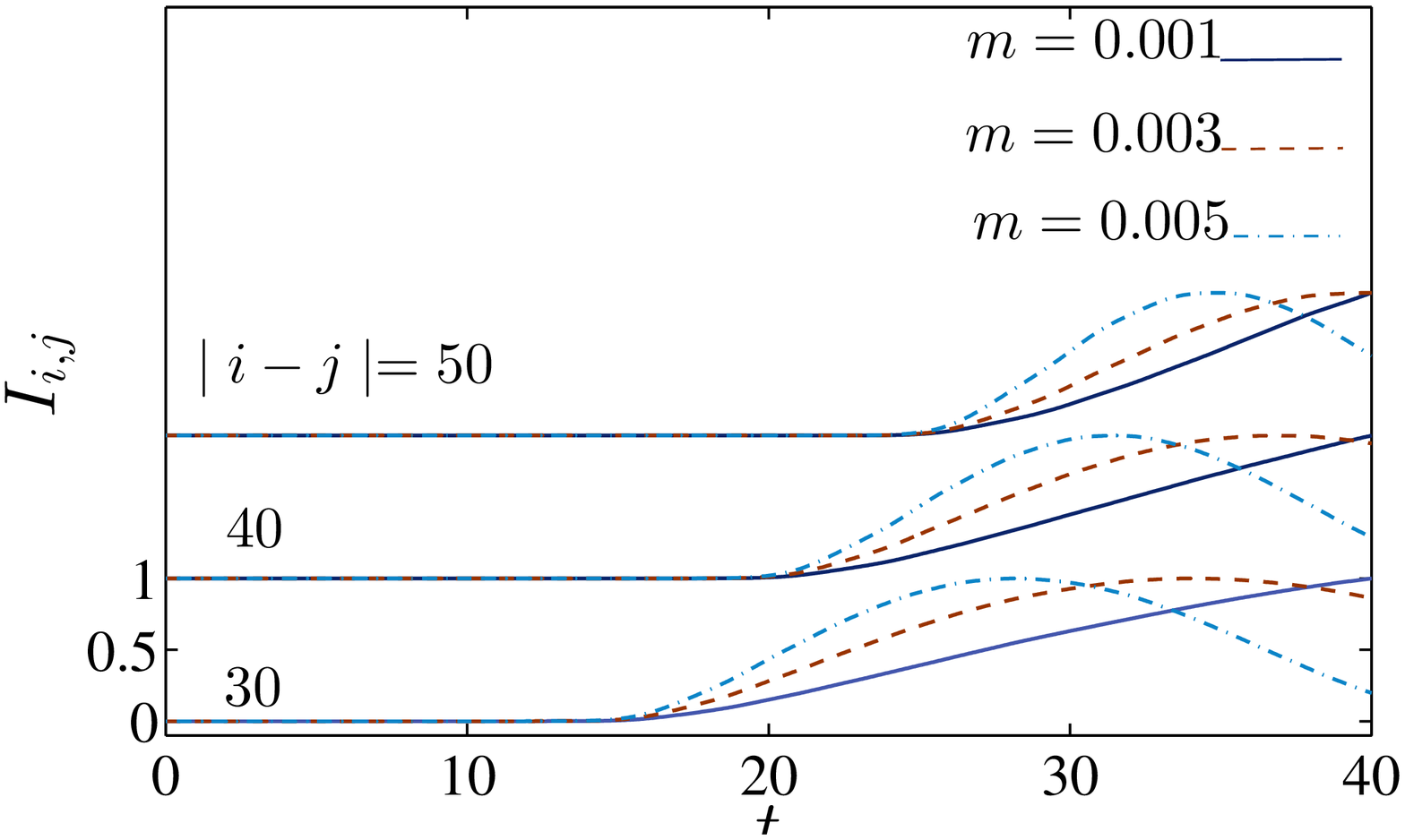}
}
\centerline{
\includegraphics[scale=0.35]{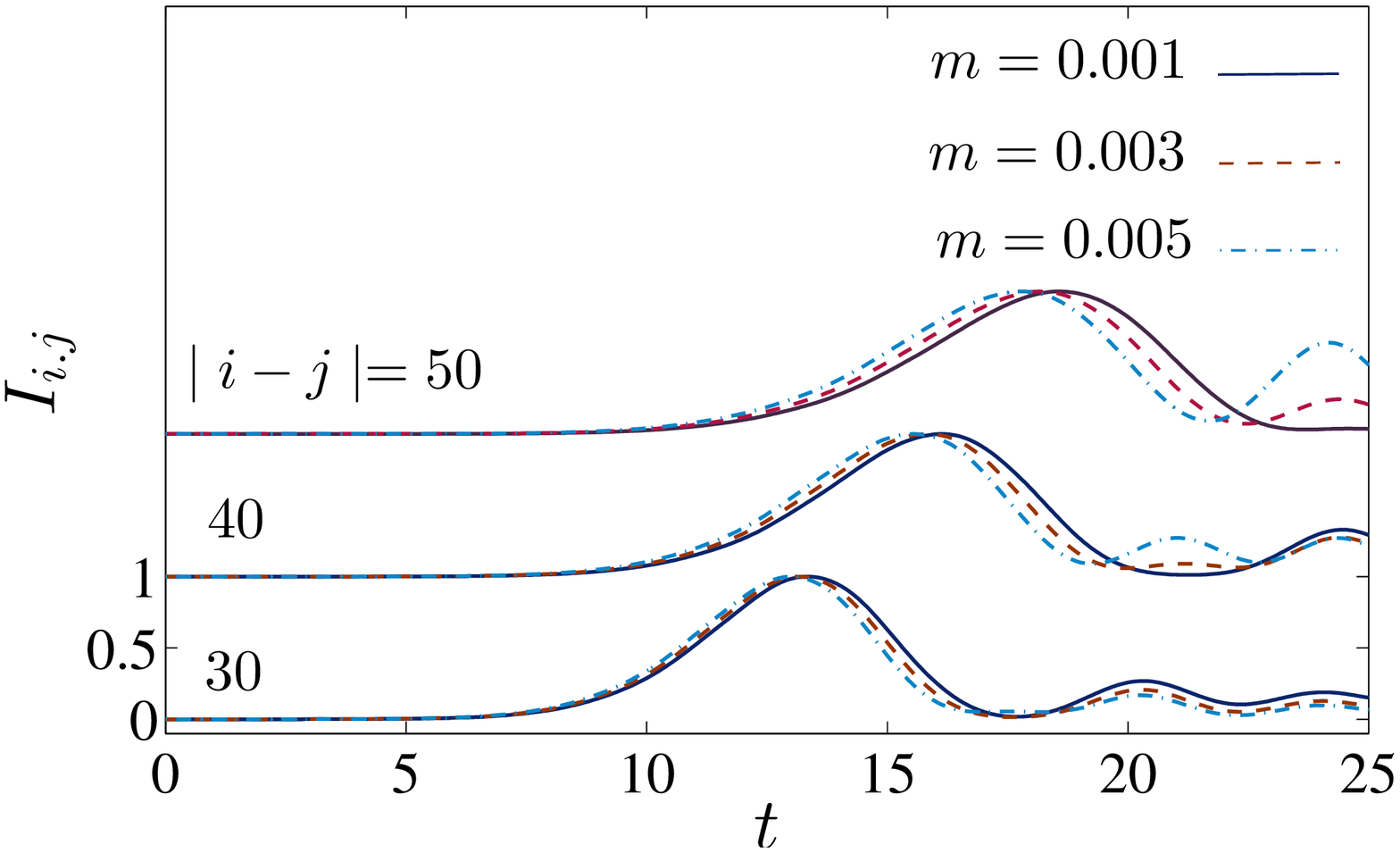}
}
\caption{(Color online) 
Top: The time evolution of the mutual information $I_{i,j}$ 
between two sites $i$ and $j$ in short range harmonic oscillators
with the total size $N = 200$, mass parameter $m_0=0.3$ and different values of
the small parameter $m$ and distance $\mid i-j\mid$.
Bottom: The same results for long-range harmonic
 oscillator ($\alpha=1.5$).
  }
\label{mutual change mprime}
\end{figure}
Let us now discuss the role of non-zero $m$ on the final results. 
It is important to note that we are not allowed to
choose arbitrary value for $m$.
 First,
 it should be smaller than the initial mass parameter $m_0$. Secondly
  one can choose $m\sim \mathcal{O}(1/N)$
  which  $N$ is the system size. It is because
  of the finite-size effect in the
   the correlation length $\xi$.
 However, the correlation length is expected to diverge
  at critical point ($m=0$), for a finite system it 
    is comparable to the system size $\xi \propto m^{-1} \sim \mathcal{O}(N)$.       
 We would also like to point out that all our results are consistent with 
 this picture as long as $m\ll m_0$ and $m\sim 1/N$. It is 
 worth mentioning that the results were affected significantly by 
 enough amount of mass term ($m\gg 1/N$). This means that the entanglement 
 dynamics is independent of the maximum group velocity 
 Eq. (\ref{max group velocity finite massive system}) for $m\gg 1/N$. 
 It is worth mentioning that the same results observed for the
  harmonic chain with long-range couplings as well as short-range harmonic 
  oscillators. In Fig. (\ref{prefactors for diffrent m}) we compare the
  numerical results of the prefactors $c_n$ and $\mathcal{A}_n$ 
  (see Eq. (\ref{renyi entropy scaling case 1}) and (\ref{renyi dynamic for LRHO}) respectively) for 
  different values of small mass parameter $m$ after quench. 
  It is clear that the numerical results are $m$-independent.

Finally we would like to note that we did the same calculations for the 
mutual information dynamics $I_{i,j}(t)$ between 
two point $i$ and $j$ of the lattice sites.
Note that for different values of the chain length $N$
  we got the previous results, as it should and the results were independent 
  from the system size $N$. In order to show the validity and reliability 
of our results, in Fig. (\ref{mutual change mprime}) we report the 
time dependent mutual information results for the short and
long-range harmonic oscillator with fixed values of $m_0 = 0.3$ and 
different values of 
the the parameter $m$. It is evident that the starting point which
 the mutual information changes from zero to non-zero value, is independent
 from $m$. We again 
  found the same result 
 $t^*_I$ for the different values of
 the small parameter $m$ which is clearly obeys the prediction
  $t^*_I = r^{\alpha/2}$ where $r = |i-j|/2$.

\end{document}